\documentclass[11pt, a4paper]{article}
\usepackage{amsmath, amssymb, amsthm, enumerate,bbm,eucal,hhline,stackrel}

\newcommand{\nalt}{\tilde{n}}
\newcommand{\kalt}{\tilde{k}}
\newcommand{\lalt}{\tilde{\ell}}
\newcommand{\etaalt}{\tilde{\eta}}
\newcommand{\domain}{{\cal D}}
\newcommand{\psym}{\bar}
\newcommand{\Alteta}{d}

\newcommand{\cu}{\textbf{u}}
\newcommand{\cv}{\textbf{v}}
\newcommand{\ce}{\textbf{e}}
\newcommand{\cE}{\textbf{E}}

\newtheorem{definition}{Definition}
\newtheorem{remark}{Remark}

\usepackage[all,cmtip]{xy}
\usepackage{cite}

\newcommand{\e}{\epsilon}  
\renewcommand{\a}{\alpha}  
\newcommand{\bR}{\mathbbm{R}}  
\newcommand{\p}{\partial}  
\newcommand{\ra}{\rightarrow}

\newtheorem{Pb}{Problem}

\newtheorem{Th}{Theorem}  
\newtheorem{Prop}{Proposition}  
  
\newtheorem{Cor}{Corollary}  
\newtheorem{Lem}{Lemma}  
\newtheorem{Def}{Definition} 
\newcommand{\bP}{\begin{Pb}\ \ } 
\newcommand{\eP}{\end{Pb}}  
\newcommand{\bt}{\begin{Th}\ \ }  
\newcommand{\et}{\end{Th}}  
\newcommand{\bp}{\begin{Prop}\ \ }  
\newcommand{\ep}{\end{Prop}}  
\newcommand{\bc}{\begin{Cor}\ \ }  
\newcommand{\ec}{\end{Cor}}  
\newcommand{\bl}{\begin{Lem}\ \ }  
\newcommand{\el}{\end{Lem}}  
\newcommand{\bd}{\begin{Def}\ \ }  
\newcommand{\ed}{\end{Def}}  
  
\newcommand{\pf}{\noindent{\it Proof:\ \ }}

\newcommand{\be}{\begin{equation}}  
\newcommand{\ee}{\end{equation}}  
  
\newcommand\re[1]{(\ref{#1})}  
\newcommand{\arr}{\begin{array}{rlll}}  
\newcommand{\ea}{\end{array}}

\begin{document}

\rightline{LTH 1049, \ ZMP-HH/15-18} 

\vskip .5 true cm  
\begin{center}  
{\Large \textbf{Special Geometry of Euclidean Supersymmetry IV: the local c-map}}\\[1em]

{V.\ Cort\'es$^1$, P.\ Dempster$^2$, T.\ Mohaupt$^3$ and O.\ Vaughan$^{1}$} \\[1em] 
$^1${Department of Mathematics and Center for Mathematical Physics\\ 
University of Hamburg\\ 
Bundesstra{\ss}e 55, 
D-20146 Hamburg, Germany\\  
vicente.cortes@math.uni-hamburg.de \\
owen.vaughan@math.uni-hamburg.de}\\[1em]  
$^2${School of Physics \& Astronomy and Center for Theoretical Physics\\
Seoul National University\\
Seoul 151-747, Korea \\
pdemp@snu.ac.kr}\\[1em]
$^3${Department of Mathematical Sciences\\ 
University of Liverpool\\
Peach Street, Liverpool L69 7ZL, UK\\  
thomas.mohaupt@liv.ac.uk}\\[.5em]

\today

\vskip .5 true cm  

\end{center}

\begin{abstract}  

\noindent  
We consider timelike and spacelike reductions of  4D, ${\cal N} = 2$ Minkowski\-an and Euclidean vector multiplets coupled to supergravity and the maps
induced on the scalar geometry. In particular, we investigate (i) the (standard) \emph{spatial $c$-map},  (ii) the \emph{temporal $c$-map}, which corresponds to the reduction of the Minkowskian theory over time, and (iii) the \emph{Euclidean $c$-map}, which corresponds to the reduction of the Euclidean theory over space. In the last
two cases we prove that the target manifold is para-quaternionic K\"ahler.

In cases (i) and (ii) we construct two integrable complex structures on the 
target manifold, one of which belongs to the quaternionic and para-quaternionic structure, respectively. In case (iii) we construct two integrable para-complex structures, one of which belongs to the  para-quaternionic structure. 

In addition we provide a new global construction of the spatial, temporal and Euclidean $c$-maps, and separately consider a description of the target manifold as a fibre bundle over a projective special K\"ahler or para-K\"ahler base.

\end{abstract}  
\newpage

\tableofcontents
\newpage

\section{Introduction and summary of results}
\subsection{Background and motivation}

This paper completes the programme started in \cite{Cortes:2003zd} and
continued in \cite{Cortes:2005uq,Cortes:2009cs}, the purpose of which is 
to describe the scalar geometries of Euclidean ${\cal N}=2$
vector and hypermultiplets both without and with coupling to supergravity.
Recall that with the standard (Minkowskian) spacetime signature 
the scalar manifolds of four-dimensional vector 
multiplets are affine special K\"ahler in the absence of
supergravity and projective special K\"ahler in the presence
of supergravity \cite{Sierra:1983cc,Gates:1983py,deWit:1984pk,Strominger:1990pd,Castellani:1990zd,DAuria:1990fj,Craps:1996vk,
Andrianopoli:1996cm,Freed:1997dp,Alekseevsky:1999ts}. The scalar manifolds  
of hypermultiplets in $d\leq 6$ space-time dimensions are
hyper-K\"ahler in the absence of supergravity and 
quaternionic K\"ahler in the presence of it 
\cite{AlvarezGaume:1981hm,Bagger:1983tt,DeJaegher:1997ka}. Together with
the affine and projective special real target manifolds of 
five-dimensional vector multiplets \cite{Gunaydin:1983bi,Alekseevsky:2001if}, 
they form a family of related
geometries which we refer to as special geometries.\footnote{ 
See also \cite{Andrianopoli:1996cm, Cortes:2001kx, Mohaupt:2011ab,
Freedman:2012zz} for reviews of special geometry.}   
In each case the corresponding special geometry exists in 
a `rigid' or `affine' version, which is realised in supersymmetric
field theories not coupled to supergravity, and a `local' or 
`projective' version, which occurs when the respective matter
supermultiplet is coupled to supergravity. 
When constructing supergravity theories using 
the so-called conformal calculus, see \cite{Freedman:2012zz} for
a review, it is manifest that the `local' versions of the 
special geometries are related to special cases of their 
`global' counterparts. In the field theoretic framework, one starts
with a field theory invariant under rigid {superconformal} transformations,
and then gauges the superconformal symmetry to obtain a theory which
is `gauge-equivalent' to a Poincar\'e supergravity theory. The scalar
geometries of the superconformal and of the Poincar\'e supergravity 
theory are related by a so-called {superconformal quotient}. 
Geometrically, the target manifolds of superconformal field theories
admit a certain homothetic action of the group $\mathbbm{R}^{>0}$, $\mathbbm{C}^*$ and $\mathbbm{H}^*/
\mathbbm{Z}_2$ for five-dimensional vector multiplets, four-dimensional
vector multiplets, and hypermultiplets, respectively. We refer to 
such affine special manifolds as {\em conical}, since their metrics 
have the form of a metric cone, at least locally. 
The corresponding `local' special geometry is then obtained by 
dividing out this group action. This motivates the terminology 
of `conic (affine)' and `projective' special geometry, which was introduced
in \cite{Alekseevsky:1999ts} and \cite{Freed:1997dp}, respectively, 
and which we will use in the following. 

Another link between the special geometries is provided by {dimensional
reduction}. 
Reducing  five-dimensional vector multiplets to four-dimensional
vector multiplets and four-dimensional vector multiplets to 
three-dimensional hypermultiplets
induces maps between their scalar manifolds. These come 
in both a rigid and local (or supergravity) version, depending on 
whether the theory is coupled to supergravity.
The (rigid/supergravity) $r$-map relates
(affine/projective) 
special real to (affine/projective) 
special K\"ahler geometry \cite{Cortes:2003zd,deWit:1991nm,Alekseevsky:2008cta},
while the (rigid/supergravity)
$c$-map relates special K\"ahler geometry 
to hyper-K\"ahler or quaternionic K\"ahler geometry
\cite{Cecotti:1988qn,Ferrara:1989ik}.

Throughout the programme \cite{Cortes:2003zd,Cortes:2005uq,Cortes:2009cs}
we have taken the approach of obtaining the
scalar geometries of the Euclidean theories by dimensional reduction 
of Minkowskian theories over time, since this automatically ensures that
the reduced theory is invariant under the Euclidean supersymmetry algebra.
Thus our programme amounts to constructing and studying new versions of
the $r$-map and $c$-map.
It is well known that the spatial and temporal reduction of a given
theory differ by relative signs in their Lagrangians, and in particular 
that temporal
reduction can lead to scalar target spaces with {\em indefinite} Riemannian
metrics. The central observation of \cite{Cortes:2003zd} was that the 
scalar geometries
of Minkowskian and Euclidean vector multiplets of the same dimension are related
systematically by replacing complex structures by {para-complex 
structures}.\footnote{Related observations were already made for ten-dimensional
IIB supergravity in \cite{Gibbons:1995vg}.} %
This is in contrast with four-dimensional Minkowskian and Euclidean hypermultiplets, which have the same target manifolds at least in the local case \cite{Sabra:2015tsa}.
The scalar geometries of four-dimensional Euclidean vector multiplets 
are affine special para-K\"ahler in the rigid case and projective 
special para-K\"ahler in the local case, as shown in \cite{Cortes:2003zd}
and \cite{Cortes:2009cs}. While para-K\"ahler manifolds had been defined
previously in the mathematical literature \cite{Libermann:1952,Libermann:1954}
(we refer to \cite{Cruc:1996} 
for a review of the history of para-complex geometry
and further references), the two
types of special para-K\"ahler geometry were described for the first time
in these references. 
As explained in \cite{Cortes:2003zd}, the natural 
expectation is that after the dimensional reduction of four-dimensional vector multiplets over time the geometry of the resulting three-dimensional Euclidean hypermultiplets is
para-hyper-K\"ahler in the rigid case and para-quaternionic K\"ahler in the
local case. While rigid hypermultiplets were dealt with in \cite{Cortes:2005uq},
it remains to consider local hypermultiplets in order to complete the programme.

As in the corresponding rigid case \cite{Cortes:2005uq}, 
we will obtain in this paper two new supergravity $c$-maps, since we can either
reduce the Minkowskian theory over time, or the Euclidean theory (which was
constructed in \cite{Cortes:2009cs}) over space. We will refer to these
constructions as the \emph{temporal} $c$-map and the \emph{Euclidean} $c$-map,
respectively. Moreover, we will also revisit the standard, `spatial',
$c$-map and thus consider all possible spacelike and timelike reductions
of both Minkowskian and Euclidean four-dimensional vector multiplets
coupled to supergravity. The reason is that as a further main result
we obtain a new global construction of the supergravity $c$-map, which we 
present in a uniform way for all three cases.

The $c$-map was first described in the context of the T-duality between
compactifications of type-IIA and type-IIB string theories with ${\cal N}=2$
supersymmetry \cite{Cecotti:1988qn}. Upon reduction to three dimensions
as an intermediate step, four-dimensional vector multiplets become
three-dimensional hypermultiplets, so that the three-dimensional
theories have two hypermultiplet sectors which only couple
gravitationally. As a result there are two different decompactification
limits, which can be used to relate the four-dimensional IIA and IIB
theories to one another. The hypermultiplet
metrics resulting from dimensional reduction were described explicitly 
in \cite{Ferrara:1989ik}, and it was shown that they are 
quaternionic K\"ahler, as predicted by supersymmetry. In the construction
of \cite{Ferrara:1989ik} it is assumed that the underlying projective
special K\"ahler manifold $\bar{M}$ is a projective special K\"ahler domain, that is defined by a single
holomorphic prepotential, which is sufficient 
to obtain a local description of the resulting quaternionic K\"ahler manifold
$\bar{N}$. This leaves open the question of how to describe the $c$-map
globally if $\bar{M}$ is not a domain, and how to characterise the
resulting quaternionic K\"ahler metric globally in terms of the
geometric data of $\bar{M}$. A global description is not only preferable
mathematically but also needed for physical questions. In particular, 
in order to understand the full non-perturbative dynamics of
${\cal N}=2$ string compactifications,
one would like to know under which conditions the resulting hypermultiplet
manifolds are complete. Some results on these global questions will be discussed 
below.  

For the rigid $r$-map and $c$-map the global geometrical description
is known. It was already observed in \cite{Cecotti:1988qn}
that the image of an affine special K\"ahler domain $M$ under the
rigid $c$-map can be interpreted as its cotangent bundle $T^*M$. 
More generally, affine special real and affine special K\"ahler 
manifolds are by definition equipped with  
a flat connection, which allows
their tangent bundle to be decomposed  into a horizontal and a vertical
distribution. This can be used to show that the tangent bundle
(equivalently, the cotangent bundle) of an affine special real or affine special K\"ahler 
manifold naturally carries the
structure of an affine special (para-)K\"ahler or of a
(para-)hyper-K\"ahler manifold, respectively 
\cite{Alekseevsky:1999ts,Cortes:2009cs,Alekseevsky:2008cta}.

Given that the affine and projective special geometries are related
by superconformal quotients, one may ask whether it is 
possible to express the supergravity $c$-map in terms of the rigid $c$-map, 
applied to the
associated conical affine special K\"ahler manifold. 
In physical 
terms this amounts to `lifting the supergravity $c$-map to the
superconformal level', which was investigated in \cite{Rocek:2005ij}
and \cite{deWit:2006gn}. Both constructions give rise to an 
off-shell realisation of the $c$-map in terms of tensor multiplets. 
Being off-shell means that supersymmetry is realised independently of
the equations of motion by the inclusion of auxiliary fields. This
has in particular the advantage that the problem of adding higher
derivative terms is tractable. Tensor multiplets are related to 
hypermultiplets by a duality transformation. The corresponding relation between
the K\"ahler and quaternionic K\"ahler metrics is as follows:
The potential for the tensor multiplet metric is related to the
prepotential of the special K\"ahler metric by a contour integral.
Performing a Legendre transform on the tensor multiplet potential
one obtains a hyper-K\"ahler potential for the hyper-K\"ahler cone
(or Swann bundle) over the quaternionic K\"ahler manifold, which
encodes the quaternionic K\"ahler metric \cite{Rocek:2005ij}. 

Another approach to relating the supergravity $c$-map to the
rigid $c$-map, and similarly, the supergravity $r$-map to the
rigid $r$-map was described in \cite{2012arXiv1205.2964A}.
Here the idea is to find a construction, dubbed `conification',
which allows one to obtain the image of the 
supergravity $c$-map (supergravity $r$-map)
by conification of the image of the 
rigid $c$-map (or $r$-map) followed by a superconformal quotient.
A general construction for the conification of 
K\"ahler manifolds and hyper-K\"ahler manifolds (satisfying 
certain technical conditions)
was given. While the conification of (pseudo-)K\"ahler manifolds
leads to a new K\"ahler/K\"ahler (`K/K') correspondence, the 
conification of (pseudo-)hyper-K\"ahler manifolds leads to a general (indefinite) version of the 
hyper-K\"ahler/quaternionic K\"ahler (`HK/QK') correspondence 
of \cite{Haydys:2008}, which was also discussed by 
\cite{Alexandrov:2011ac} and \cite{Hitchin:2013iga, 2013arXiv1306.4241H, Macia:2014oca, Macia:2014nka}.
Moreover one obtains
a new explicit expression for the quaternionic K\"ahler metric, which allows one to recover 
the explicit form of the $c$-map metric of \cite{Ferrara:1989ik}  and its one-loop deformation  \cite{RoblesLlana:2006ez} as a special case,
see \cite{2012arXiv1205.2964A,Alekseevsky:2013nua} for details. This method 
provides a direct proof that these metrics are quaternionic K\"ahler, which is independent of supersymmetry
or the proofs in the undeformed case given in \cite{Ferrara:1989ik, Hitchin:2008}. 
As a consequence, one recovers the earlier result  of  \cite{Alexandrov:2011ac},  obtained using twistor methods, 
that applying the QK/HK correspondence (inverse to the HK/QK-correspondence) to the Ferrara-Sabharwal metric one obtains the 
rigid $c$-map metric. 

We remark that since every (para-)quaternionic K\"ahler manifold has an
associated twistor or para-twistor 
space \cite{Gunaydin:2007qq,Alekseevsky:2008}, 
one can also approach the geometry of the $c$-map through the
corresponding twistor spaces. For this approach we refer to the
literature, see in particular \cite{Gunaydin:2007qq}.

Another approach to the global description of the $c$-map is to cover the initial projective special K\"ahler manifold 
by projective special K\"ahler domains, to which one applies the supergravity $c$-map as formulated in 
\cite{Ferrara:1989ik}, and then to check that the resulting quaternionic K\"ahler domains can be consistently 
glued to a quaternionic K\"ahler manifold. It was shown in \cite{Cortes:2011aj} that the quaternionic K\"ahler 
domains take the form $\bar{N} = \bar{M} \times G$,
where $G$ is a solvable Lie group, and that the quaternionic
K\"ahler metric is a bundle metric $g_{\bar{N}} = g_{\bar{M}}
+ g_G (p)$, where $g_G(p)$ is a family of left-invariant metrics
on $G$ parametrised by $p\in \bar{M}$.\footnote{It was already 
observed in \cite{Ferrara:1989ik} that by fixing a point
$p\in \bar{M}$ one obtains a K\"ahler metric.} This was used to prove
that the quaternionic K\"ahler domains obtained by applying the
supergravity $c$-map domain-wise can be glued together such that
resulting manifold has a well-defined 
quaternionic K\"ahler structure. Moreover, it was proved 
in \cite{Cortes:2011aj}, that both the supergravity $r$-map and
the supergravity $c$-map preserve completeness of the Riemannian metrics. 
While complete projective special real curves and surfaces
were classified in \cite{Cortes:2011aj} and \cite{2013arXiv1302.4570C}
respectively, a necessary and sufficient condition for the completeness of a projective special real manifold 
was obtained more recently in 
\cite{2014arXiv1407.3251C}. In fact, it was shown that a projective special real manifold 
$\mathcal{H} \subset \bR^{n+1}$ is complete if and only if it is closed as a subset of $\bR^{n+1}$, a condition
which can be easily checked in many examples. 
Moreover it was shown that any projective special real manifold respecting a generic 
regularity condition  on its boundary is complete. Therefore the composed
$r$- and $c$-map can be used to construct many new examples of
non-homogeneous complete quaternionic K\"ahler manifolds.

Yet another description of the spatial and temporal $c$-map was
obtained in \cite{Mohaupt:2011aa}, where the objective was 
to find a formulation of the temporal $c$-map which is adapted 
to lifting three-dimensional Euclidean supergravity solutions
(`instantons') to four-dimensional stationary supergravity 
solutions (black holes and other `solitons') 
\cite{Mohaupt:2011aa,Errington:2014bta,Dempster:2015xqa}. To maintain 
the symplectic covariance of the four-dimensional theory, 
dimensional reduction was performed without taking the
superconformal quotient in the four-dimensional theory, 
which resulted in the description 
of the (para-)quaternionic K\"ahler manifold $\bar{N}$ in terms 
of a $U(1)$ principal bundle $P \rightarrow \bar{N}$. In this paper we will
extend the local description given in \cite{Mohaupt:2011aa} to the
Euclidean $c$-map. Moreover, we will give a global construction 
of the bundle $P$ and show that it is obtained with all data 
needed to define the (para-)quaternionic structure of $\bar{N}$
in a natural way from the underlying projective special 
(para-)K\"ahler manifold $\bar{M}$ as a one-dimensional extension
of the tangent bundle $TM$ of the associated conical affine
special (para-)K\"ahler manifold $M$. Another approach, left to the 
future, would be to adapt the HK/QK-correspondence to encompass
para geometries.

One ingredient of \cite{Mohaupt:2011aa} which will be 
useful in the present paper is to employ {\em special real coordinates}
for the conical special (para-)K\"ahler manifold $M$. 
Special real coordinates make explicit the flat 
symplectic (rather then holomorphic) aspects of special 
K\"ahler manifolds 
\cite{Cecotti:1988qn,Freed:1997dp,1999math......1069H,Alekseevsky:1999ts}.
From the affine point of view the existence of special real
coordinates is related to the fact that any simply connected
affine special K\"ahler manifold can be realised as a
parabolic affine hypersphere \cite{1999math.....11079B}, 
while the natural $S^1$ bundle
over the associated projective special K\"ahler manifold
carries the structure of a proper affine hypersphere endowed
with a Sasakian structure \cite{Baues:2003}. Analogously,
affine special para-K\"ahler manifolds are intrinsically 
improper affine hyperspheres \cite{Cortes:2005}.
Real coordinates play a central role in the analysis of
black hole partition functions and their
relation to the topological string 
\cite{Ooguri:2004zv,LopesCardoso:2006bg,Cardoso:2010gc,Cardoso:2014kwa}.
The formalism of \cite{LopesCardoso:2006bg,Mohaupt:2011aa} uses
special real coordinates on a conical affine special K\"ahler manifold
to describe the underlying projective special K\"ahler manifold. 
A different approach where special real coordinates are introduced
directly on the projective special K\"ahler manifold was described
in \cite{Ferrara:2006at} (see also \cite{Ferrara:2006js} for a review
of special real coordinates in the affine case).

One aim of our programme is to make explicit the fact that Minkowskian 
and Euclidean theories can be presented in a uniform way.
In \cite{Cortes:2003zd} it was noted that in suitable coordinates
the Lagrangian and supersymmetry transformations of vector multiplets
take exactly the same form in either signature, 
and are only distinguished by interpreting the involution 
$z \mapsto \overline{z}$ as
complex conjugation in the Minkowskian and as para-complex conjugation 
in the Euclidean case. Starting from \cite{Cortes:2009cs} a unified
$\varepsilon$-complex notation was used, where $\varepsilon=-1$
corresponds to the complex and $\varepsilon=1$ to the para-complex
case. This notation will also be used in the present paper. Since,
apart from choosing to reduce a Minkowskian theory 
over space or over time, we can choose to start with a Euclidean theory
in four dimensions, we will need a further refinement of our notation.
Our convention is that whenever we talk about complex/para-complex
manifolds or structures in a generic way, we will use the
symbol $\varepsilon=\pm 1$, whereas $\epsilon_1=\pm 1$ refers to the
geometry of the four-dimensional theory we start with, while 
$\epsilon_2=\pm 1$ distinguishes between reduction over time and reduction
over space.
We will explain more about this notation in the next subsection.

The temporal $c$-map has been studied before in various publications,
mostly in relation to constructing stationary solutions by lifting
Euclidean solutions over time. In \cite{Gunaydin:2005mx} a list of
the symmetric spaces resulting from applying the temporal $c$-map
to symmetric projective special K\"ahler manifolds was given. As 
mentioned in \cite{Gunaydin:2005mx}, these symmetric spaces are
indeed para-quaternionic K\"ahler. This can be proved by either analysing 
the holonomy representation, or by comparing with the classification
of pseudo-Riemannian symmetric para-quaternionic K\"ahler manifolds 
of \cite{Cortes:2004aa}.\footnote{See also Section 3.6 of \cite{MR2681600}.}

\subsection{Main results}

Recall that given a projective special K\"ahler domain $\bar{M}$ 
of dimension $2n$, defined
by a holomorphic prepotential $F$ that is homogeneous of degree two, 
the supergravity $c$-map
assigns a quaternionic K\"ahler metric $g_{\bar{N}}$ on a manifold $\bar{N}$ of dimension $4n + 4$. The target metric is induced by the dimensional reduction of 4D, ${\cal N}=2$ supergravity coupled to $n$ vector multiplets over a spacelike dimension, and was first computed explicitly by Ferrara and Sabharwal in \cite{Ferrara:1989ik}. Henceforth we shall refer to this construction specifically as the {\em spatial $c$-map}. 
It turns out that this metric can be defined even if the
projective special K\"ahler manifold is not defined by a single holomorphic
prepotential, but is rather covered by domains on which such 
prepotentials exist \cite{Cortes:2011aj}. The total space $\bar{N}$ is then interpreted as a bundle over $\bar{M}$, the fibres of which are solvable Lie groups isomorphic to the Iwasawa subgroup of $SU(1,n+2)$.

The main purpose of this paper is to generalise the spatial $c$-map construction.
We will give a different description of the total space $\bar{N}$
as an $S^1$-quotient $\bar{N} = P / S^1$, where $P=TM \times \mathbbm{R}$ 
is the product of the tangent bundle of the $(2n+2)$-dimensional conical affine special K\"ahler manifold $(M,J,g,\nabla,\xi)$ 
underlying $\bar{M}$ with the real line. 
We will assume that $M$ is simply connected in which case one may 
identify $TM = M \times {\mathbbm R}^{2n+2}$ using the flat connection and $P = M \times {\mathbbm R}^{2n+3}$. 
The principal $S^1$-action on $P$ corresponds to the
$U(1)$ subgroup of the natural ${\mathbbm C}^*$-action on the first factor. 
It is locally generated by the trivial extension $Z_P$ to $P$ of the
Killing vector field $J\xi$ on $M$.
An advantage of this construction is that it does not place any restrictions on the
projective special K\"ahler manifold $\bar{M}$, only that the underlying conic affine special K\"ahler manifold $M$ is simply connected.
It can also be adapted to the following two new cases, which is the main goal of this paper: 
\begin{enumerate}[(i)]
	\item \emph{The temporal $c$-map.} \\ 
	This assigns to every projective special K\"ahler manifold of dimension $2n$ a para-quater\-nio\-nic K\"ahler manifold of dimension $4n + 4$. It is induced by the reduction of $4D,\,{\cal N} = 2$ supergravity coupled to vector multiplets over a timelike dimension. 
	\item \emph{The Euclidean $c$-map.} \\ 
	This assigns to every projective special para-K\"ahler manifold of dimension $2n$ a para-quater\-nio\-nic K\"ahler manifold of dimension $4n + 4$. It is induced by the reduction of $4D,\,{\cal N} = 2$ Euclidean supergravity coupled to vector multiplets over a spacelike dimension. 
\end{enumerate}	
This information is summarised in Table \ref{table:Cmap}.
While the explicit form of the target metric of the temporal and Euclidean $c$-maps can  be easily adapted from the case of the spatial $c$-map, 
it is not obvious that the metrics are para-quaternionic K\"ahler. In order to 
prove this we will explicitly compute the Levi-Civita connection and show that it is compatible with an $Sp(2)\cdot Sp(2n, {\mathbbm R})$-structure. 
We will see that the reduced scalar curvature for all $c$-map target manifolds%
\footnote{In the conventions of \cite{Ferrara:1989ik} the reduced scalar curvature is $-1$ for the spatial $c$-map.}
 is equal to $-2$.

It was observed in \cite{Cortes:2011ut} that the target manifold of the spatial $c$-map 
admits a complex structure which is part of the quaternionic K\"ahler structure. We will show
that it also admits a second complex structure which is not part of the quaternionic K\"ahler structure. Similarly, the temporal $c$-map admits two complex structures,
one of which is part of the para-quaternionic K\"ahler structure, and the Euclidean $c$-map 
admits two para-complex structures, one of which is part of the para-quaternionic K\"ahler structure.

\begin{table}
	\centering
		\begin{tabular}{|c|c|c|c|}
			\hline
			 & Base & Target & Spacetime signature \\
			\hhline{|=|=|=|=|}
			spatial & projective special & quaternionic & $(3 + 1)\to(2 + 1)$ \\
			 $c$-map& K\"ahler & K\"ahler &    \\
			\hline
			temporal & projective special & para-quaternionic & $(3 + 1)\to (3 + 0)$ \\
			 $c$-map & K\"ahler & K\"ahler & \\ 
			\hline
			Euclidean & projective special & para-quaternionic & $(4 + 0)\to (3 + 0)$ \\
			 $c$-map & para-K\"ahler & K\"ahler &  \\
			\hline
		\end{tabular}
	\caption{
	Summary of spatial, temporal and Euclidean $c$-maps. For a base manifold of dimension $2n$ the target manifold has dimension $4n + 4$.} \label{table:Cmap}
\end{table}

Let us give a brief summary of our construction for the spatial $c$-map.
In order to define the quaternionic K\"ahler metric 
we must first recall some facts concerning conical affine special K\"ahler manifolds
that can be found in \cite{Alekseevsky:1999ts, Mohaupt:2011aa}. 
Let $(M,J,g,\nabla,\xi)$ be a conic affine special K\"ahler manifold of complex dimension $n + 1$. 
We will assume that $M$ is simply connected and therefore there exists a conic holomorphic nondegenerate Lagrangian immersion $\phi:M \to T^*{\mathbbm C}^{n + 1}$ that is unique up to symplectic transformations. On $M$ there exist $2n + 2$ globally-defined real functions $(x^0 = \text{Re} \, Z^0, \ldots, x^{n} = \text{Re} \, Z^{n}, y_0 = \text{Re} \, W_0, \ldots, y_{n} = \text{Re} \, W_{n})$, where $(Z^I,W_I)_{I=0,\ldots, n}$ are complex linear coordinates on $T^*{\mathbbm C}^{n + 1}$, that satisfy $\omega = g(\cdot, J \cdot) = 2 dx^I \wedge dy_I$ and locally form a $\nabla$-affine coordinate system about any point of $M$ \cite[Thm 9]{Alekseevsky:1999ts}. 
Since the functions $(q^a)_{a=0, \ldots, 2n+1}:=(x^I,y_I)_{I=0, \ldots,
n}$ are unique up to linear symplectic transformations one may uniquely define the following global one-forms on $TM$:
\begin{equation}
q^a \Omega_{ab} dq^b \;,\;\;\; 
\hat{q}^a \Omega_{ab} dq^b \;,\;\;\; 
q^a \Omega_{ab} d\hat{q}^b  \;,\;\;\; 
\hat{q}^a \Omega_{ab} d\hat{q}^b \;, \notag
\end{equation}
where 
\[
(\Omega_{ab}) =\left( \begin{array}{cc}
0 & \mathbbm{1} \\
- \mathbbm{1} & 0 \\
\end{array} \right)
\]
is two times the Gram matrix 
$\omega(\frac{\partial}{\partial q^a}, \frac{\partial}{\partial q^b})$
of $\omega$, i.e.\
$\omega = \Omega_{ab} dq^a \wedge dq^b$, and $(q^a, \hat{q}^a)$ 
are global functions $TM$ associated with the functions $(q^a)$ on $M$. 
The special K\"ahler metric 
$g$ on $M$ is given by the Hessian of the function $H=\frac{1}{2} 
g(\xi, \xi)<0$:
\begin{equation}
g = \nabla d H = H_{ab} dq^a dq^b \;, \notag
\end{equation}
where $H_{ab} = \frac{\partial^2 H}{\partial q^a \partial q^b}$. 
The function $H$ is 
homogeneous of degree two with respect to the
functions $(q^a)$. It is called the {\em Hesse potential} and,
in the real formulation of special K\"ahler geometry, 
plays a role analogous to the holomorphic prepotential. 
The projective special K\"ahler metric $\bar{g}$ is related to $H$
by \cite{Mohaupt:2011aa}
\begin{equation}
h = h_{ab}dq^adq^b := \pi^* \bar{g} = - \frac{1}{2H} \nabla d H + \frac{1}{4H^2} (dH)^2
+ (v)^2
= \nabla d \tilde{H} - (d\tilde{H})^2 
+ (v)^2 \;, \notag
\end{equation}
where $\tilde{H} = -\frac{1}{2} \log(-2H)$ and
$v =  -\frac{1}{H} q^a \Omega_{ab} dq^b = - \frac{1}{2H} J^* d H = 
J^* d \tilde{H}$. Here we have denoted by $\pi: M \rightarrow \bar{M}=M/\mathbbm{C}^*$ 
the canonical projection of the $\mathbbm{C}^*$-action on $M$, which is locally generated 
generated by the vector fields $\xi$ and $J\xi$.

We will now construct the quaternionic K\"ahler metric on $\bar{N}=
P/S^1$. We first remark that the symmetric $(0,2)$-tensor field 
\begin{equation}
h + (d\tilde{H})^2 = 
\nabla d \tilde{H} + (v)^2 \notag ,
\end{equation}
on $M$ 
has one-dimensional kernel $\mathbbm{R} J \xi$. 
Using the canonical projection $TM 
 \rightarrow M$ we may consider any covariant
tensor field (such as $h,H, \ldots $) on $M$ as a tensor field on $TM$. 
Similarly, any covariant tensor field on $TM$ can be considered 
as a tensor field on $P$ by means of the canonical projection 
$P = TM \times \mathbbm{R} \rightarrow TM$. 
In particular we will consider the one-forms on $P$
\begin{align*}
u^1 &= -d\tilde{H} = \frac{1}{2H} H_a dq^a\;, &
\hat{u}^1 &= -\frac{1}{2H} H_a d\hat{q}^a \;, \\
u^2 &= -\frac{1}{2H} \left(d\tilde{\phi} + 2\hat{q}^a \Omega_{ab} d \hat{q}^b\right)  \;, &
\hat{u}^2 &= -\frac{1}{H} q^a \Omega_{ab} d \hat{q}^b\;,
\end{align*}
where $\tilde{\phi}$ is the coordinate in the second factor of $P = TM \times \mathbbm{R}$. 
Let us define on $P$ the symmetric $(0,2)$-tensor field 
\begin{equation}
g' = h_{ab} \left( dq^a dq^b + d\hat{q}^a d\hat{q}^b  \right) 
			+ \left(u^1\right)^2 + \left(u^2\right)^2 + \left(\hat{u}^1\right)^2 + \left(\hat{u}^2\right)^2 \;, \label{eq:gprime}
\end{equation}
which has kernel ${\mathbbm R} Z_P$ and is invariant under the circle group 
$S^1_{Z_P}$. It induces a pseudo-Riemannian metric $g_{\bar{N}}$ on $\bar{N}=P/S^1$, 
which is positive definite if the projective special 
K\"ahler metric $\bar{g}$ is positive definite. 
We will verify later that this metric can be brought to the
standard form of the Ferrara-Sabharwal metric, and is therefore
pseudo-quaternionic K\"ahler.

When we consider the cases of the temporal and Euclidean $c$-map we will find that the
tensor field $g'$ on $P$ differs from the case of the spatial $c$-map \eqref{eq:gprime} only by
certain sign-flips. It is convenient to introduce the parameters $\epsilon_1,\epsilon_2 \in \{+1, -1\}$ which are determined by 
the rule 
\begin{equation}
	(\epsilon_1, \epsilon_2) = 
	\begin{cases}
		(-1,-1) & \text{spatial $c$-map} \\
		(-1,+1) & \text{temporal $c$-map} \\
		(+1,\pm1) & \text{Euclidean $c$-map} \;.
	\end{cases}
	\label{eq:Epsilons}
\end{equation}
When we are not specifically discussing the $c$-map we will use the symbol
\[
	\varepsilon = \text{`generic' epsilon, which can be either $\pm1$}\;.
\]
One may interpret the parameters $\epsilon_1,\epsilon_2$ physically as follows: The choice $\epsilon_1 = -1$ corresponds to 
starting with a theory of 4D, ${\cal N} = 2$ supergravity coupled to vector multiplets with Lorentzian spacetime signature, and $\epsilon_1 = 1$ to the same theory with Euclidean spacetime signature. If $\epsilon_1 = -1$ then $\epsilon_2 = -1$ corresponds to the dimensional reduction of this theory over a spacelike dimension, and $\epsilon_2 = 1$ to dimensional reduction over a timelike dimension. If $\epsilon_1 = 1$ then one must necessarily reduce over a spacelike dimension,  
which corresponds to $\epsilon_2 = -1$. However, as we will explain later, if one chooses instead $\epsilon_2 = 1$ then the resulting target manifold is globally isometric to the case $\epsilon_2 = -1$, and so both choices are mathematically equivalent. Using this notation one may write various expressions in a unified way for all three $c$-maps. For example,
the expression for $g'$ can be written for all $c$-maps as 
\begin{equation}
g' = h_{ab} \left( dq^a dq^b  - \epsilon_2 d\hat{q}^a d\hat{q}^b  \right) 
			+ \left(u^1\right)^2 - \epsilon_1 \left(u^2\right)^2 - \epsilon_2 \left(\hat{u}^1\right)^2 + \epsilon_1 \epsilon_2 \left(\hat{u}^2\right)^2 \;. \notag 
\end{equation}
Note that when $\epsilon_1 = 1$ the tensor $h$ on $M$ is of split-signature on any subspace complementary to its kernel. It is therefore clear from the above expression that $g'$ induces a positive definite metric on $\bar{N}$ only when the metric $\bar{g}$ on $\bar{M}$ is positive definite and $\epsilon_1 = \epsilon_2 = -1$.
For all other choices of $\epsilon_1$ and $\epsilon_2$ it induces a metric of split-signature.

We will also discuss a complementary approach to describing $c$-map target spaces locally as the product 
\[
	\bar{N} = \bar{M} \times G \;,
\]
where $\bar{M}$ is  (a domain in) the original projective special $\epsilon_1$-K\"ahler base manifold and $G$ is
the Iwasawa subgroup of $SU(1, n + 2)$. With respect to this decomposition the metric on $\bar{N}$ can be written as
\[
	g_{\bar{N}} = \bar{g} + g_G(p) \;,
\]
where $\bar{g}$ is the metric on $\bar{M}$ and $g_G(p)$ is a family of left-invariant 
metrics on $G$ that depends on the point $p \in \bar{M}$. We will explicitly show that for fixed $p$ the metric $g_G$ is a symmetric $\epsilon_1$-K\"ahler metric of constant 
$\epsilon_1$-holomorphic sectional curvature.

This paper is organised as follows. 
We begin in Section \ref{sec:KandQKgeom} with a review of background material. 
In Section \ref{sec:eKmfs} we discuss $\varepsilon$-complex vector spaces, spaces of $\varepsilon$-complex lines and how these can be represented as symmetric spaces and realised as solvable Lie groups, and special $\varepsilon$-K\"ahler manifolds. 
In Section \ref{sec:QK} we discuss $\varepsilon$-quaternionic K\"ahler structures on  
vector spaces and $\varepsilon$-quaternionic K\"ahler manifolds.
The physical aspects of the $c$-map construction are dealt with in Section \ref{sec:physics}.
We discuss theories of 4D, ${\cal N} = 2$ supergravity coupled to vector
multiplets with either Lorentzian or Euclidean spacetime signature, and the reduction 
of such theories to
three dimensions over a spacelike or timelike circle. This provides the motivation for the
choice of metric on the $c$-map target manifold. 
In Section \ref{sec:cmap} we present our construction of the $c$-map. 
We provide a detailed description of the target space topology, metric and $\varepsilon$-quaternionic structure. 
The explicit calculation of the Levi-Civita connection is postponed until Section \ref{sec:LCint}, where we discuss each $c$-map on a case-by-case basis. In this section we also
prove the existence of two integrable $\varepsilon$-complex structures on the $c$-map target manifold. Finally, in Section \ref{Sec:fibre_geometry} we discuss the complementary 
approach to describing $c$-map target manifolds locally as group bundles.
Throughout this paper we will use the index conventions
\begin{align*}
		A,B,C,\ldots &= 1, \ldots, n \;, & m,p,q, \ldots &= 1, \ldots, 2n \;,  \\
		I,J,K,\ldots &= 0, \ldots, n \;, & a,b,c, \ldots &= 0, \ldots, 2n + 1  \;.
\end{align*}

\section{$\varepsilon$-K\"ahler and $\varepsilon$-quaternionic K\"ahler geometry}
\label{sec:KandQKgeom}

\subsection{$\varepsilon$-K\"ahler manifolds}
\label{sec:eKmfs}

In this section we review $\varepsilon$-complex and 
$\varepsilon$-K\"ahler manifolds, and provide some examples
which we will use later in the paper. The concepts of $\varepsilon$-complex
geometry allow us to talk about complex and para-complex geometry
in parallel. 
Intuitively, para-complex geometry differs from complex geometry by 
replacing the field of complex numbers $\mathbbm{C}$ by the ring of
para-complex numbers $C=\mathbbm{R} \oplus e \mathbbm{R}$, where $e$ is
the para-complex imaginary unit satisfying
\[
e^2 = 1 \;,\;\;\;\bar{e} = -e \;.
\]
We assume that the reader is familiar with the definitions and the relevant properties of para-complex, 
para-Hermitian and para-K\"ahler manifolds, which can be found, for instance, in 
\cite{Cortes:2003zd}. As in \cite{Cortes:2009cs} we will use
a unified $\varepsilon$-complex notation and terminology, where $\varepsilon=-1$
refers to the complex case and $\varepsilon=1$ to the para-complex
case. Thus, for example, we will use the symbol $i_\varepsilon$ to denote
the complex imaginary unit $i$ in the case $\varepsilon = -1$ and the 
para-complex imaginary unit $e$ in the case $\varepsilon = 1$ such that 
\[ i_\varepsilon^2 = \varepsilon 1,\quad \bar{i}_\varepsilon = - i_\varepsilon .\]
We denote by $\mathbbm{C}_\varepsilon = \mathbbm{R}[i_\varepsilon]$ the ring of
$\varepsilon$-complex numbers. Similarly, an almost $\varepsilon$-complex 
structure $J$ on a real differentiable
manifold $M$ is a field of endomorphisms of the tangent bundle $TM$
such that $J^2 = \varepsilon \mathbbm{1}$, and such that the 
eigendistributions of $J$ have equal rank. Our convention for the
relation between the $\varepsilon$-complex structure $J$, 
pseudo-Riemannian metric $g$ and $\varepsilon$-K\"ahler form $\omega$
on an
$\varepsilon$-K\"ahler manifold $(M,J,g)$ is 
\begin{equation}
\omega = \varepsilon g( J \cdot, \cdot) \;.
\label{eq:MetricKahler}
\end{equation}

Among the simplest examples of $\varepsilon$-K\"ahler manifolds 
are spaces of constant $\varepsilon$-holomorphic sectional curvature,
which are always  (pseudo-)Riemann\-ian locally symmetric spaces. As we will see later
$c$-map spaces are fibre bundles  over special $\varepsilon$-K\"ahler
manifolds with fibres of constant $\varepsilon$-holomorphic sectional
curvature. Therefore we will now discuss these spaces in some detail.

\subsubsection{$\varepsilon$-complex vector spaces \label{sec:vector_spaces}}

The construction of $\varepsilon$-K\"ahler metrics that we 
are going to present is a generalisation of the well-known
Fubini-Study metric on complex projective spaces $\mathbbm{C}P^n$. 
Consider the vector space 
$\mathbbm{R}^{n+1} \oplus \mathbbm{R}^{n+1} = \mathbbm{R}^{2n+2}$
with coordinates
\[
({\bf x}, {\bf y}) = \left(x^I, y^J\right) \;,\;\;\;I,J=0,\ldots, n\;.
\]
We introduce a scalar product
\[
\langle ({\bf x}, {\bf y}), ({\bf u}, {\bf v}) \rangle =
\eta({\bf x}, {\bf u}) -\varepsilon \eta({\bf y}, {\bf v}) \;,
\]
on $\mathbbm{R}^{2n+2}$, 
where $\eta(\cdot, \cdot)$ is a scalar product of signature
$(k,\ell)$ on $\mathbbm{R}^{n+1}$. In terms of orthonormal coordinates
on $\mathbbm{R}^{n+1}$ we have
\[
\eta({\bf x}, {\bf u}) = \eta_{IJ} x^I u^J \;,
\]
where $(\eta_{IJ}) = \mbox{diag}( {\mathbbm 1}_k, -{\mathbbm 1}_\ell )$. 
Thus $\langle \cdot, \cdot \rangle$ is a scalar product of
signature $(2k,2\ell)$ if $\varepsilon = -1$ and of signature $(n+1,n+1)$ 
if $\varepsilon =1$. 

Next, we define an  $\varepsilon$-complex structure on $\mathbbm{R}^{2n+2}$ by
\[
J ({\bf x},{\bf y})  = (\varepsilon {\bf y} , {\bf x}) \;.
\]
Note that $J$ is skew with respect to $\langle \cdot , \cdot \rangle$,
so that $(\mathbbm{R}^{2n+2}, J, \langle \cdot, \cdot \rangle)$ is 
an $\varepsilon$-Hermitian vector space, that is a  (pseudo-)Hermitian vector space\footnote{Since we will frequently 
deal with indefinite scalar products, we will usually omit the 
prefix `pseudo-'.} if $\varepsilon =-1$ and a para-Hermitian vector space
if $\varepsilon=1$.

We identify $(\mathbbm{R}^{2n+2}, J)$ with the standard
$\varepsilon$-complex vector space
$\mathbbm{C}^{n+1}_\varepsilon = \mathbbm{R}^{n+1} \oplus
i_\varepsilon \mathbbm{R}^{n+1}$ by means of the isomorphism 
\[ \mathbbm{C}^{n+1}_\varepsilon \ni 
{\bf z} = {\bf x} + i_\varepsilon {\bf y} \mapsto ({\bf x}, {\bf y}) \in \mathbbm{R}^{2n+2}\;.
\]
This identifies $J$ with the standard $\varepsilon$-complex structure on $\mathbbm{C}^{n+1}_\varepsilon$, that is 
$J {\bf z} = i_\varepsilon  {\bf z}.$
On $\mathbbm{C}^{n+1}_\varepsilon$ we consider the $\varepsilon$-Hermitian form \begin{equation}
\label{gamma1}
\gamma({\bf z}, {\bf w}) = \eta_{IJ} z^I \bar{w}^J \;,
\end{equation}
which is of complex signature
$(k,\ell)$ if $\varepsilon= -1$. 
Using the isomorphism $\mathbbm{C}^{n+1}_\varepsilon \simeq \mathbbm{R}^{2n+2}$, 
we can write it as  
\be \label{gamma2} \gamma = \langle \cdot , \cdot \rangle +\varepsilon i_\varepsilon \langle J \cdot, \cdot \rangle .\ee
In coordinates it is given by
\begin{align*}
\gamma(({\bf x},{\bf y}), ({\bf u}, {\bf v})) &=  \eta_{IJ} x^I u^J -\varepsilon \eta_{IJ} y^I v^J + i_\varepsilon \left( \eta_{IJ} y^I u^J -
\eta_{IJ} x^I v^J \right) \\
&= \eta_{IJ} z^I \bar{w}^J =\gamma({\bf z}, {\bf w})\;,
\end{align*}
where ${\bf z} = {\bf x} + i_\varepsilon {\bf y}$, ${\bf w}= {\bf u} + i_\varepsilon {\bf v}$.

\subsubsection{Spaces of $\varepsilon$-complex lines}
We continue to consider 
the $\varepsilon$-complex vector space $\mathbbm{C}_\varepsilon^{n+1}= 
(\mathbbm{R}^{2n+2},J)$,
equipped with the $\varepsilon$-Hermitian form \re{gamma2}.

Consider the open subset
\[
\domain = \domain_\varepsilon =\begin{cases} \{ v \in \mathbbm{R}^{2n+2} | \langle v, v \rangle > 0 \} & \mbox{if}\; \varepsilon = -1 \\
\{ v \in \mathbbm{R}^{2n+2} | \langle v, v \rangle \neq 0 \}  &\mbox{if}\; \varepsilon = +1\;,\end{cases}
\]
and define the space of $\varepsilon$-complex lines
\[
P(\domain) = \domain / \sim \;,\;\;\;\mbox{where}\;\;\;
v \sim v' \Leftrightarrow (\mathbbm{R} + \mathbbm{R} J) v =
(\mathbbm{R} + \mathbbm{R} J) v'  \;.
\]
This can be viewed as taking a quotient with respect to the 
natural action $v \mapsto z v$ of the group of units 
\[
\mathbbm{C}^*_\varepsilon = \{ z \in \mathbbm{C}_\varepsilon | z \;\;\;\mbox{invertible} \} 
\subset \mathbbm{C}_\varepsilon \;,
\] 
of the ring $\mathbbm{C}_\varepsilon$. Since this group will play some role in 
the following, let us make some remarks. 
\begin{remark} 
In the complex case
$\mathbbm{C}^*_\varepsilon$ is the multiplicative group $\mathbbm{C}^*$ of non-zero complex
numbers, which is connected. In contrast, the para-complex
numbers $z=x+ey$ that are not invertible are precisely those which are located on the light cone $x^2 - y^2 = 0$, 
and the group of para-complex units $C^*$ has four
connected components:
\[
C^* = C^*_0 \cup e C^*_0 \cup - C^*_0 \cup -e C^*_0 \;,
\]
where $C^*_0$ is the connected component of unity. 
\end{remark}

\begin{remark}
Note that when defining $\domain$ we have excluded not only the zero vector but all null vectors. 
This is done for two reasons. In fact,  in the case $\varepsilon =1$ there exist
non-zero singular vectors, that is vectors $v$ such that the orbit $C^*v$ is of lower
dimension than $2$. In order to obtain a free action of $C^*$ we therefore need
to exclude such vectors. This is ensured by excluding null vectors. Another reason is 
that, as we will see below, in order to define the induced metric on $P(\domain)$ we will 
have to divide by $\langle v, v\rangle$. Finally, in the case $\varepsilon=-1$, to avoid jumping of the signature
of the metric on the quotient we restrict to spacelike complex lines. The restriction to spacelike lines is no loss of generality, since
we can always multiply the metric by $-1$. Notice that in the case $\varepsilon = 1$
multiplication by $e$ maps spacelike to timelike vectors and vice-versa, and therefore there is no
notion of spacelike (nor timelike) para-complex lines.  
\end{remark}

The group $\mathbbm{C}^*_\varepsilon$ acts freely and properly on $\domain$ by $\varepsilon$-holomorphic transformations. Therefore, $P(\domain)$ is a smooth
$\varepsilon$-complex manifold and  $\pi : \domain \rightarrow P(\domain)$ is
an $\varepsilon$-holomorphic  $\mathbbm{C}^*_\varepsilon$-principal bundle. 

Using the $\varepsilon$-Hermitian form $\gamma$ on $\mathbbm{C}_\varepsilon^{n+1}$, we define
an $\varepsilon$-Hermitian form $\bar{\gamma}$ on $P(\domain)$ by
\[
{\gamma}'( u, v)_p = 
\psym{\gamma}_{\pi(p)} ( d\pi_p u, d\pi_p v) =
\frac{ \gamma(u,v) \gamma(p,p) - \gamma(u,p) \gamma(p,v) }{\gamma(p,p)^2}\;,
\]
where $p\in \domain$ and $u,v\in T_p \domain \simeq \mathbbm{R}^{2n+2}$. 
In terms of $\varepsilon$-complex coordinates, 
this sesquilinear form corresponds to the following tensor field on $\domain$:
\[
{\gamma}' = \frac{ \left(\eta_{KL} z^K \bar{z}^L\right) \eta_{IJ}  
- (\eta \bar{z})_I (\eta z)_J }{ \left(\eta_{KL} z^K \bar{z}^L\right)^2}
dz^I \otimes d\bar{z}^J \;.
\]
To see that this defines an $\varepsilon$-Hermitian metric on $P(\domain)$, 
we first note that ${\gamma}'$ is manifestly invariant under
$\mathbbm{C}_\varepsilon^*$. Moreover it is easy to see that
${\gamma}'(\xi, \cdot ) = 0 = {\gamma}'(J\xi , \cdot)$, 
where $\xi = z^I \frac{\partial}{\partial z^I} + \bar{z}^I 
\frac{\partial}{\partial \bar{z}^I}$ is the 
position vector field on $\domain$. 
Thus ${\gamma}'$ is also horizontal with respect 
to the $\mathbbm{C}^*_\varepsilon$-action and hence can be pushed down to $P(\domain)$.
Since the kernel of ${\gamma}'$ is spanned by $\xi, J\xi$, it 
defines a non-degenerate $\varepsilon$-Hermitian metric on $P(\domain)$. 

Consequently the real part 
\[
{g}'(u,v)_p = \frac{ \langle u,v \rangle \langle p, p \rangle 
- \langle u,p \rangle \langle v,p \rangle 
+ \varepsilon \langle u, Jp \rangle \langle v , Jp \rangle}
{\langle p,p \rangle^2} ,
\]
of ${\gamma}'$ defines a metric $\bar{g}$ on $P(\domain)$ such that ${g}'=\pi^*\psym{g}$. 
The degenerate tensor field ${g}'$ on $\domain$, when expressed 
in $\varepsilon$-complex coordinates, is
\[
{g}' = {g}_{I\bar{J}} dz^I d\bar{z}^J = 
\frac{1}{ \left(\eta_{KL} z^K \bar{z}^L \right)^2 }
\left( \left(\eta_{KL} z^K \bar{z}^L\right) \eta_{IJ} -   \eta_{IK}\bar{z}^K
\eta_{JL} z^L \right) dz^I d\bar{z}^J  \;.
\]
This symmetric tensor field can be locally expressed using 
a potential $K$, which is given by the logarithm of the squared length of
the position vector field:
\[
{g}_{I\bar{J}} =\frac{\partial^2 K}{\partial z^I \partial \bar{z}^J} \;,
\]
where
\[
	K = \log |\eta_{KL} z^K \bar{z}^L|  = \log |\langle \xi, \xi \rangle |  \;.
\]
We can also describe the metric $\psym{g}$ using inhomogeneous coordinates on $P(\domain)$, instead of
using the symmetric tensor field ${g}'$ on $\domain$. 
If we identify $P(\domain)$ locally with the hypersurface $z^0=1$ the associated 
inhomogeneous coordinates are $z^A$, $A=1,\ldots, n$.  In terms of these coordinates,
\[
\psym{g} = \left(1 - \psym{\eta}_{CD} z^C \bar{z}^D \right)^{-2}
\left(-  \left(1-\psym{\eta}_{CD} z^C \bar{z}^D\right) \psym{\eta}_{AB} - \psym{\eta}_{AC}\psym{z}^C
\psym{\eta}_{BD} z^D \right) dz^A d\bar{z}^B \;.
\]
For later convenience we have taken $\eta_{00}=1$, which will cover
all cases relevant to us, and defined
$\psym{\eta}_{AB} = - \eta_{AB}$.
Thus $\psym{\eta}_{AB}$ has signature $(\ell,k-1)$. 
The tensor $\psym{g}$  is an $\varepsilon$-K\"ahler metric on $P(\domain)$ 
with $\varepsilon$-K\"ahler potential 
\begin{equation}
\label{Kprime}
\psym{K} = \log  |1- \psym{\eta}_{AB} z^A \bar{z}^B | \;.
\end{equation}
It is straightforward to check that the $\varepsilon$-K\"ahler metric $\bar{g}$ has constant $\varepsilon$-holomorphic
sectional curvature, that is the sectional curvature of a 
$J$-invariant plane does not depend on the chosen plane. We will recover this later using an alternative
description of these spaces (with the exception of $\mathbbm{C}P^n$) in terms of open orbits of solvable Lie groups. It is known\footnote{This can be proven as in the case
of (positive definite) K\"ahler manifolds of constant holomorphic sectional curvature, see \cite[Ch.\ IX, Thm.\ 7.9]{KNII}.}  
 that $\varepsilon$-K\"ahler spaces with constant $\varepsilon$-holomorphic sectional 
curvature $c$ are locally symmetric and locally uniquely determined by the value of the constant $c$.  
Next we discuss in more detail the spaces $P(\domain)$ as globally symmetric $\varepsilon$-K\"ahler spaces, 
which we  represent as coset spaces.

\subsubsection{Representation as symmetric spaces \label{sec:symmetric_spaces}}

The space $P(\domain)$ is the space of $\varepsilon$-complex lines
in an open subset $\domain$ of the $\varepsilon$-complex vector space
$\mathbbm{C}_\varepsilon^{n+1}$. We will now describe it as a symmetric space.
Let $G\subset GL(2n+2,\mathbbm{R})$ be the $\varepsilon$-unitary group of $\mathbbm{C}_\varepsilon^{n+1}= \mathbbm{R}^{2n+2}$, 
that is the group of $\varepsilon$-complex
linear transformations which leave 
the scalar product $\langle \cdot, 
\cdot \rangle$ invariant. For $\varepsilon=-1$ this is the
(pseudo-)unitary group $U(k,\ell)$, while for $\varepsilon=1$,  
$G$ is the para-unitary group, which is isomorphic to $GL(n+1,\mathbbm{R})$  \cite{Cortes:2003zd}.  
More precisely, the representation of the para-unitary group on $\mathbbm{R}^{2n+2}$ is equivalent to the sum of the 
standard $(n+1)$-dimensional representation of  $GL(n+1,\mathbbm{R})$ and its dual. 
 Since the group $G$ acts transitively on $P(\domain)$ 
we can identify $P(\domain)$ with the corresponding homogeneous space
\[
P(\domain) \simeq G/H \;,
\]
where  $H$ is the stabiliser of an $\varepsilon$-complex line. We notice that 
already the special (pseudo-)unitary group $SU(k,\ell)$, respectively the special para-unitary group
$SL(n+1,\mathbbm{R})$, acts transitively on $P(\domain)$. For notational convenience we prefer
to work with the full $\varepsilon$-unitary group. 
Let us consider the possible cases in turn.

\begin{enumerate}
\item
For $\varepsilon=-1$ and $\eta_{IJ} = \delta_{IJ}$ the
Hermitian form is invariant under $U(n+1)$ and $\psym{\eta}_{AB} = -\delta_{AB}$ . 
The stabiliser of a complex line in $\mathbbm{C}^{n+1}$  
is $U(1) \times U(n)$.
The resulting complex projective space is the symmetric space 
\[
P(\domain) = \mathbbm{C}P^n \simeq U(n+1)/(U(1) \times U(n))\;,
\]
and the corresponding K\"ahler metric is the Fubini-Study metric:  
\[
\psym{K} = \log \left( 1 + \delta_{AB} z^A \bar{z}^B \right) \;,\;\;\;
\psym{g} =  \frac{ \left(1+\delta_{CD} z^C \bar{z}^D\right) \delta_{AB} - \bar{z}_A z_B}{
\left(1+\delta_{CD} z^C \bar{z}^D\right)^2} dz^A d\bar{z}^B \;,
\]
where $z_A = \delta_{AB} z^B=z^A$. 

\item
For $\varepsilon=-1$ and 
$(\eta_{IJ}) = \mbox{diag}(1,-\mathbbm{1}_n)$, the Hermitian
form is invariant under $U(1,n)$  and $\psym{\eta}_{AB}=\delta_{AB}$. Complex lines are stabilised by 
$U(1) \times U(n)$. The resulting symmetric space is the
complex hyperbolic space
\[
P(\domain) = \mathbbm{C}H^n \simeq U(1,n)/(U(1) \times U(n)) \;,
\] 
which is the dual, in the sense of Riemannian symmetric spaces, of 
$\mathbbm{C}P^n$.
We remark that both spaces are real forms 
of $GL(1+n,\mathbbm{C})/\linebreak(GL(1,\mathbbm{C})
\times GL(n,\mathbbm{C}))$. The K\"ahler metric $\psym{g}$ defined in the previous section 
is negative definite and coincides with the complex hyperbolic metric up to sign: 
\[
\psym{K} = \log  \left( 1 - \delta_{AB} z^A \bar{z}^B \right),\;\;
\psym{g} = - \frac{  \left(1-\delta_{CD}z^C \bar{z}^D \right) \delta_{AB} + \bar{z}_A z_B}
{\left(1-\delta_{CD}z^C \bar{z}^D\right)^2} dz^A d\bar{z}^B ,
\]
where $z_A = \psym{\eta}_{AB} z^B = z^A$.

\item
For $\varepsilon=-1$ and 
$(\eta_{IJ}) = \mbox{diag}(\mathbbm{1}_k, -\mathbbm{1}_\ell)$,  
the Hermitian form is invariant under $U(k,\ell)$ and lines are
stabilised by $U(1) \times U(k-1,\ell)$ and $(\psym{\eta}_{AB})=\mbox{diag}(-\mathbbm{1}_{k-1} ,\mathbbm{1}_{\ell})$. The resulting symmetric 
spaces
\[
P(\domain) = \mathbbm{C}H^{(k-1,\ell)} \simeq 
U(k,\ell)/(U(1)\times U(k-1,\ell)) \;,
\]
are indefinite signature versions 
of the Hermitian symmetric spaces $\mathbbm{C}P^n$ and $\mathbbm{C}H^n$. 
We remark that they are again real forms of  $GL(1+n,\mathbbm{C})/(GL(1,\mathbbm{C})
\times GL(n,\mathbbm{C}))$. The resulting pseudo-K\"ahler metric
has complex signature $(k-1,\ell)$:
\[
\psym{K} = \log  \left( 1 - \psym{\eta}_{AB} z^A \bar{z}^B \right) ,\;\;
\psym{g} = - \frac{  \left(1-\psym{\eta}_{CD}z^C \bar{z}^D\right) \psym{\eta}_{AB} + \bar{z}_A z_B}
{\left(1-\psym{\eta}_{CD}z^C \bar{z}^D\right)^2} dz^A d\bar{z}^B ,
\]
since $(\psym{\eta}_{AB})$ has signature $(\ell,k-1)$, and where $z_A =\psym{\eta}_{AB} z^B$. 

\item
We finally consider the para-complex case, $\varepsilon=1$. The stabiliser of a point of $P(\domain)$ under the para-unitary group $GL(n+1,\mathbbm{R})$ is $GL(1,\mathbbm{R}) \times GL(n,\mathbbm{R})$.  The resulting space 
\[
P(\domain) = C{H}^n \simeq GL(n+1,\mathbbm{R})/ GL(1,\mathbbm{R}) \times GL(n,\mathbbm{R}) \;,
\]
is the para-complex analogue of any of the above spaces, which for convenience is referred to as para-complex
hyperbolic space. The corresponding symmetric space is 
yet another real form of $GL(n+1,\mathbbm{C})/(GL(1,\mathbbm{C}) \times
GL(n,\mathbbm{C}))$. The resulting para-K\"ahler metric has 
real signature $(n,n)$ irrespective of the signature of
$(\eta_{IJ}) = (1, -(\psym{\eta}_{AB}))$:
\[
\psym{K} = \log | 1 - \psym{\eta}_{AB} z^A \bar{z}^B | ,\;\;
\psym{g} = - \frac{  \left(1-\psym{\eta}_{CD}z^C \bar{z}^D\right) \psym{\eta}_{AB} + \bar{z}_A z_B}
{(1-\psym{\eta}_{CD}z^C \bar{z}^D)^2} dz^A d\bar{z}^B ,
\]
where $z_A = \psym{\eta}_{AB} z^B$.

\end{enumerate}

\subsubsection{Realisation as a solvable Lie group \label{sec:solvable_Lie_group}}

Recall that the \emph{Iwasawa subgroup} $L$ of a non-compact semi-simple group $G$ is 
the maximal triangular (and, hence, solvable) Lie subgroup of $G$, which is unique up to conjugation.  
As a consequence of the Iwasawa decomposition it acts simply transitively on the 
corresponding Riemannian 
symmetric space of the non-compact type $G/H$, which is defined as the quotient of $G$ by its maximal
compact subgroup $H$, which is unique up to conjugation. Standard examples include hyperbolic 
spaces such as  
$G/H= SU(1,\nalt + 2)/S(U(1) \times U(\nalt + 2))$. This allows us
to identify $G/H$ with $L$ and to compute geometric quantities on $G/H$, 
such as the Levi-Civita connection and curvature, purely algebraically
on the Lie algebra $\mathfrak{l}$ of $L$. On pseudo-Riemannian symmetric spaces 
$G/H$ the group $L$, in general, no longer acts transitively, but it may still act 
at least with open orbit such that we can still identify the symmetric space
with $L$ locally and perform computations on $\mathfrak{l}$. This is indeed the case for 
all non-compact symmetric spaces of constant $\varepsilon$-holomorphic sectional curvature
considered in the previous subsection. In fact, in this
section, we show explicitly that the Iwasawa subalgebra $\mathfrak{l}\subset \mathfrak{su}(1,\nalt + 2)$ 
can be equipped with a scalar product $\langle \cdot , \cdot \rangle$
and $\varepsilon$-complex structure $J$, such that the metric on $L$ obtained
by left-invariant extension of the scalar product is $\varepsilon$-K\"ahler
and has constant $\varepsilon$-holomorphic sectional curvature. Depending
on our choice of scalar product, this provides a local description of 
$\mathbbm{C}H^{\nalt + 2}$, $\mathbbm{C}H^{(\kalt-1,\lalt)}$ where $\kalt + \lalt = \nalt + 3$, or $CH^{\nalt + 2}$ in terms of a 
solvable Lie group equipped with a left-invariant $\varepsilon$-K\"ahler
metric.

We start by reviewing the standard realisation of the Lie algebra 
$\mathfrak{l}$ of the Iwasawa subgroup $L\subset SU(1,\nalt+2)$. 
The $(2\nalt+4)$-dimensional Lie algebra $\mathfrak{l}$ admits the decomposition
\[
\mathfrak{l} = V \oplus \mathbbm{R}Z_0 \oplus \mathbbm{R}D \;,
\]
where $V=\mathbbm{R}^{2\nalt+2}$ is the abelian Lie algebra of
dimension $2\nalt+2$. 
The non-trivial commutators are
\begin{equation}
\label{RelationsIwasawa}
[X,Y] =\omega(X,Y) Z_0 \;,\;\;\;
[D,X] = \frac{1}{2} X \;,\;\;\;
[D,Z_0] = Z_0 \;,
\end{equation}
where $X, Y\in V$, and where $\omega$ is a non-degenerate symplectic
form on $V$. Thus $Z_0$ extends $V$ into the standard Heisenberg Lie
algebra of dimension $2\nalt+3$ on which $D$ acts as a derivation. 
We choose an $\omega$-skew-symmetric $\varepsilon$-complex structure $J$ on $V$ which is 
extended to $\mathfrak{l}$ by setting
\[
JD =  -Z_0 \;,\;\;\;J Z_0 = -\varepsilon D\;.
\]
On $V$ this determines  the (possibly indefinite) scalar product  $\langle \cdot ,\cdot \rangle =  \omega (J\cdot , \cdot )$ 
which we extend orthogonally to
$\mbox{span}\{Z_0, D\}$ by
\[
\langle D, D \rangle = 1 \;,\;\;\; \langle D, Z_0 \rangle = 0 \;,\;\;\; 
\langle Z_0, Z_0 \rangle = - \varepsilon \;.
\]
This also determines the extension of the symplectic form 
to $\mathfrak{l}$ by $\omega(D,Z_0)=1$. 
Since the $\varepsilon$-complex structure is skew-symmetric with respect to the scalar product,
$J \in \mathfrak{so}(\mathfrak{l}, \langle \cdot, \cdot \rangle)$,
we can express it in terms of bivectors as
\[
J = D \wedge Z_0  + J \circ \mbox{pr}_V\;,
\]
using the convention that
\[
(X \wedge Y )(Z) = X \langle Y, Z \rangle - Y \langle X, Z \rangle \;,
\]
for $X,Y,Z \in \mathfrak{l}$. Here $\mbox{pr}_V$ denotes the projection onto $V$.

By identification of $\mathfrak{l}$ with the Lie algebra of left-invariant vector
fields on the associated Lie group $L$, we obtain a left-invariant
metric $g_L$, $\varepsilon$-complex structure $J$, 
and symplectic form $\omega$ on $L$.
To compute the Levi-Civita 
connection $\nabla$ on $L$ we use the 
Koszul formula:
\begin{eqnarray}
2 \langle \nabla_X Y, Z\rangle &=& 
X  \langle Y, Z \rangle + Y \langle X, Z \rangle
- Z  \langle X,Y \rangle  \nonumber \\
&& + \langle [X,Y], Z \rangle - \langle X, [Y,Z] \rangle 
- \langle Y , [X,Z] \rangle \;,
\end{eqnarray}
where $X,Y,Z$ are vector fields on $L$. It is sufficient to
evaluate for left-invariant vector fields, in which case 
the first three terms on the right hand side are zero. 
The remaining terms can be evaluated using the scalar product 
and commutator relations of $\mathfrak{l}$. 

The result can be summarised as
follows:
\begin{eqnarray}
\nabla_D &=& 0 \;, \nonumber \\
\nabla_{Z_0} &=& D \wedge Z_0  + \frac{1}{2} J \circ \mbox{pr}_V \;,
\nonumber \\
\nabla_X &=& \frac{1}{2} D \wedge X  + \frac{\varepsilon}{2} Z_0 \wedge JX  \;,
\;\;\;\forall X \in V \;. \nonumber
\end{eqnarray}
Here $\nabla_X$, with $X\in \mathfrak{l}$, is interpreted as an
endomorphism of $\mathfrak{l}$. 

It is straightforward to verify that
\[
\nabla_X (JY) = J \nabla_X Y \;\;\;\forall X,Y \in \mathfrak{l} \;.
\]
Thus the $\varepsilon$-complex structure is parallel, $\nabla J=0$, 
and in particular integrable.  
We conclude that the metric $g_L$ on $L$ is a left-invariant
$\varepsilon$-K\"ahler metric.

The curvature of the connection $\nabla$ is computed by the 
formula
\[
R(X,Y) = [ \nabla_X, \nabla_Y] - \nabla_{[X,Y]} \;.
\]
$R(X,Y)$ can be interpreted as a skew endomorphism of $\mathfrak{l}$, 
and thus be computed on $\mathfrak{l}$. 
When evaluating commutators of skew endomorphisms, the following
formula is useful:
\[
[X\wedge Y, Z \wedge W ] = (X\wedge W) \langle Y,Z \rangle
+ (Y\wedge Z) \langle X,W \rangle
- (X\wedge Z) \langle Y, W \rangle
- (Y \wedge W) \langle X, Z \rangle \;.
\]

It is straightforward to show that $R(X,Y)$ takes the canonical form
\[
R(X,Y) = R(X,Y)_{\rm can}
= - \frac{1}{4} \left( X \wedge Y - \varepsilon
JX \wedge JY + 2  \omega(X,Y) J \right) \;,
\]
and one easily verifies that the $\varepsilon$-holomorphic sectional
curvature is $-1$:
\[
\frac{
\langle R(X,JX)_{\rm can} JX , X \rangle}{
\langle X \wedge JX, X \wedge JX \rangle} = - 1 \;.
\]

For later applications is useful to introduce a Darboux basis
$(Y^i, X_j)_{i,j = 0,\ldots,\nalt}$ of $V$:
\[
\omega(Y^i, X_j)  = \delta^i_j 
\Rightarrow
[Y^i, X_j] = \delta^i_j Z_0 \;.
\]
The Gram matrix of the scalar product $\langle \cdot , \cdot \rangle$ on $V$ 
with respect to this basis is given by 
\[
\langle Y^i, Y^j \rangle = \etaalt^{ij} \;,\;\;\;
\langle X_i, X_j \rangle = -\varepsilon \etaalt_{ij} \;,\;\;\;
\]
where $\etaalt^{ij}$ has signature $(\lalt -1, \kalt -1)$,
(with $\kalt + \lalt = \tilde{n}+3$, where $\dim V = 2 \tilde{n}+2$)
and where $\etaalt^{ik} 
\etaalt_{kj}
= \delta^i_j$. 
This determines the expression of $J$ on $V$ in this basis:
\[
J Y^i = -\etaalt^{ij} X_j \;,\;\;\;
J X_i = -\varepsilon \etaalt_{ij} Y^j \;.
\]
We can choose the Darboux basis such that $\etaalt^{ij}$ is diagonal
with entries $\pm 1$. 

Now, to finish this subsection, we would like to indicate
which scalar products on $V$ correspond to which of the symmetric spaces
of constant $\varepsilon$-holomorphic sectional curvature discussed in the 
previous subsection. At this point we anticipate some results 
which will be proven in Section \ref{Sec:fibre_geometry}. 
We will see in Section \ref{Sec:fibre_geometry}  
that $c$-map spaces are fibre bundles over 
projective special $\varepsilon$-K\"ahler manifolds, where the fibre is  precisely
the solvable Iwasawa group $L$ of $SU(1,\tilde{n}+2)$ equipped with a left-invariant 
$\varepsilon$-K\"ahler metric which depends on the base point. (Here $\tilde{n}$ is 
the complex dimension of the base manifold.)  
The $c$-map will provide coordinates on $L$, which easily allow
one to find the associated $\varepsilon$-K\"ahler potential. 
By comparing to the $\varepsilon$-K\"ahler potentials listed 
in Section \ref{sec:symmetric_spaces} we will then be able to
identify the symmetric spaces that actually occur in the context
of the $c$-map. For convenience
we already summarise the result here:
\begin{enumerate}
\item
For $\varepsilon=-1$ and $\etaalt_{ij} = \delta_{ij}$, we obtain
the complex hyperbolic space $\mathbbm{C}H^{\nalt + 2}$ equipped with the
positive definite metric $-\psym{g}$.
\item
For $\varepsilon=-1$ and $\etaalt_{ij}$ a matrix of signature
$(\lalt-1,\kalt-1)$ we obtain the indefinite complex hyperbolic space 
$\mathbbm{C}H^{(\lalt,\kalt-1)}$ equipped with the metric $-\psym{g}$ of
complex signature $(\lalt,\kalt-1)$.
\item
For $\varepsilon=1$ and any signature of $\etaalt_{ij}$ we 
obtain the para-complex hyperbolic space $CH^{\nalt + 2}$, equipped with the
metric $-\psym{g}$ of real signature $(\nalt + 2,\nalt +2)$. 
\end{enumerate}
Note that when choosing $\varepsilon=-1$ and a positive definite
scalar product on $V$, we do not obtain a metric on the compact
space $\mathbbm{C}P^{\nalt + 2}$, but a positive definite metric on its
non-compact dual $\mathbbm{C}H^{\nalt + 2}$. We will see in Section
\ref{Sec:fibre_geometry} why the compact space $\mathbbm{C}P^{\nalt + 2}$
cannot arise in the context of the $c$-map.

\subsection{Special $\varepsilon$-K\"ahler manifolds}
\label{Sec:SpecialK}

For later use we now review special $\varepsilon$-K\"ahler manifolds, following
the definitions and theorems stated in \cite{Cortes:2009cs}. 

\begin{definition}
	An affine special $\varepsilon$-K\"ahler (AS$\varepsilon$K) 
manifold $(M, J, g, \nabla)$ is an $\varepsilon$-K\"ahler manifold $(M,g,J)$ endowed with a flat torsion-free connection $\nabla$ such that
	\begin{enumerate}
		\item $\nabla$ is symplectic with respect to the $\varepsilon$-K\"ahler form, i.e.\ $\nabla \omega = 0$.
		\item $\nabla J$ is a symmetric (1,2)-tensor field, i.e.\ $(\nabla_X J)Y = (\nabla_Y J)X$ for all $X, Y$.
\end{enumerate}
\end{definition}

Every simply connected AS$\varepsilon$K manifold admits a canonical realisation 
as an immersed Lagrangian submanifold of  the 
$\varepsilon$-complex symplectic vector space $T^*\mathbbm{C}^{n+1}_\varepsilon=\mathbbm{C}^{2n+2}_\varepsilon$, 
such that the special geometry of $M$ is induced by the immersion,  
where $n + 1$ is the $\varepsilon$-complex dimension of $M$. 
From this one obtains 
the local characterisation of an AS$\varepsilon$K manifold in terms of
an $\varepsilon$-holomorphic prepotential. For any given $p\in M$ one can choose linear 
symplectic coordinates 
\[
	\left(X^I, W_J\right) = \left(x^I + i_\varepsilon u^I, y_J + i_\varepsilon v_J \right) \;, \qquad I,J = 0,\ldots, n ,
\]
in $\mathbbm{C}^{2n+2}_\varepsilon$ such that the symplectic form is given by $dX^I\wedge dW_I$ and the 
functions $X^I$ restrict  to a system of local $\varepsilon$-holomorphic
coordinates near $p$, which we call \emph{special $\varepsilon$-holomorphic coordinates}. The Lagrangian submanifold 
is then defined by equations of the form $W_I = F_I(X):=\frac{\partial F(X)}{\partial X^I}$, where $F(X)=F(X^0,\ldots,X^n)$ is 
an $\varepsilon$-holomorphic function, which is called 
the \emph{prepotential}. The metric is given by
\begin{equation}
\label{g}
	g = N_{IJ} \, dX^I d\bar{X}^J \;,
	\qquad
	N_{IJ} = -i_{\varepsilon}(F_{IJ} - \bar{F}_{IJ}) \;,
\end{equation}
where $F_{IJ}$ are the second derivatives of the prepotential $F$, and a K\"ahler potential is therefore
\[
	K = -i_{\varepsilon}(X^I \bar{F}_I - F_I \bar{X}^I) \;.
\]
On $M$ the $2n + 2$ globally-defined real functions 
\[
\left(q^a\right) = \left(x^I,y_J\right)\;, \qquad a = 0,\ldots, 2n + 1\;, 
\]
form a local system of $\nabla$-affine coordinates about any point, which we call \emph{special real coordinates}. 
Both special  $\varepsilon$-holomorphic and special real coordinates are useful when investigating AS$\varepsilon$K geometry, although many of the new results in this paper will be presented in terms of the latter. 
The K\"ahler form and metric are given by 
the following globally-defined expressions 
\begin{align} \label{Omegaab}
	\omega &= \Omega_{ab}\, dq^a \wedge dq^b =2  dx^I\wedge dy_I, 
	&
	(\Omega_{ab}) &= \left( \begin{array}{cc} 0 & {\mathbbm 1}_{{n} + 1} \\ -{\mathbbm 1}_{{n} + 1} & 0 \end{array} \right) ,\nonumber
	\\ 
	g &= H_{ab}\, dq^a \otimes dq^b=H_{ab}\, dq^a dq^b,
	&
	H_{ab} &= \frac{\partial^2}{\partial q^a \partial q^b} H \;,
\end{align}
where $H$ is a globally-defined real function called the Hesse potential.

The $\varepsilon$-complex structure is determined by the metric and K\"ahler form according to (\ref{eq:MetricKahler}) 
\begin{equation}
J = -\tfrac12 \Omega^{ac}H_{cb} \frac{\partial }{\partial q^a} \otimes dq^b \;,
	\label{eq:ASKCx}
\end{equation}
and $J^2=\varepsilon Id$ ensures that
\begin{equation}
	H_{ab} \Omega^{bc} H_{cd} = \varepsilon 4 \Omega_{ad} \;. \label{eq:id3}
\end{equation}
The matrix $(H_{ab})$ is related to $(F_{IJ})$ by
\begin{equation}
	(H_{ab}) = 
	\left( \begin{array}{cc}
	N - \varepsilon R N^{-1} R & \varepsilon2 RN^{-1} \\
	\varepsilon 2 N^{-1} R & - \varepsilon 4 N^{-1} \\
	\end{array} \right) \;,
	\label{eq:Hab}
\end{equation}
where 
\[
F_{IJ} =: \frac{1}{2} (R_{IJ} - \varepsilon i_{\varepsilon} N_{IJ}) \;.
\]
The imaginary part of the $\varepsilon$-holomorphic prepotential is related to the Hesse potential by a Legendre transform $H(x, y) = -\varepsilon ( 2 \text{Im}F(X(x,y)) - 2y_I u^I(x,y))$, which replaces the $u^I$ with $y_I$ as independent functions \cite{Cortes:2001aa}.

\begin{definition}
A conic affine special $\varepsilon$-K\"ahler (CAS$\varepsilon$K) manifold 
$(M,J,g,\nabla, \xi)$ is an AS$\varepsilon$K manifold 
$(M,J,g,\nabla)$ equipped with a vector field $\xi$ such that 
\[
\nabla \xi = D\xi = \text{Id}\;,
\]
 where $D$ is the Levi-Civita connection.
\end{definition}
The definition implies that $L_\xi g = 2 g$ and $L_{J\xi} g =0$,
so that while $\xi$ acts homothetically, $J\xi$ acts isometrically. 
Moreover the vector field $\xi$ and, hence, $J\xi$ preserves $J$
and the two vector fields generate an infinitesimal action of a two-dimensional
abelian Lie algebra. 
The corresponding condition on the Hesse potential for an AS$\varepsilon$K manifold to be conical 
is that it must be homogeneous of degree two, once we have restricted the special 
real coordinates such that $\xi$ is the 
corresponding Euler field, $\xi = q^a \frac{\partial}{\partial q^a}$. Such special coordinates
are called \emph{conical}, and it is understood in the following that 
special coordinates are conical. 

As in \cite{Cortes:2009cs}  we will always assume that $g(\xi,\xi) = 2 H$ does not vanish on $M$,
which will be used in (\ref{eq:PSKInducedMetric}). In addition, we will assume for simplicity that $M$ is simply connected and impose the following regularity assumption 
on CAS$\varepsilon$K manifolds
in order to discuss projective special $\varepsilon$-K\"ahler manifolds in a convenient way. 
We assume that the infinitesimal action generated by $\xi$ and $J\xi$ is induced by a 
principal $\mathbbm{C}_\varepsilon^*$-action on $M$ and that the Lagrangian
immersion $M \ra \mathbbm{C}_\varepsilon^{2n+2}$ is $\mathbbm{C}_\varepsilon^*$-equivariant.

Consider the (0,2)-tensor field
\begin{align}
	h &= - \frac{g}{g(\xi, \xi)} + \frac{g(\cdot, \xi)\otimes g(\cdot, \xi) - \varepsilon g(\cdot, J\xi) \otimes g(\cdot, J\xi)}{g(\xi,\xi)^2}  \label{eq:PSKInducedMetric} 
\end{align}
on a regular CAS$\varepsilon$K manifold, which is $\mathbbm{C}^*_\varepsilon$-invariant and degenerate along the orbits of the $\mathbbm{C}^*_\varepsilon$-action.
In terms of special $\varepsilon$-complex coordinates $h$ is given by
\be \label{gIJ}
	g_{IJ}\, dX^I d\bar{X}^J = \left( - \frac{N_{IJ}}{XN\bar{X}} + \frac{(N \bar{X})_I (NX)_J}{(XN\bar{X})^2} \right) dX^I d\bar{X}^J \;,
\ee
where $(NX)_I = N_{IJ} X^J$ and $XN\bar{X} = N_{IJ} X^I \bar{X}^J$, whilst in terms of special real coordinates it is given by
\be \label{eq:hab}
	h_{ab}\, dq^a dq^b = \left( -\frac{1}{2H}H_{ab} + \frac{1}{4H^2} H_a H_b - \varepsilon \frac{1}{H^2} \Omega_{ac}q^c \Omega_{bd} q^d \right) dq^a dq^b \;,
\ee
where 
\be \label{eq:Ha} (H_a) = \left(\frac{\partial }{\partial q^a}H \right) = (-\varepsilon 2v_I, \varepsilon 2u^J)^T. \ee

The requirement that a CAS$\varepsilon$K manifold $M$ is regular ensures that the projection $\pi: M \to \bar{M}$ onto the space of $\mathbbm{C}_\varepsilon^*$-orbits 
is the quotient map of a holomorphic principal bundle over an $\varepsilon$-complex manifold, and that the (0,2)-tensor field $h$ on $M$ induces an $\varepsilon$-K\"ahler metric $\bar{g}$ on $\bar{M}$, such that $\pi^* \bar{g} = h$. The $\varepsilon$-K\"ahler manifold $(\bar{M},\bar{J},\bar{g})$ is called a \emph{projective special $\varepsilon$-K\"ahler (PS$\varepsilon$K) manifold.\footnote{In the case $\varepsilon=1$ one may define instead a projective special para-K\"ahler manifold as  
the quotient $\hat{M}$ of $M$ by the action of the connected group $C^*_0$ which is related to 
$\bar{M}$ by the four-fold covering $\hat{M}\rightarrow \bar{M}$, see \cite{Cortes:2009cs}.}}

The following remark will be used later. 
\begin{remark}
Note in the case $\varepsilon=1$ that the action of 
$i_\varepsilon= e \in \mathbbm{C}^*_\varepsilon=C^*$ induces an anti-isometry of the CAS$\varepsilon$K manifold that sends 
$(M, J, g, \nabla, \xi)$ to $(M, J, -g, \nabla, \xi)$ but preserves the $C^*$-invariant tensor $h$. 
\end{remark}

The relation between a CAS$\varepsilon$K manifold and the associated
PS$\varepsilon$K manifold is via an $\varepsilon$-K\"ahler quotient and
generalises the Fubini-Study-type constructions of the previous section. 
In terms of special coordinates $(X^I)$ on $M$, the degenerate 
and $\mathbbm{C}^*_\varepsilon$-invariant $(0,2)$
tensor $h$ has a potential of the form
\[
K_h = - \log |-i_\varepsilon(X^I \bar{F}_I - F_I \bar{X}^I)| \;,
\]
where  $g_{IJ} = \frac{\partial^2 K_h}{\partial X^I \partial \bar{X}^J}$ in the expression (\ref{gIJ}) for $h$. One can 
describe $(\bar{M},\bar{g})$ using homogeneous special $\varepsilon$-holomorphic coordinates $(X^I)$ 
and the tensor $h$. Alternatively, one can introduce, for $X^0\not=0$, 
inhomogeneous
special $\varepsilon$-holomorphic coordinates $z^A = \frac{X^A}{X^0}$, where $A = 1,\ldots,n$, and define an
associated prepotential 
${\cal F}(z^1,\ldots, z^n)$
by
\[
F(X^0, \ldots, X^n) = (X^0)^2 {\cal F}(z^1, \ldots, z^n) \;.
\]
Then the $\varepsilon$-K\"ahler metric $\bar{g}$ of $\bar{M}$ 
has the $\varepsilon$-K\"ahler potential 
\begin{equation}
\label{Kbarg}
K_{\bar{g}} = - \log |Y| \;,
\qquad
Y =   i_\varepsilon \left( 
2 ({\cal F}-\bar{\cal F}) - (z^A - \bar{z}^A)({\cal F}_A 
+ \bar{\cal F}_A)\right) \;,
\end{equation}
where ${\cal F}_A = \frac{\partial {\cal F}}{\partial z^A}$. 
We note that we can identify $\bar{M}$ locally with the submanifold 
$\{ X^0=1 \} $ of $M$, in particular $K_{\bar{g}}$ agrees with the
pull-back of $K_h$ to $\{ X^0=1 \}$ up to an 
$\varepsilon$-K\"ahler transformation.

The simplest class of examples is provided by models with 
a quadratic prepotential
\[
F = \frac{\varepsilon i_\varepsilon}{4} \eta_{IJ} X^I X^J \;,
\]
where we take $\eta_{IJ}$ to be real and non-degenerate.
The potential for the tensor $h$ is
\[
K_h = - \log | X^I \eta_{IJ} \bar{X}^J | \;.
\]
Evaluating this on the hypersurface $X^0=1$, taking $\eta_{00}=1$,
and setting $\psym{\eta}_{AB} = - \eta_{AB}$, 
we obtain the following $\varepsilon$-K\"ahler potential on $\bar{M}$:
\[
K_{\bar{g}} = -\log | 1 - \psym{\eta}_{AB} z^A \bar{z}^B | \;.
\]
This agrees, up to an overall sign, with the 
$\varepsilon$-K\"ahler potentials for the metrics on the
spaces $P(\domain)$ of $\varepsilon$-complex lines given in (\ref{Kprime}).

\subsection{$\varepsilon$-quaternionic K\"ahler manifolds}
\label{sec:QK}

\subsubsection{$\varepsilon$-quaternionic structure on a vector space}

Recall that an {\em $\varepsilon$-quaternionic structure} on a 
real vector space $V$ (of dimension $4n$) 
is a Lie subalgebra $Q \subset \mbox{End}(V)$ spanned by three
pairwise anti-commuting endomorphisms $J_1, J_2, J_3$ that satisfy 
the $\varepsilon$-quaternion algebra
	\[
		J_1{}^2 = J_2{}^2 = -\varepsilon J_3{}^2 = \varepsilon \;, \qquad J_1J_2 = J_3 \;.
	\]
Such a triple $(J_1, J_2, J_3)$ is called a {\em standard basis} 
of $Q$.   
We call $Q$ a quaternionic structure if $\varepsilon  = - 1$ and a para-quaternionic structure if $\varepsilon = 1$.
If $V$ is endowed with a pseudo-Euclidean scalar product $\langle \cdot, \cdot \rangle$ then $Q$ is called {\em skew-symmetric} if $Q$ consists of skew-symmetric endomorphisms. This implies that the signature of $\langle \cdot, \cdot \rangle$ is of the form $(4k, 4\ell)$, where $n=k+\ell$, 
if $\varepsilon = -1$ and $(2n,2n)$ if $\varepsilon = 1$.  
An orthonormal frame of $(V,\langle \cdot, \cdot \rangle, Q)$ is called an {\em $\varepsilon$-quaternionic frame} if it is of the form 
\begin{equation}
\label{StandardBasis}
(e_1, \ldots, e_n, J_1 e_1, \ldots, J_1 e_n, 
J_2 e_1, \ldots, J_2 e_n, 
J_3 e_1, \ldots, J_3 e_n) \;,
\end{equation}
where $(J_1, J_2, J_3)$ is any standard basis of $Q$ and
\[
\langle e_1, e_1 \rangle = \cdots =\langle e_k , e_k \rangle = 1
\;\;\;\mbox{and}\;\;\; \langle e_{k+1}, e_{k+1}\rangle = \cdots = \langle e_{k+\ell} , 
e_{k+\ell} \rangle = -1 \;.
\] 
The Gram matrix of the basis (\ref{StandardBasis}) defines a canonical scalar product $\langle \cdot, \cdot \rangle_{\rm can}$   on $\mathbbm{R}^{4n}$ of signature $(4k,4\ell)$ if $\varepsilon = -1$ or $(2n,2n)$ if $\varepsilon = 1$.
We will denote by $O_\varepsilon(4k,4\ell )$ the pseudo-orthogonal group with respect to $\langle \cdot, \cdot \rangle_{\rm can}$, and by $\mathfrak{so}_\varepsilon (4k,4\ell)$ its Lie algebra.

Let us denote by $J_\alpha^{\rm can}\in \mathfrak{so}_\varepsilon (4k,4\ell)$ 
the matrix which represents the endomorphism $J_\alpha$ with
respect to the basis (\ref{StandardBasis}). Then 
\[Q^{\rm can}  := 
\mbox{span} \{ J_\alpha^{\rm can} | \alpha =1,2,3 \} \;,
\]
is a skew-symmetric
$\varepsilon$-quaternionic structure on $(\mathbbm{R}^{4n},\langle \cdot, 
\cdot \rangle_{\rm can})$. The triple $(\mathbbm{R}^{4n},  
\langle \cdot,  \cdot \rangle_{\rm can}, Q^{\rm can})$ is our standard model 
for a pseudo-Euclidean vector space endowed with a skew-symmetric
$\varepsilon$-quaternionic structure. We denote by $Sp_\varepsilon (1)$ the group generated by 
the Lie algebra $\mathfrak{sp}_\varepsilon (1) := Q^{\rm can}$ and by $Sp_\varepsilon (k,\ell)$
the centraliser of $Sp_\varepsilon (1)$  in $O_\varepsilon (4k,4\ell)$. The Lie algebra 
of that centraliser is a real form of the complex Lie algebra
of type $C_n$. The inner product and $\varepsilon$-quaternionic structure are preserved by the group 
\begin{align*}
	Sp_\varepsilon (1) \cdot Sp_\varepsilon (k,\ell) =
	\begin{cases}
		Sp(1) \cdot Sp(k,\ell) \subset SO(4k,4\ell)  &\mbox{if } \; \varepsilon = -1 \\
		Sp(2,{\mathbbm R}) \cdot Sp(2n, {\mathbbm R}) \subset SO(2n,2n)  &\mbox{if } \; \varepsilon = +1 \;.
	\end{cases}
\end{align*}
Notice that our notation is such that $Sp(k,\ell )$, $k+\ell =n$, and $Sp(2n,{\mathbbm R})$ are real forms of the 
same complex Lie group $Sp(2n,{\mathbbm C})$.

\subsubsection{The $H\otimes E$ formalism}

Let $E = {\mathbbm C}^{2n}$ with standard basis $B_E = ({\cal E}_1,\ldots, {\cal E}_{2n})$. On $E$ one may define an anti-linear complex structure $j_E$ and non-degenerate skew-symmetric bilinear form $\omega_E$ that satisfy the following formulae
\begin{align*}
	&j_E {\cal E}_A =  {\cal E}_{A + n} \;, 
	\qquad
	j_E {\cal E}_{A + n} = -{\cal E}_{A} \;, 
	\qquad
	A = 1,\ldots,n \\
	&\omega_E =  \sum_{A,B =1}^{n} \eta_{AB} {\cal E}^A \wedge {\cal E}^{B + n} \;,
\end{align*}
where $B_E^* = ({\cal E}^1,\ldots, {\cal E}^{2n})$ is the basis of ${\mathbbm C}^{2n*}$ dual to $B_E$ and $(\eta_{AB}) = \mbox{diag}(\mathbbm{1}_k,-\mathbbm{1}_\ell)$, with $n = k + \ell$. 
Complex conjugation is denoted by $\rho_E$. The group $Sp(k,\ell)$  preserves both $j_E$ and $\omega_E$, the group $Sp(2n,{\mathbbm R})$ preserves both $\rho_E$ and $\omega_E$, and the symplectic form satisfies the following reality condition: $j_E^* \omega_E = \bar{\omega}_E( = \rho_E^* \omega_E)$.  

Let $H= {\mathbbm C}^{2}$ denote a specific case of the above construction, where the standard basis is denoted by $B_H = (h_1,h_2)$, the anti-linear complex structure by $j_H$, complex conjugation by $\rho_H$ and the bilinear form by $\omega_H$.

Consider the $4n$-dimensional complex vector space $H \otimes E$ with standard basis $(h_{\cal A} \otimes {\cal E}_\mu)_{{\cal A} =1,2; {\cal \mu} = 1,\ldots,2n}$. On $H \otimes E$ we may define the following:
\begin{enumerate}[(i)]
	\item Two real structures $j_H \otimes j_E$ and $\rho_H \otimes \rho_E$.
	\item A $\mathbbm{C}$-bilinear scalar product $\langle \cdot, \cdot \rangle^{\mathbbm C} = \omega_H \otimes \omega_E$. 	\item Three skew-symmetric endomorphisms $J_1,J_2,J_3$ that satisfy the $\varepsilon$-quaternion algebra and act according to  
	\begin{equation}
			J^*_\alpha(h^{\cal A} \otimes {\cal E}^\mu) = \left(J^H_\alpha\right)^{\cal A}_{\;\;\,{\cal B}}h^{\cal B} \otimes {\cal E}^\mu \;,
			\quad \;\;
			J^H_\alpha = 
			\begin{cases}
					-i\sigma_\alpha &\mbox{if } \; \varepsilon = -1\\
					\tau_\alpha &\mbox{if } \; \varepsilon = +1 ,
			\end{cases}
			\label{eq:CxStrsHE}
		\end{equation}
		where $\sigma_\alpha$ are the Pauli matrices 
		\[
					\sigma_1 = \left(\begin{array}{cc} 0 & 1 \\ 1 & 0 \end{array}\right) \;, \qquad
					\sigma_2 = \left(\begin{array}{cc} 0 & -i \\ i & 0 \end{array}\right) \;, \qquad
					\sigma_3 = \left(\begin{array}{cc} 1 & 0 \\ 0 & -1 \end{array}\right) \;,
		\]
		and
		\[
					\tau_1 = \left(\begin{array}{cc} 1 & 0 \\ 0 & -1 \end{array}\right) \;, \qquad
					\tau_2 = \left(\begin{array}{cc} 0 & 1 \\ 1 & 0 \end{array}\right) \;, \qquad
					\tau_3 = \left(\begin{array}{cc} 0 & 1 \\ -1 & 0 \end{array}\right) \;.
		\]
\end{enumerate}
One may use the above data to construct an example of an $\varepsilon$-quaternionic Hermitian vector space $(V,\langle \cdot, \cdot \rangle, Q)$ of real dimension $4n$ given by
\[
	V = (H \otimes E)^\rho \;,
	\qquad
			\rho =
		\begin{cases}
			j_H \otimes j_E &\mbox{if } \; \varepsilon = -1 \\ 
			\rho_H \otimes \rho_E &\mbox{if } \; \varepsilon = +1 \;,
		\end{cases}
\]
\begin{equation}
	\langle \cdot, \cdot \rangle = \langle \cdot, \cdot \rangle^{\mathbbm C}\Big|_V \;,
	\qquad
	Q = \text{span}\{J_1,J_2,J_3\}\Big|_V \;.
	\label{eq:HErealVS}
\end{equation}
Since all $\varepsilon$-quaternionic Hermitian vector spaces of a given dimension are isomorphic we may state the following proposition:

\bp
	Let $(V,\langle \cdot, \cdot \rangle, Q)$ be an $\varepsilon$-quaternionic Hermitian vector space of real dimension $4n$. Then we can identify $V^{\mathbbm C} = H \otimes E$ 
	such that the properties (\ref{eq:HErealVS}) are satisfied.
\label{prop:HE}
\ep

Indeed, a standard co-frame of $H^* \otimes E^*$ may be matched with an $\varepsilon$-quaternionic co-frame of $V^*$ through the expressions 
\begin{align*}
	&h^{\cal A} \otimes {\cal E}^\mu = \\
	&\frac{1}{\sqrt{2}}\left( \begin{array}{cccccc}
	e^1 + i J_3^* e^1 & \ldots & e^n + i J_3^* e^n & J_2^* e^1 + i J_1^* e^1 & \ldots & J_2^* e^n + i J_1^* e^n  \\
	-J_2^* e^1 + i J_1^* e^1 & \ldots & -J_2^* e^n + i J_1^* e^n & e^1 - i J_3^* e^1 & \ldots & e^n - i J_3^* e^n 
	\end{array} \right)
\end{align*}
if $\varepsilon = -1$ and
\begin{align*}
	&h^{\cal A} \otimes {\cal E}^\mu = \\
	&\frac{1}{\sqrt{2}}\left( \begin{array}{cccccc}
	e^1 + J_1^* e^1 & \ldots & e^n + J_1^* e^n & -J_3^* e^1 + J_2^* e^1 & \ldots & -J_3^* e^n + J_2^* e^n  \\
	J_3^* e^1 + J_2^* e^1 & \ldots & J_3^* e^n + J_2^* e^n & e^1 - J_1^* e^1 & \ldots & e^n - J_1^* e^n 
	\end{array} \right)
\end{align*}
if $\varepsilon = +1$.

\subsubsection{$\varepsilon$-quaternionic structure on the tangent bundle}

The above notions can be easily transferred to vector bundles. For instance, a {\em (fibre-wise) $\varepsilon$-quaternionic structure} in a vector bundle $E\rightarrow M$ is a subbundle $Q \subset \mbox{End}(E)$ such that $Q_p \subset \mbox{End}(E_p)$ is an $\varepsilon$-quaternionic structure on $E_p$ for all $p\in M$. 
One may introduce pairwise anti-commuting  local sections $J_1,J_2,J_3$ of $Q$ defined over an open subset $U \subset M$ satisfying the $\varepsilon$-quaternion algebra, such that $Q_p = \mbox{span}\{ (J_\alpha)_p |\, \alpha =1,2,3 \}$ for all $p \in U$. A fibre-wise $\varepsilon$-quaternionic structure on the vector bundle $TM$ is called an {\em almost $\varepsilon$-quaternionic structure} on $M$.
An almost $\varepsilon$-quaternionic structure $Q$ on $M$ is called an {\em $\varepsilon$-quaternionic structure} if it is parallel with respect to a torsion-free connection, which can be characterised by the property that the covariant derivative of any section of $Q$ in the direction of any vector field is again a section of $Q$. 
If $M$ is endowed with a pseudo-Riemannian metric then an almost $\varepsilon$-quaternionic structure on $M$ is called  {\em Hermitian} if it consists of skew-symmetric endomorphisms. A pseudo-Riemannian manifold of real dimension $4n > 4$ with almost $\varepsilon$-quaternionic Hermitian structure $Q$ is called {\em $\varepsilon$-quaternionic K\"ahler} if $Q$ is parallel with respect to the Levi-Civita connection.

On a pseudo-Riemannian manifold $(M,g)$ with almost $\varepsilon$-quaternionic Hermitian structure we may use Proposition \ref{prop:HE} to make the local identification $TM^{\mathbbm C} = H \otimes E$, where $H$ and $E$ are (at least locally defined) complex vector bundles of dimension 2 and $2n$ respectively, such that the metric and $\varepsilon$-quaternionic structure satisfy (\ref{eq:HErealVS}). 
We call a local complex co-frame of the form $({\cal U}^{{\cal A} \mu}) = (h^{\cal A} \otimes\, {\cal E}^\mu)$ 
an {\em $\varepsilon$-quaternionic vielbein}. 
The metric takes the form $g = \epsilon_{{\cal A} {\cal B}} \, \rho_{\mu \nu} \, {\cal U}^{{\cal A} \mu} \otimes {\cal U}^{{\cal B} \nu}$ and on $T^*M$ an $\varepsilon$-quaternionic vielbein is subject to the reality condition
\[
	\bar{{\cal U}}^{{\cal A} \mu} = 
	\begin{cases}
			\epsilon_{{\cal A} {\cal B}} \, \epsilon_{\mu \nu} \, {\cal U}^{{\cal B} \nu}  &\mbox{if } \; \varepsilon = -1 \\
			{\cal U}^{{\cal A} \mu} &\mbox{if }\; \varepsilon = +1 \;,
	\end{cases}
\]
where 
\[
	(\epsilon_{{\cal A} {\cal B}}) = \left( \begin{array}{cc} 0 & 1 \\ -1 & 0 \end{array} \right) \;, 
	\;\;\; \;\;
	(\epsilon_{\mu \nu}) = \left( \begin{array}{cc} 0 & {\mathbbm 1}_n \\ -{\mathbbm 1}_n & 0 \end{array} \right) \;, 
	\;\;\; \;\;
	(\rho_{\mu \nu}) = \left( \begin{array}{cc} 0 & \eta \\ -\eta & 0 \end{array} \right) \;. 
\]
Recall that $\eta = (\eta_{AB}) = \mathrm{diag}(\mathbbm{1}_k,-\mathbbm{1}_\ell)$, $k+\ell =n$. 
An $\varepsilon$-quaternionic vielbein may be identified with an $\varepsilon$-quaternionic co-frame through the expressions given below Proposition \ref{prop:HE}.

On a manifold with almost $\varepsilon$-quaternion structure we call an {\em adapted connection} a connection on $TM$ for which the almost $\varepsilon$-quaternionic structure is parallel. It is well-known from the theory of $G$-structures that, with respect to a frame of the $G$-structure, the connection one-form of an adapted connection takes values in the Lie algebra $\mathfrak{g}$. An almost $\varepsilon$-quaternionic Hermitian structure corresponds to a $G$-structure with Lie group $G = Sp_\varepsilon (1) \cdot Sp_\varepsilon (k,\ell)$ and therefore the connection one-form of an adapted connection takes values in $\mathfrak{sp}_\varepsilon(1) \oplus \mathfrak{sp}_\varepsilon(k,\ell)$. Since this is a subalgebra of $\mathfrak{so}_\varepsilon (4k,4\ell)$ an adapted connection is automatically metric compatible. In an $\varepsilon$-quaternionic vielbein basis the connection one-form $\Omega$ of an adapted connection takes the form
\begin{equation} 
	\Omega = p \otimes {\mathbbm 1}_{2n} + {\mathbbm 1}_2 \otimes \left( \begin{array}{cc} q & t \\ -s & -\eta q^T \eta \end{array}\right) \;,
	\label{eq:QK_connection}
\end{equation}
where $p$ is a $2\times 2$ matrix satisfying 
\begin{align*}
	p &\in 
		\begin{cases}
			\mathfrak{sp}(1) \;\; \Leftrightarrow \;\;  \text{Tr}(p) = 0  \,,\;\;p^\dagger + p = 0  &\mbox{if } \; \varepsilon = -1   \\
			\mathfrak{sp}(2,{\mathbbm R})  \;\; \Leftrightarrow \;\;  \text{Tr}(p) = 0\,,\; p=\bar{p}   &\mbox{if } \; \varepsilon = +1 \;,
	\end{cases} 
\end{align*}
and $q,s,t$ are $n\times n$ matrices satisfying 
\begin{align*}
	&\left( \begin{array}{cc} q & t \\ -s & -\eta{q}^T\eta \end{array}\right) \in \\
			&\hspace{2em}\begin{cases}
			\mathfrak{sp}(k,\ell ) \;\; \Leftrightarrow \;\; t = \eta t^T \eta \,,\; s =\bar{t} \,,\;  q^\dagger \eta + \eta q = 0   &\mbox{if } \; \varepsilon = -1   \\
			\mathfrak{sp}(2n,{\mathbbm R}) \;\; \Leftrightarrow \;\;  t = \eta t^T\eta =\bar{t}\,,\; s = \eta s^T \eta =\bar{s}\,,\; q=\bar{q}&\mbox{if } \; \varepsilon = +1  \;.
	\end{cases}
\end{align*}
The coefficients of the torsion tensor are given by
\[
	T^{{\cal A} \mu} = d {\cal U}^{{\cal A} \mu} + \Omega^{{\cal A} \mu}_{\;\;\;\;\;{{\cal B} \nu}} \wedge  {\cal U}^{{\cal B} \nu} \;.
\]
Notice that the connection matrix $(\Omega^{{\cal A} \mu}_{\;\;\;\;\;{{\cal B} \nu}})$ has the following structure, see \re{eq:QK_connection}:
\[ \Omega^{{\cal A} \mu}_{\;\;\;\;\;{{\cal B} \nu}} = \Omega^A_B\delta^\mu_\nu + \delta^A_B\Omega^\mu_\nu,\]
where $(\Omega^A_B) \in \mathfrak{sp}_\varepsilon(1)$ and $(\Omega^\mu_\nu )\in \mathfrak{sp}_\varepsilon(k,\ell)$. 
If the torsion vanishes then the adapted connection coincides with the Levi-Civita connection and the manifold is $\varepsilon$-quaternionic K\"ahler. Alternatively, if the Levi-Civita connection one-form takes values in $\mathfrak{sp}_\varepsilon(1) \oplus \mathfrak{sp}_\varepsilon(k,\ell)$ when written in an $\varepsilon$-quaternionic vielbein basis then the manifold is  $\varepsilon$-quaternionic K\"ahler.

\section{Dimensional reduction of four-dimensional vector multiplets}
\label{sec:physics}

\subsection{Four-dimensional vector multiplets}
\label{sec:3.1}

Our starting point is the bosonic part of the Lagrangian for 
$n=n_V^{(4)}$ ${\cal N}=2$ vector multiplets coupled to supergravity,
as given by (7.9) of \cite{Cortes:2009cs}:
\begin{equation}
\label{VMLagrangian}
{\bf e}_4^{-1} {\cal L}_4^{(\epsilon_1)} = \frac{1}{2} R_4 
- \bar{g}_{AB} \partial_{\hat{\mu}} z^A \partial^{\hat{\mu}} \bar{z}^B + \frac{1}{4}
{\cal I}_{IJ} F^I_{\hat{\mu} \hat{\nu}} F^{J|\hat{\mu} \hat{\nu}} + \frac{1}{4} {\cal R}_{IJ}
F^I_{\hat{\mu} \hat{\nu}}  \tilde{F}^{J|\hat{\mu} \hat{\nu}} \;.
\end{equation}
In the following we will explain
each term appearing in this expression.
$R_4$ and ${\bf e}_4$ are the four-dimensional Ricci scalar and
vielbein, and $\hat{\mu}, \hat{\nu}, \ldots$ are four-dimensional space-time
indices. 
We employ a notation which applies to standard (Lorentzian) and 
Euclidean supergravity simultaneously. 
The main difference between Euclidean and standard
vector multiplets is that 
the complex structure of the scalar manifold $\bar{M}$ 
is replaced by a para-complex
structure \cite{Cortes:2003zd,Cortes:2009cs}, and thus we use
the $\varepsilon$-complex notation introduced previously. From now on 
the parameter $\epsilon_1$ distinguishes between 
Lorentzian space-time ($\epsilon_1=-1$) and Euclidean space-time
($\epsilon_1 =1$).

The $\epsilon_1$-complex 
scalar fields $z^A$ are local coordinates of a PS$\epsilon_1$K manifold
$M$ with metric $\bar{g} = \bar{g}_{AB} dz^A d\bar{z}^B$, where
$\bar{g}_{AB}$ is the $\epsilon_1$-K\"ahler metric with 
$\epsilon_1$-K\"ahler potential $K_{\bar{g}}$ given in 
(\ref{Kbarg}).  
For $\epsilon_1=-1$ this is the well known projective special K\"ahler
geometry of vector multiplets in the `new conventions' of \cite{deWit:1995zg}, 
while for $\epsilon_1=1$ this is the projective special para-K\"ahler geometry 
of Euclidean vector multiplets
which was defined in \cite{Cortes:2009cs}. The scalar metric $\bar{g}$
has positive signature $(2n,0)$ for $\epsilon_1=-1$ and 
split-signature $(n,n)$ for $\epsilon_1=1$.

The original construction of the vector multiplet Lagrangian 
in Lorentzian signature was performed using the superconformal
calculus \cite{deWit:1984pk}. 
This employs an auxiliary theory of $n+1$ rigid superconformal
vector multiplets with complex scalars  $X^I$, $I=0, \ldots, n$,
which are local coordinates of a CASK manifold $M$. 
After gauging the superconformal transformations
the theory becomes gauge equivalent to a theory of $n$ vector multiplets
coupled to Poincar\'e supergravity. This construction is reviewed
in \cite{Freedman:2012zz}. 
The scalar metric $\bar{g}$ is obtained from the scalar metric
$g=N_{IJ}dX^I d\bar{X}^J$ of the scalar manifold $M$ 
of the associated superconformal theory by gauge fixing the local 
symmetry group $\mathbbm{C}^* \simeq \mathbbm{R}^{>0} \times U(1)$,
where $\mathbbm{R}^{>0}$ are dilatations, while the chiral 
$U(1)$ transformations are part of the R-symmetry group $U(1) \times SU(2)$
of the ${\cal N}=2$ supersymmetry algebra.
In \cite{Cortes:2009cs} it was shown how this procedure can be 
adapted to Euclidean vector multiplets, where the scalar manifold 
$M$ is a conical affine special para-K\"ahler manifold, and where the
symmetry group
is $\mathbbm{R}^{>0} \times SO(1,1) \subset C^*=\mathbbm{R}^{>0}\times O(1,1)$ 
\cite{Cortes:2003zd}. While in this paper 
we find it convenient to define projective special para-K\"ahler manifolds by
dividing out the full group $C^*$, only the subgroup $SO(1,1) \subset
O(1,1)$ is part of the R-symmetry group $SO(1,1)\times SU(2)$ of the
Euclidean supersymmetry algebra. Consequently only the group
$\mathbbm{R}^{>0} \times SO(1,1)$ is a symmetry of the superconformal
Lagrangian.  But as explained previously, dividing out the group $\mathbbm{R}^{>0} \times SO(1,1)$ leads to the same scalar manifold $\bar{M}$, provided 
that we restrict to the subset on which the function $-i_{\epsilon_1} (X^I \bar{F}_I - F_I \bar{X}^I) = -N_{IJ}X^I \bar{X}^J$ 
is positive. 
We remark that $SO(1,1)\simeq GL(1,\mathbbm{R})$ is the para-unitary group
$U_{\epsilon_1}(1)$, $\e_1=1$.

The relations between the superconformal theories and 
the supergravity theories are given by $\epsilon_1$-complex versions
of the standard formulae of special K\"ahler geometry,
which were presented in Section \ref{Sec:SpecialK}. 
It is possible to rewrite the scalar term using $\epsilon_1$-complex
scalar fields which are local coordinates of the associated CAS$\epsilon_1$K
manifold $M$:
\be \label{eq:gbarg}
\bar{g}_{AB} \partial_{\hat{\mu}} z^A \partial^{\hat{\mu}} \bar{z}^B =
\left. g_{IJ} \partial_{\hat{\mu}} X^I \partial^{\hat{\mu}} \bar{X}^J \right|_{D} \;,
\ee
where the D-gauge
\[
-i_{\epsilon_1} (X^I \bar{F}_I - F_I \bar{X}^I) = 1
\]
has been imposed. Here $g_{IJ}$ are the coefficients of the 
pullback $h=\pi^* \bar{g}$ of the PS$\epsilon_1$K metric $\bar{g}$
to $M$, which are given by (\ref{gIJ}). The D-gauge
restricts the scalars $X^I$ to a real hypersurface $S\subset M$,
and since the right hand side is in addition invariant under local 
$U_{\epsilon_1}(1)$ transformations,
the $n+1$ $\epsilon_1$-complex
fields $X^I$ represent as many physical degrees of freedom as the
fields $z^A$.
While it is possible to gauge-fix the residual local $U(1)_{\epsilon_1}$
symmetry too, 
we prefer not to do so at this point, because this allows us to keep
all expressions manifestly covariant under symplectic transformations.
The field equations of ${\cal N}=2$ supergravity are invariant under
electric-magnetic duality transformations, which act by 
$Sp(2n+2, \mathbbm{R})$ transformations.\footnote{We refer to 
\cite{deWit:1996ix} for a review of symplectic transformations
for ${\cal N}=2$ vector multiplets.} Under these transformations
$(X^I, F_I)^T$ transforms as a vector, while the transformation 
of $z^A = X^A/X^0$ is non-linear. 

The remaining two terms in (\ref{VMLagrangian}) involve the
abelian field strengths $F^I_{\hat{\mu} \hat{\nu}}$ and their Hodge-duals 
$\tilde{F}^I_{\hat{\mu} \hat{\nu}}$. As with the scalar term, the couplings
${\cal I}_{IJ}$ and ${\cal R}_{IJ}$ can be expressed in terms
of the prepotential $F(X^0, \ldots, X^n)$. The relevant formula
is
\[
{\cal N}_{IJ} = {\cal R}_{IJ} + i_{\epsilon_1} {\cal I}_{IJ} 
= \bar{F}_{IJ}({\bar z}) - \epsilon_1 i_{\epsilon_1}
\frac{(Nz)_I (Nz)_J}{zNz} \;,
\]
where we defined $z^0=1$, and where
$N_{IJ}$ are the coefficients 
of the metric $g$ on the CAS$\epsilon_1$K manifold 
$M$, which are given by (\ref{g}). We use a short-hand notation
where $(Nz)_I := N_{IJ}z^J$ and $zNz:=z^IN_{IJ}z^J$.

The negative imaginary part $- {\cal I}_{IJ}$ of the vector coupling
matrix ${\cal N}_{IJ}$ determines the kinetic terms for the vector fields.
Therefore it must be positive definite in Lorentzian space-time signature. 
It is known that by choosing $g=N_{IJ}dX^I d\bar{X}^J$ to have 
signature $(2n,2)$, the scalar couplings $\bar{g}_{AB}$ and vector 
couplings $- {\cal I}_{IJ}$ are positive definite. We remark
that both 
$-{\cal I}_{IJ}dX^I d\bar{X}^J$ and 
$g=N_{IJ}dX^I d\bar{X}^J$ can be viewed as metrics
on the scalar manifold $M$, and are related to one another by 
a simple geometric operation which flips the signature along a
complex one-dimensional subspace \cite{Cortes:2011aj}.

In the Euclidean case the metric
$-{\cal I}_{IJ} dX^I d\bar{X}^J$ always has split-signature,
irrespective of the signature of the real matrix
$-{\cal I}_{IJ}$. If the Euclidean theory has been obtained
by dimensional reduction of a five-dimensional theory with
respect to time, then the matrix 
$-{\cal I}_{IJ}$ has Lorentz signature $(n,1)$,
with the negative definite direction corresponding to the 
Kaluza-Klein vector.
The metrics $g$ and 
$\bar{g}$ are para-K\"ahler and have 
split-signature $(n+1,n+1)$ and $(n,n)$, respectively. 
  
Electric magnetic duality acts on the gauge fields through 
the linear action of $Sp(2n+2,\mathbbm{R})$ on the
vector $(F^I_{\hat{\mu} \hat{\nu}}, {\cal N}_{IJ} F^J_{\hat{\mu} \hat{\nu}})^T$.

\subsection{Reduction to three dimensions}\label{3dSec}

We now carry out the reduction of the four-dimensional 
vector multiplet Lagrangian (\ref{VMLagrangian}) to 
three dimensions. This type of calculation is standard,
so we will not give many details, though we need to specify
our notation and conventions. 
If we start with Lorentzian signature ($\epsilon_1=-1$) 
in four-dimensions we
have the option to either reduce over space, or over time, 
which will be distinguished by a new parameter $\epsilon_2$, 
where $\epsilon_2=-1$ for spacelike reduction and $\epsilon_2=1$
for timelike reduction. If we start with a Euclidean theory ($\epsilon_1=1$), 
then we can only reduce over space, so $\epsilon_2=-1$. All three
cases will be treated simultaneously up to a certain point. 

The reduction is performed along the lines of \cite{Mohaupt:2011aa},
with the following modifications: (i) we now include the reduction
of four-dimensional Euclidean theories, (ii) some fields have been
renamed, (iii) the definition of the Riemann tensor has been changed
by an overall sign. For completeness, we 
briefly review the relation between 
the four-dimensional and three-dimensional quantities. 
Four-dimensional indices $\hat{\mu}, \hat{\nu}, \ldots$ are split into 
three-dimensional indices $\mu,\nu,\ldots$ and the index $y$, which refers
to the dimension we reduce over.
We decompose the four-dimensional metric 
\[
ds_4^2 = 
-\epsilon_2 e^\phi (dy + V_\mu dx^\mu)^2 + e^{-\phi} g_{\mu\nu} dx^\mu dx^\nu
\]
into a three-dimensional metric $g_{\mu\nu}$, the Kaluza-Klein vector
$V_\mu$ and the Kaluza-Klein scalar $\phi$. The four-dimensional 
vector fields have been decomposed into a scalar part $\zeta^I = A^I_y$
and a vector part $A^I_\mu - \zeta^I V_\mu$, with the second term 
restoring manifest gauge invariance. The three-dimensional field
strengths $V_{\mu\nu}$ and $F^I_{\mu\nu}$ have then been dualised into
scalars $\tilde{\phi}$ and $\tilde{\zeta}_I$, see \cite{Mohaupt:2011aa}
for details. Instead of the four-dimensional scalars $z^A$ we are
using the corresponding superconformal scalars $X^I$ and the 
degenerate tensor $g_{IJ}$.

The resulting three-dimensional Lagrangian is
\begin{eqnarray}
\label{3dLagrangian}
&& {\bf e}_3^{-1} {\cal L}_3^{(\epsilon_1, \epsilon_2)}  =     \\
&& \frac{1}{2} R_3 - \left. g_{IJ} \partial_\mu X^I \partial^\mu \bar{X}^J
\right|_D 
-\frac{1}{4} \partial_\mu \phi \partial^\mu \phi \nonumber \\
&& +\epsilon_1 e^{-2\phi} \left[ \partial_\mu \tilde{\phi} + \frac{1}{2}
\left( \zeta^I \partial_\mu \tilde{\zeta}_I 
- \tilde{\zeta}_I \partial_\mu \zeta^I \right) \right]
\left[ \partial^\mu \tilde{\phi} + \frac{1}{2}
 \left( \zeta^I \partial^\mu \tilde{\zeta}_I 
- \tilde{\zeta}_I \partial^\mu \zeta^I \right) \right]
\nonumber \\ 
&& -\frac{\epsilon_2}{2} e^{-\phi} \left[ {\cal I}_{IJ} \partial_\mu \zeta^I
\partial^\mu \zeta^J - \epsilon_1 {\cal I}^{IJ}
\left(\partial_\mu \tilde{\zeta}_I - {\cal R}_{IK} \partial_\mu \zeta^K\right)
\left(\partial^\mu \tilde{\zeta}_J - {\cal R}_{JL} \partial^\mu \zeta^L\right)
\right] \;. \nonumber
\end{eqnarray}
For $\epsilon_1=\epsilon_2=-1$ this 
agrees, up to conventional choices, with \cite{Ferrara:1989ik},
and for $\epsilon_1=- \epsilon_2=-1$ this agrees, up to the above mentioned 
changes in conventions, with \cite{Mohaupt:2011aa}.

As explained in \cite{Mohaupt:2011aa}, one can absorb the Kaluza-Klein
scalar $\phi$ into the scalar fields $X^I$ by the field redefinition
$Y^I = e^{\phi/2} X^I$. These fields 
are  now related to the Kaluza-Klein scalar
$\phi$ by the D-gauge condition
\[
e^\phi = -i_{\epsilon_1}(Y^I \bar{F}_I - F_I \bar{Y}^I)\;.
\]
So $\phi$ will be now considered as a function of the independent variables $Y^I$. Geometrically,
we interpret $\phi$ as a coordinate
along the orbit of the homothetic action 
of $\mathbbm{R}^{>0}$ on $M$. Using homogeneity, we can rewrite
the scalar terms as follows:
\[
\left. g_{IJ}(X) \partial_\mu X^I \partial^\mu \bar{X}^J \right|_{D} + 
\frac{1}{4} \partial_\mu \phi \partial^\mu \phi =
g_{IJ}(Y) \partial_\mu Y^I \partial^\mu \bar{Y}^J + 
\left. \frac{1}{4} \partial_\mu \phi \partial^\mu \phi \right|_{\phi = 
\phi(Y)}\;.
\] 
Note that while both expressions 
take the same form, on the left hand side the fields $X^I$ are
subject to the D-gauge, while $\phi$ is an independent field. 
In contrast on the right hand side $\phi$ is considered to be 
a dependent field, which can be expressed in terms of the $Y^I$. 
Since both sides of the equation are invariant under local 
$U_{\epsilon_1}(1)$ transformations, both sets of fields 
represent the same $2n+1$ independent physical real degrees of freedom.

Now we interpret the $Y^I$ as $\epsilon_1$-holomorphic
special coordinates on $M$. We can therefore rewrite the theory 
in terms of the associated special real coordinates $q^a$, defined
by decomposing
\[
Y^I = x^I + u^I(x,y) \;,\;\;\;
F_I = y_I + v_I(x,y) \;,
\]
and setting $(q^a) = (x^I, y_I)^T$. Note that in this parametrisation 
the Kaluza-Klein scalar is expressed in terms of $q^a$ by 
\be \label{eq:phi}  e^\phi =  -i_{\epsilon_1}(Y^I \bar{F}_I - F_I \bar{Y}^I)=- 2H,\ee 
where  we recall that $H$ denotes the Hesse potential. 
We also define
$\hat{q}^a =\frac{1}{2} ( \zeta^I, \tilde{\zeta}_I)^T$ and 
remark that both $q^a$ and $\hat{q}^a$ are symplectic vectors while
the dualised Kaluza-Klein vector
$\tilde{\phi}$ is a symplectic scalar. As explained in detail 
in \cite{Mohaupt:2011aa}, the Lagrangian (\ref{3dLagrangian})
can be written in terms of the $4n+5$ fields
$(q^a, \hat{q}^a,\tilde{\phi})$ with 
all couplings expressed using the Hesse potential $H$, 
the tensor field 
\[
\tilde{H}_{ab} = \frac{\partial^2}{\partial q^a \partial q^b}
\left[ - \frac{1}{2} \log (-2H) \right] =
- \frac{1}{2H} H_{ab} + \frac{1}{2H^2} H_a H_b \;,
\]  
and the constant matrix $\Omega_{ab}$ representing the symplectic form (\ref{Omegaab}): 
\begin{eqnarray}
\label{3dLagrangianH}
&& {\bf e}_3^{-1} {\cal L}_3^{(\epsilon_1,\epsilon_2)}  =     \\
&& \frac{1}{2} R_3 - \tilde{H}_{ab} 
\left( \partial_\mu q^a \partial^\mu q^b - 
\epsilon_2 \partial_\mu \hat{q}^a \partial^\mu \hat{q}^b \right) \nonumber \\
&&+ \epsilon_1 \frac{1}{H^2} 
\left( q^a \Omega_{ab} \partial_\mu q^b \right)
\left( q^a \Omega_{ab} \partial^\mu q^b \right) \nonumber \\
&& - \epsilon_1\epsilon_2 \frac{2}{H^2} 
\left( q^a \Omega_{ab} \partial_\mu \hat{q}^b \right)
\left( q^a \Omega_{ab} \partial^\mu \hat{q}^b \right) \nonumber \\
&& + \epsilon_1 \frac{1}{4H^2} 
\left( \partial_\mu \tilde{\phi} + 2 \hat{q}^a \Omega_{ab} \partial_\mu \hat{q}^b
\right)
\left( \partial^\mu \tilde{\phi} + 2 \hat{q}^a \Omega_{ab} \partial^\mu \hat{q}^b
\right) \;. \nonumber
\end{eqnarray}
Since all local degrees of freedom have been converted into scalars,
the Lagrangian (\ref{3dLagrangianH}) is a
non-linear sigma model coupled to gravity. 
The $4n+5$ real scalar fields $(q^a, \hat{q}^a, \tilde{\phi})$
are local coordinates of its target space $P$. 
Due to the local $U_{\epsilon_1}(1)$ symmetry, there are only 
$4n+4$ physical degrees of freedom, and the symmetric tensor
field on $P$ defined by the Lagrangian is invariant and degenerate
along the orbits of the $U_{\epsilon_1}(1)$-action. 
By gauge-fixing this symmetry
we can obtain a sigma model with a $(4n+4)$-dimensional target manifold
$\bar{N}$, equipped with a non-degenerate metric. 
Since $U_{\epsilon_1}(1)$ acts on the symplectic vector $q^a$,
while $\hat{q}^a$ and $\tilde{\phi}$ are invariant, such a gauge 
fixing will break the manifest symplectic invariance of the 
sigma model with target $P$. Therefore 
it is advantageous to describe $\bar{N}$ in terms of the larger
space $P$.

In the following sections we will show that $P$ is a principal
$U_{\epsilon_1}(1)$-bundle over $\bar{N}$, and that the
degenerate symmetric tensor on $P$ defined by the 
Lagrangian (\ref{3dLagrangianH}) projects down to
an $\varepsilon$-quaternionic K\"ahler metric on $\bar{N}$. 
For $\epsilon_1=\epsilon_2=-1$ we recover the result of
\cite{Ferrara:1989ik}, in which it was shown that $\bar{N}$ is
quaternionic K\"ahler. For 
$\epsilon_1=1$, $\epsilon_2=-1$ and $\epsilon_1=-1$, 
$\epsilon_2=1$, we prove that the manifold $\bar{N}$ is
para-quaternionic K\"ahler.

\section{Global construction of the $c$-map}
\label{sec:cmap}

\subsection{Geometric data on a conic affine special $\varepsilon$-K\"ahler manifold}
\label{sec:PSKbase}

The starting point for our construction of the $c$-map will be a regular, simply connected CAS$\varepsilon$K manifold $M$, see Section \ref{Sec:SpecialK}. The purpose of this section is to introduce a global orthogonal co-frame on $M$ and to express certain geometrical data in terms of this co-frame. We are specifically interested in the cubic tensor $C = \nabla g$, the difference tensor $S = D - \nabla$, and the pull-back of the Levi-Civita connection one-form on the corresponding PS$\varepsilon$K manifold $\sigma = \pi^* \bar{\sigma}$. The necessary expressions are given by \eqref{eq:C}, \eqref{eq:S} and \eqref{eq:LConM} respectively. 

Let  $(M, J, g, \nabla, \xi)$ be a regular, simply connected CAS$\varepsilon$K manifold of dimension $2n + 2$, which in the case $\varepsilon = -1$ has  signature $(2k, 2\ell + 2)$, $k + \ell = n$. Recall that $M$ is a principal $\mathbbm{C}_\varepsilon^*$-bundle over a PS$\varepsilon$K manifold $(\bar{M},\bar{J},\bar{g})$ of dimension $2n$, with fundamental vector fields
\[
	\xi = q^a \frac{\partial}{\partial q^a} \;, \qquad J\xi = \tfrac{1}{2} H_b \Omega^{ba} \frac{\partial}{\partial q^a} \;.
\]
For the case $\varepsilon = -1$ the signature of $\bar{M}$ is $(2k,2\ell)$.
The tangent space at any point $p \in M$ decomposes into horizontal and vertical parts $T_pM  = {\cal H}_p \oplus \text{span}\{\xi_p,J\xi_p\}$, where the horizontal space 
${\cal H}_p$ is defined as the orthogonal complement of  the vertical space of $\pi : M \ra \bar{M}$ and we identify ${\cal H}_p \simeq  T_{\pi(p)} \bar{M}$ by the 
projection.
Recall that there exists a global $\nabla$-affine symplectic frame on $M$ given by ${\cal B} = \left(\frac{\partial}{\partial q^0}, \ldots, \frac{\partial}{\partial q^{2n + 1}} \right)$, which is unique up to $Sp(2n + 2, {\mathbbm R})$ transformations. Recall also that the projective special $\varepsilon$-K\"ahler metric on $\bar{M}$ is induced by the 
degenerate tensor $h$ defined in equation (\ref{eq:PSKInducedMetric}). Therefore the horizontal lift of an orthonormal
frame defined over an open subset $U\subset \bar{M}$ yields an $h$-orthonormal  frame ${\cal B}' = \left( e_1, \ldots, e_{2n}\right)$  
of ${\cal H}$, defined over $\pi^{-1}(U)$, such that 
\[ h(e_m,e_p) =\eta_{mp}\,,\quad  m,p = 1, \ldots, 2n \;,\]
where
\begin{equation}
	(\eta_{mp})  = \left( 
		\begin{array}{cc}
			\Alteta & 0 \\
			0 & -\varepsilon \Alteta
		\end{array}
	\right) \;,
	\qquad
	\Alteta = 
	\begin{cases} 
	\mbox{diag}({\mathbbm 1}_{k}, -{\mathbbm 1}_\ell) \;, & \varepsilon = -1 \\ 
	\mbox{diag}({\mathbbm 1}_n) \;, & \varepsilon = +1  \;.
	\end{cases} 
	\label{eq:eta}
\end{equation}
Moreover, by choosing the orthonormal frame on $U\subset \bar{M}$ adapted to the
$\varepsilon$-complex structure we can further assume that 
\begin{align*}
	J(e_{A}) &= e_{A + n}\:, 
	\qquad
	J (e_{A + n}) = \varepsilon e_A \;,
	\qquad 
	A = 1,\ldots,n \;.
\end{align*}
In such a frame the $\varepsilon$-complex structure $J(e_p) = J^m_{\;\;\;p} e_m$ is represented by the constant matrix
\begin{equation}
	(J^m_{\;\;\;p}) = \left( \begin{array}{cc} 0 & \varepsilon {\mathbbm 1} \\ {\mathbbm 1} & 0  \end{array}\right) \;.
	\label{eq:Jmatrix}
\end{equation}
Such a choice of frame is not unique, with any two choices differing by a gauge transformation with 
values in  
\[ {U}_{\varepsilon}(k,\ell) \subset SO_\varepsilon(2k,2\ell) := \begin{cases} 
	SO(2k ,2\ell) \;, & \varepsilon = -1 \\ 
	SO({n},{n}) \;, & \varepsilon = +1  \;, \quad n=k+\ell\;.
	\end{cases}  \]

We consider the frame ${\cal B}'{}^* = (e^1,\ldots,e^{2{n}})$  of  ${\cal H}^* = \mbox{span}\{\xi, J\xi\}^0\subset T^*M$ dual to ${\cal B}'$; that is $e^m(e_p) = \delta^m_p$. 
(Here the superscript $0$ refers to the annihilator.) Note that the $(0,2)$-tensor $h$ takes the form $h = \eta_{mp} e^m \otimes e^p$.
Similarly, on span$\{\xi,J\xi\}^* = {\cal H}^0$ one may introduce the frame $(u,v)$ given by
\begin{equation}\label{eq:uv} 
	u = \frac{1}{2H} dH \;, \qquad v = \varepsilon J^*u= -\frac{1}{H} {q}^a \Omega_{ab} dq^b \;, 
\end{equation}
which is dual to $(\xi , J\xi )$.

Consider the orthogonal projection $\varphi:T M \to{\cal H}$. Using the frame  ${\cal B}$ of $TM$ and the local frame ${\cal B}'$ of ${\cal H}$ this projection is represented by the matrix
\[
	M_{{\cal B},{\cal B}'}(\varphi) = {\cal P} = ({\cal P}^m_{\;\;\;a})_{a = 0,\ldots, 2n + 1}^{m = 1,\ldots, 2n} \;.
\]
One may also consider the corresponding dual map $\varphi^*: {\cal H}^* \rightarrow T^* M$, which with the above identifications is simply the inclusion map. Using the dual basis ${\cal B}^* = (dq^0,\ldots,dq^{2n + 1})$ of $T^*M$ and ${\cal B}'{}^*$ of ${\cal H}^*$ this map is represented by the transposed matrix
\[
	M_{{\cal B}'{}^*,{\cal B}^*}(\varphi^*) = {\cal P}^t = ({\cal P}_a^{\;\;m})_{a = 0,\ldots, 2n + 1}^{m = 1,\ldots, 2n} \;,
\]
in other words 
\be \label{eq:edq} e^m=	{\cal P}_a^{\;\;m}dq^a\,.\ee

It is useful to consider also the inclusion map $\iota :{\cal H} \rightarrow TM$, which is characterised by 
\be \label{varphi^perp}\varphi \circ \iota = \text{Id}_{{\cal H}}\,,\quad  \varphi^\perp \circ \iota =0,\ee
where $\varphi^\perp : TM \rightarrow {\mathcal H}^\perp = \mathrm{span}\{ \xi , J\xi \}$ is the orthogonal projection onto ${\cal H}^\perp$. 

\bl
\label{lem:Projection} The matrix $M_{{\cal B}',{\cal B}}(\iota )$ representing the inclusion $\iota : {\cal H} \rightarrow TM$ is given by 
\be   \label{eq:Tab} {\cal T} = \left( {\cal T}^{a}_{\;\;m} \right)_{a = 0,\ldots, 2n + 1}^{m = 1,\ldots, 2n}\,,\quad  {\cal T}^{a}_{\;\;m} =-2H H^{ab}  {\cal P}_{\;\;b}^{p} \eta_{pm}.\ee
We have the equation
\be \label{eq:en} e_m = {\cal T}^{a}_{\;\;m}\frac{\partial }{\partial q^a}\; ,\ee 
and the matrix ${\cal P}^t$ has the following properties:
	\begin{enumerate}[(i)]
	\item \hspace{2em} 
	$
		{\cal P}_a^{\;\;m} q^a = {\cal P}_a^{\;\;m} \Omega^{ab} H_b = 0 \;.
	$
	\item \hspace{2em} 
	$
		\eta_{mp}({\cal P}_a^{\;\;m} {\cal P}_b^{\;\;p}) = h_{ab}=-\dfrac{1}{2H}H_{ab} + \dfrac{1}{4H^2} H_a H_b - \varepsilon \dfrac{1}{H^2} \Omega_{ac}q^c \Omega_{bd} q^d \;.
	$
	\item \hspace{2em} 
	$
		-2H {\cal P}_a^{\;\;m} H^{ab}\, {\cal P}_b^{\;\;p} = \eta^{mp} \;.
	$
	\item \hspace{2em} 
	$
		H({\cal P}_a^{\;\;m} \Omega^{ab} {\cal P}_b^{\;\;q})\eta_{qp}  = J^*{}_{p}^{\;\;m} = J^m_{\;\;\;p} \;, 
	$
	\end{enumerate}
	where $(J^m_{\;\;\;p})$ is the constant matrix \eqref{eq:Jmatrix} representing the tensor $J|_{\cal H} : {\cal H}\rightarrow {\cal H}$ in the frame $(e_m)$. 
\el
\pf
	Part (i) follows immediately from the fact that $\ker \varphi = \text{span}\{\xi,J\xi\}$. For part (ii) one may use the fact that 
\begin{align*}
	h 	&
		= \varphi ^*h
		= \eta_{mp}\left({\cal P}_a^{\;\;m} {\cal P}_b^{\;\;p} \right) dq^a \otimes dq^b \;,
\end{align*}
and therefore $\eta_{mp}({\cal P}_a^{\;\;m} {\cal P}_b^{\;\;p}) = h_{ab}$, cf.\  (\ref{eq:hab}). 
For part (iii) we note that $-\frac{1}{2H}g$ and $h$ coincide when restricted to ${\cal H}$
\[
	 -\frac{1}{2H} g \big|_{{\cal H}} = h \big|_{{\cal H}}  \;.
\]
Since this is non-degenerate on ${\cal H}$ we can invert this formula 
\[
	-2H g^{-1} \circ (\varphi^*,\varphi^*) =-2H g^{-1}\big|_{{\cal H}^*} = -2H (g|_{\cal H})^{-1}= (h \big|_{{\cal H}} )^{-1}=\eta^{mn} e_m \otimes e_n\;.
\]
Plugging in $({e}^m,{e}^p)$ gives expression (iii). Using (iii) one can easily check that 
${\cal T}$ satisfies the equation 
\[  {\cal P}^m_{\;\;a}{\cal T}^a_{\;\;p} = \delta^m_p.\]
Using (i) one can also check that the vectors ${\cal T}^a_{\;\;m}\frac{\partial}{\partial q^a}$ are perpendicular to $\xi$ and $J\xi$. 
In view of the characterisation (\ref{varphi^perp}), this proves that the matrix ${\cal T}$ represents the  inclusion map $\iota :{\cal H} \rightarrow TM$. 
The latter property implies (\ref{eq:en}). 

For part (iv) we note that $\varphi^* \circ  {J}^* = J^* \circ \varphi^*: {\cal H}^* \to T^*M$. Acting on ${e}^m$ this gives
\begin{equation}
	 J^*{}^{\;\;m}_{p} e^p={\cal P}_a^{\;\;m} J^*{}^{\;\,a}_{b} dq^b \;. \label{eq:PhiHol}
\end{equation}
 Plugging $e_n$ into this expression and using (\ref{eq:en}), (\ref{eq:Tab}) and (\ref{eq:ASKCx}) gives the desired result.
\qed

Let us now turn our attention to the cubic tensor $C = \nabla g = H_{abc} dq^a \otimes dq^b \otimes dq^c$, where $H_{abc}$ are the triple derivatives of the Hesse potential. 
The cubic tensor is related to the difference tensor 
\[ S=D-\nabla = \frac{\varepsilon}{2}J\nabla J\;, \] 
by the formula 
\be \label{eq:CS} C(X,Y,Z) = 2g(S_XY,Z),\ee
which is  immediately obtained from $g = \omega (J \cdot ,\cdot )$. 
Differentiating the equation $g(JX,JY)=\varepsilon g(X,Y)$ with respect to $\nabla$ and using the equation (\ref{eq:CS})
one can prove that 
\begin{equation}
	C (\cdot, J\cdot, J\cdot) = \varepsilon C (\cdot,\cdot,\cdot) \;. 
	\label{eq:CJJJ}
\end{equation}	
The cubic tensor is degenerate  with kernel containing $\mathrm{span}\{ \xi,J\xi \}$ but not $\mathbbm{C}_{\varepsilon}^*$-invariant, and therefore not well-defined on $\bar{M}$. 
In the above  local frame of ${\cal H}^*\subset T^*M$ we may write $C  = C_{mpq} \, e^m \otimes e^p \otimes e^q$, where the components are symmetric and satisfy
\begin{equation}
	 C_{mpq} =  \sum_{a,b,c = 0}^{2{n} + 1} {\cal T}_{\;\;m}^{a}{\cal T}_{\;\;p}^{b}{\cal T}_{\;\;q}^{c} H_{abc} \;. 	\label{eq:C}
\end{equation}
Due to (\ref{eq:CS}) the components of the difference tensor $S = S^m_{\;\;\;pq}  \, e_m \otimes e^p \otimes e^q$ are given by
\[
	S^m_{\;\;\;pq} = -\frac{1}{4H} \eta^{mr} C_{rpq} \;,
\]
and the one-forms $X\mapsto S^m_{\;\;\;\,p}(X) = e^m(S_Xe_p)$  by 
\begin{equation}
	S^m_{\;\;\;\,p} 
	= \sum_{q = 1}^{2{n}}S^m_{\;\;\;pq} e^q
	= \sum_{q = 1}^{2{n}} \sum_{a,b = 0}^{2{n} + 1} -\frac{1}{4H} \eta^{mq} {\cal T}_{\;\;q}^{a}{\cal T}_{\;\;p}^{b} d H_{ab}\;. \label{eq:S}
\end{equation}
It follows from (\ref{eq:CJJJ}) that 
	\begin{align}
		\big(S^A_{\;\;\,B}\big) &= \Alteta \big(S^A_{\;\;\,B}\big)^T \Alteta \;, & 
		S^A_{\;\;\,B} &= -S^{A + {n}}_{\phantom{{A + {n}}} B + {n}}\;, \notag \\
		\big(S^{A + {n}}_{\phantom{A + {n}}B}\big) &= \Alteta \big(S^{B + {n}}_{\phantom{B + {n}}A} \big)^T \Alteta \;,  &	
		S^{A + {n}}_{\phantom{{A + {n}}} B} &= -\varepsilon S^A_{\;\;\,{B + {n}}} \;. 
		\label{eq:Scomps}
	\end{align}

We end this section by computing the pull-back to $M$ of the Levi-Civita connection one-form on the corresponding PS$\varepsilon$K manifold $\bar{M}$.
\bl
Let $\sigma \in \Omega^1(M,{\mathfrak so}_\varepsilon (k,\ell))$ denote the pull-back of the Levi-Civita connection one-form $\bar{\sigma}$ on $\bar{M}$. The components of ${\sigma}$ in the above  local frame of ${\cal H}^*\subset T^*M$ are given by 
\begin{equation}
 \sigma^m_{\;\;\;p} = -v \, J^m_{\;\;\;p}  + H \left(d{\cal P}_a^{\;\;m} H^{ab}\, {\cal P}_b^{\;\;q}  - {\cal P}_a^{\;\;m} H^{ab}\, d{\cal P}_b^{\;\;q} \right) \eta_{qp}  \;,
	\label{eq:LConM}
\end{equation}
where the one-form $v$ was defined in (\ref{eq:uv}).  
\el
\begin{proof}
The Levi-Civita connection one-form $\bar{\sigma}$ on $\bar{M}$ is uniquely determined by the requirement that it is metric compatible and torsion-free. In terms of the pull-back $\sigma$ the metric compatibility condition implies that 
	\begin{equation}
		\left(\sigma^m_{\;\;\;p}\right) \; \in \; 
		\begin{cases} 
		\Omega^1({M},\mathfrak{so} (2k,2\ell)) \;, & \varepsilon = -1 \\ 
		\Omega^1({M},\mathfrak{so} (n,n))  \;, & \varepsilon = +1  
		\end{cases} 
		\;\;\;\; \Leftrightarrow \;\;\;\;
		\sigma^{mp} = \sigma^m_{\;\;\;q} \eta^{qp} = -\sigma^{pm} \;, \label{eq:MetricConnection}
	\end{equation}
	which is easily seen to be satisfied by (\ref{eq:LConM}). 
	The  torsion-free condition implies that 
	\begin{equation}
		de^m + \sigma^m_{\;\;\;p} \wedge e^p = 0 \;, \label{eq:TorsionFree}
	\end{equation}
	which we will now show is satisfied by (\ref{eq:LConM}). Using Lemma \ref{lem:Projection} (iii) and (iv) we have 
\begin{align*}
	\sigma^m_{\;\;\;p} \wedge e^p 
		&= - (dq \Omega q)  \wedge {\cal P}_a^{\;\;m} \Omega^{ab} {\cal P}_b^{\;\;p} \eta_{pq} e^q - \frac{1}{2H} dH \wedge e^m \\
		&\hspace{2em}+ \;\underbrace{2H d{\cal P}_a^{\;\;m} H^{ab} {\cal P}_b^{\;\;p} \eta_{pq}  \wedge e^q}_{(*)}  \;+\; \underbrace{H {\cal P}_a^{\;\;m} dH^{ab} {\cal P}_b^{\;\;p} \eta_{pq} \wedge e^q}_{S^m_{\;\;\;p} \wedge e^p = 0} \;.
\end{align*}
To calculate the last term we have used (\ref{eq:S}) and (\ref{eq:Tab}). It vanishes in virtue of the symmetry of $S$.  
Calculating $(*)$ individually using (\ref{eq:id3}), (\ref{eq:edq}) and Lemma \ref{lem:Projection}  (i--ii) we find 
\begin{align*}
	(*) &= 2H d{\cal P}_a^{\;\;m} H^{ab}  \wedge \left[ -\frac{H_{bc}}{2H} + \frac{H_b H_c}{4H^2}  - \varepsilon \frac{\Omega_{bd}q^d \Omega_{ce}q^e}{H^2} \right] dq^c  \\
	&= -d{\cal P}_a^{\;\;m} \wedge dq^a + \frac{1}{2H} d{\cal P}_a^{\;\;m} q^a \wedge d H - \varepsilon \frac{2}{H} d{\cal P}_a^{\;\;m} {H^{ab} \Omega_{bc} q^c} \wedge (dq \Omega q)  \\
	&= - de^m + \frac{1}{2H} dH \wedge e^m + (dq \Omega q) \wedge {\cal P}_a^{\;\;m} \Omega^{ab} {\cal P}_b^{\;\;p}\eta_{pq}e^{q} \;,
\end{align*}
and therefore expression (\ref{eq:TorsionFree}) is satisfied.

By Proposition \ref{prop:Connection} of Section \ref{sec:LCint} the solution to (\ref{eq:MetricConnection}) and (\ref{eq:TorsionFree}) is unique, and, moreover, it is precisely the pull-back of the Levi-Civita connection one-form on $\bar{M}$.
\end{proof}

For an $\varepsilon$-K\"ahler manifold the Levi-Civita one-form satisfies $J^* \circ \sigma = \sigma \circ J$. It follows that
	\begin{align}
		\left(\sigma^A_{\;\;\,B} \right) &= - \Alteta \big( \sigma^A_{\;\;\,B}\big)^T \Alteta \;, & 
		\sigma^A_{\;\;\,B} &= \sigma^{A + {n}}_{\phantom{{A + {n}}} B + {n}} \;, \notag \\
		\big(\sigma^{A + {n}}_{\phantom{A + {n}}B}\big) &= \Alteta \big( \sigma^{A + {n}}_{\phantom{B + {n}} A}  \big)^T \Alteta \;, &
		\sigma^{A + {n}}_{\phantom{{A + {n}}} B} &= \varepsilon \sigma^A_{\;\;\,{B + {n}}} \;. \label{eq:SigmaComps}
	\end{align}

\subsection{The $c$-map for various spacetime signatures}
\label{sec:cmapVar}

In this section we will construct the $c$-map target manifold $(\bar{N},g_{\bar{N}},Q_{\bar{N}})$. We will present this construction for the spatial, temporal and Euclidean $c$-maps in a unified way using the $(\epsilon_1,\epsilon_2)$-notation introduced previously. We will begin with the topological data on $\bar{N}$, before moving on to the metric $g_{\bar{N}}$ and $\varepsilon$-quaternionic structure $Q_{\bar{N}}$.

Consider a regular, simply connected CAS$\epsilon_1$K manifold $M$ of dimension $2n + 2$. 
Given $M$ one may construct the ($4n + 5$)-dimensional manifold $P = TM\times {\mathbbm R}$, that is the product of the tangent bundle of $M$ with the real line. On $P$ we have $4n + 5$ global real functions $(q^a,\hat{q}^b,\tilde{\phi})$ which are defined as follows. We start with the globally-defined  functions  $(q^a)$ on $M$, 
introduced before, which restrict to special real coordinates in a neighbourhood $U$ of any point of $M$. The function $\hat{q}^b$ on $TM$ is defined by the property that it 
takes the value $v^b$ on the vector $v^a\frac{\partial}{\partial q^a}$.  As we have natural projections 
$P\rightarrow TM$ and $TM \rightarrow M$,  the functions $q^a,\hat{q}^b,$ can be considered as functions on $P$. 
Finally $\tilde{\phi}$ is the coordinate on the ${\mathbbm R}$-factor. Notice that the functions $(q^a,\hat{q}^b,\tilde{\phi})$ 
restrict to a local coordinate system in a neighbourhood of any point of $P$, 
in fact, they give coordinates on the open set 
$pr^{-1}(U)\times \mathbbm{R} \subset P$. 

The principal  $\mathbbm{C}^*_{\epsilon_1}$-action on $M$ may be lifted to a principal $\mathbbm{C}^*_{\epsilon_1}$-action on $P$ in the following way. 
Using the global frame $(\frac{\partial}{\partial q^a})$ we can identify the vector bundle $TM$ with the trivial bundle $M\times \mathbbm{R}^{2n+2}$
and extend the principal $\mathbbm{C}_\varepsilon^*$-action on $M$ trivially to a principal $\mathbbm{C}_\varepsilon^*$-action on $TM$ and $P=TM\times \mathbbm{R}$. 
The actions $\varphi_\lambda^M$ and $\varphi_\lambda^P$ of an element $\lambda\in \mathbbm{C}_\varepsilon^*$ on $M$ and $P$ are related by
\be  \label{eq:action} (\varphi_\lambda^P)^*q^a =   pr^*(\varphi_\lambda^M)^*(q^a|_{M}),\quad (\varphi_\lambda^P)^*\hat{q}^a =\hat{q}^a,\quad (\varphi_\lambda^P)^*\tilde{\phi}=
\tilde{\phi}.\ee
On $P$ one may consider the principal action of the subgroup $U_{\epsilon_1}(1) \subset \mathbbm{C}^*_{\epsilon_1}$. In this way, one may interpret $P$ as a principal $U_{\epsilon_1}(1)$ bundle over a manifold $\bar{N}$. Let $Z_P$ be the vector field generating the infinitesimal $U_{\epsilon_1} (1)$-action on $P$. This is precisely the horizontal lift of the vector field $J\xi$ on $M$, and is given  by  
\[ Z_P = \frac12 H_a \Omega^{ab} \frac{\partial}{\partial q^b}.\] 
We define the $c$-map target manifold as the orbit space 
\[ \bar{N} = P/U_{\epsilon_1}(1)\;,\] 
which by construction has dimension $4n + 4$. This information is summarised in Figure \ref{fig:GlobalCmap}.  Notice that in the case $\epsilon_1=1$ the manifold $\bar{N}$ 
has at least two connected components distinguished by the sign of the Hesse potential $H$. 
\begin{figure}
\[
	\xymatrix{
		\stackrel[2n + 2]{}{M} \ar[d]^{\mathbbm{C}_{\epsilon_1}^*} &  \stackrel[4n + 4]{}{TM} \ar[l]  & \stackrel[4n + 5]{}{P} \ar[l]  \ar[d]^{U_{\epsilon_1}(1)} \\
  		\stackrel[2n]{}{\bar{M}} \ar@{|-{>}} [rr] ^{\text{$c$-map}} &  & \stackrel[4n + 4]{}{\bar{N}}
	}
\]
\caption{The global construction of the $c$-map target manifold.}
\label{fig:GlobalCmap}
\end{figure}

The non-linear sigma model of the dimensionally reduced Lagrangian (\ref{3dLagrangianH}) defines on $P$ the symmetric bilinear form
\begin{align}
	{g}' \;=\;\; &\tilde{H}_{ab} \left(d q^a \otimes dq^b - \epsilon_2\, d\hat{q}^a \otimes d \hat{q}^b \right) \notag \\ 
			&- \epsilon_1 \,\frac{1}{H^2} \left( q^a \Omega_{ab} d q^b \right) \otimes \left( q^a \Omega_{ab} d q^b \right) + \epsilon_1 \epsilon_2\, \frac{2}{H^2} \left( q^a \Omega_{ab} d \hat{q}^b \right) \otimes \left( q^a \Omega_{ab} d \hat{q}^b \right)  \nonumber \\
			&- \epsilon_1 \,\frac{1}{4 H^2} \left( d \tilde{\phi} + 2\hat{q}^a \Omega_{ab} d \hat{q}^b  \right) \otimes \left( d \tilde{\phi} + 2\hat{q}^a \Omega_{ab} d \hat{q}^b  \right) \;, \label{eq:gprimeepsilons}
\end{align}
where $\epsilon_1$ and $\epsilon_2$ are determined by the different $c$-maps according to the rule (\ref{eq:Epsilons}) and where $H\neq 0$ is now allowed to change
sign and the PS$\epsilon_1$K metric is allowed to be indefinite. The bilinear form ${g}'$ has a one-dimensional kernel ${\mathbbm R} Z_P$ and is invariant under the $U_{\epsilon_1} (1)$-action on $P$. It may therefore be pushed-down to give a well-defined metric $g_{\bar{N}}$ on $\bar{N}$. This procedure makes sense even in the case $\epsilon_1 =\epsilon_2=1$, which we have not given a physical interpretation so far. As the next proposition
shows,  this case gives a metric equivalent to the one in the case $\epsilon_1 =-\epsilon_2=1$.

\bp
	For the case $\epsilon_1 = 1$ the pull-back of ${g}'$ under the  action of $e\in C^*$ is given by
	\[
		(\varphi^P_e)^* \left(  {g}'\big|_{(\epsilon_1,\epsilon_2) = (1,1)} \right) = {g}'\big|_{(\epsilon_1,\epsilon_2) = (1,-1)} \;.
	\]
\ep
\pf
	The pull-back of the functions $(q^a,\hat{q}^b,\tilde{\phi})$ are given by
	\[
		(\varphi^P_e)^* q^a = \epsilon_1 J^a_{\;\;b} q^b \;,
		\qquad
		(\varphi^P_e)^* \hat{q}^a = \hat{q}^a \;,
		\qquad
		(\varphi^P_e)^* \tilde{\phi} = \tilde{\phi} \;,
 	\]
	cf.\ {(\ref{eq:action}). 
	In fact, the first term is computed as follows using (\ref{eq:Ha}):   
\[	 (\varphi^P_e)^*\left( \begin{array}{c} x^I\\ y_J
		\end{array} \right)  = \left( \begin{array}{c} u^I\\ v_J
		\end{array} \right) = \left( -\epsilon_1 \frac12 \Omega^{ab}H_b\right) = \left( -\epsilon_1 \frac12 \Omega^{ab}H_{bc}q^c\right) = (\epsilon_1 J^a_{\;\;b}q^b)
		\]
	where $J^a_{\;\;b}$ are the components of the para-complex structure on $M$, pulled-back to $P$. Notice that from this 
	calculation we also obtain
	\[   (\varphi^P_e)^*H_a = -\epsilon_1 2\Omega_{ab}q^b. \] 
	Using these formulae together with the identities 
$(\varphi^P_e)^*H=-H$ and \sloppy $(\varphi^P_e)^*H_{ab}=H_{ab}$, which follow from the
fact that $e$ acts anti-isometrically on the metric $g$ of $M$, 
	the result is easy to check.
\qed

Recall that the manifold $(\bar{N},g_{\bar{N}})$ is obtained by taking the quotient of $P$ with respect to the action of $U_{\epsilon_1}(1) \subset \mathbbm{C}^*_{\epsilon_1}$. In the case $\epsilon_1 = 1$ the action of $e \notin U_{\epsilon_1}(1)$ on $P$ induces an involution on $\bar{N}$ which interchanges the connected components
of $\bar{N}$. This $\mathbbm{Z}_2$-action does not preserve the metric $g_{\bar{N}}$, but maps $\left(\bar{N}, g_{\bar{N}}|_{(\epsilon_1,\epsilon_2) = (1,-1)}\right)$ to $\left(\bar{N},g_{\bar{N}}|_{(\epsilon_1,\epsilon_2) = (1,1)}\right)$ and therefore both manifolds are globally isometric.  For this reason one may take either $\epsilon_2 = -1$ or $\epsilon_2 = +1$ for the Euclidean $c$-map at the expense of working with a manifold $\bar{N}$ which is not connected but naturally contains
both possible choices.

\subsubsection{A co-frame of $P$ adapted to the pull back of the $c$-map metric}
On a local patch of $P$ it is convenient to introduce the following $4n + 4$ linearly independent one-forms:
\begin{align}
	e^m &= {\cal P}_a^{\;\;m} dq^a\;,  &  \hat{e}^m &= {\cal P}_a^{\;\;m} d\hat{q}^a\;,  \notag \\
	 u^1 &= \frac{1}{2H} H_a dq^a \;,  & \hat{u}^1 &= -\frac{1}{2H} H_a d\hat{q}^a \;,  \notag \\
	u^2 &= -\frac{1}{2H} \left(d\tilde{\phi} + 2 \hat{q}^a \Omega_{ab} d\hat{q}^b \right) \;,& \hat{u}^2 &= -\frac{1}{H} {q}^a \Omega_{ab} d\hat{q}^b \,. \label{eq:RealViel} 
\end{align}
We will refer to the collection $L^* = (e^m,u^1,u^2,\hat{e}^n,\hat{u}^1,\hat{u}^2)_{m,n=1,\ldots,2n}$ as a local {\em partial co-frame} on $P$. Note that $Z_P^0 = \mbox{span}\, L^*$.
The globally-defined  one-forms $u^1,u^2,\hat{u}^1,\hat{u}^2 \in L^*$ are independent of the choice of the functions $q^a$ and therefore uniquely defined. The one-forms $e^m,\hat{e}^n \in L^*$  are unique up to $U_{\epsilon_1}(k,\ell) \subset SO_{\epsilon_1}(2k,2\ell)$ gauge transformations, which act according to
\begin{equation}
	e^m \mapsto A^m_n e^n\;, 
	\qquad
	\hat{e}^m \mapsto A^m_n \hat{e}^n \;,
	\qquad (A^m_n) \in U_{\epsilon_1}(k,\ell)  \;.
	\label{eq:Lchoice}
\end{equation}
The bilinear form is written in terms of the partial co-frame $L^*$ as 
\begin{equation}
	{g}' =  \eta_{mp} e^m \otimes e^p + u^1 \otimes u^1 - \epsilon_1 u^2 \otimes u^2 - \epsilon_2 \Big[  \eta_{mp} \hat{e}^m \otimes \hat{e}^p + \hat{u}^1 \otimes \hat{u}^1 - \epsilon_1 \hat{u}^2 \otimes \hat{u}^2 \Big] \;,  \label{eq:GP} 
\end{equation}
where $(\eta_{mp})$ is given by (\ref{eq:eta}).
Consider the globally-defined one-form 
\begin{equation}
	v = - \frac{1}{H} q^a \Omega_{ab} dq^b \;. \label{eq:v}
\end{equation}
This satisfies $v(Z_P) = 1$ and is invariant with respect to the $U_{\epsilon_1} (1)$-action on $P$. Therefore it may be interpreted as a connection on the principal $U_{\epsilon_1}(1)$-bundle $P \to \bar{N}$. We extend the partial local co-frame $L^*$ to a  local co-frame $(L^*, v)$ on $P$.

It is important to note that although $g'$ is invariant under the $U_{\epsilon_1}(1)$-action this is not necessarily true for the individual one-forms in $L^*$. In fact, only ${e}^m, {u}^1, {u}^2$ are invariant under the action of $U_{\epsilon_1}(1)$, with the remaining one-forms transforming according to 
\begin{align*}
	{\cal L}_{Z_P} \hat{e}^A = -\epsilon_1 \hat{e}^{A+n},  \;\;\;\; {\cal L}_{Z_P} \hat{e}^{A + n} = - \hat{e}^{A} , \;\;\;\;
	{\cal L}_{Z_P} \hat{u}^1 =  \epsilon_1 \hat{u}^2 ,  \;\;\;\; {\cal L}_{Z_P} \hat{u}^2 =  \hat{u}^1 \,.
\end{align*}

The following lemma can be directly calculated using the results of Section \ref{sec:PSKbase}. It will be used later to extract
the Levi-Civita connection one-form on $(\bar{N},g_{\bar{N}})$.
\bl \label{lem:extder}
The exterior derivatives of  the one-forms in the co-frame $(L^*, v)$ are given by
\begin{align}
	de^m &= -\sigma^m_{\;\;\;p} \wedge e^p \;, \notag \\ 
	du^1 &= 0\;,  \notag \\
	du^2 &= 2 u^2 \wedge u^1 + 2 \hat{u}^2 \wedge \hat{u}^1 +  2 \Alteta_{AB} \hat{e}^{A + n}  \wedge \hat{e}^{B} \;,\notag \\
	d\hat{e}^m &= \hat{e}^m \wedge u^1 + e^m \wedge \hat{u}^1  +  \hat{u}^2 \wedge J^m_{\;\;\;p} e^{p} - (\sigma^m_{\;\;\;p} + v J^m_{\;\;\;p}) \wedge \hat{e}^p  \notag + S^{m}_{\phantom{m} p} \wedge \hat{e}^p \;, \notag \\ 
	d\hat{u}^1 &= \hat{u}^1 \wedge u^1 - \epsilon_1 \hat{u}^2 \wedge v + \Alteta_{AB}  {e}^A \wedge \hat{e}^B  - \epsilon_1 \Alteta_{AB}  {e}^{A + n} \wedge \hat{e}^{B + n} \;, \notag \\
	d\hat{u}^2 &= \hat{u}^2 \wedge u^1 + v \wedge \hat{u}^1 + \Alteta_{AB}  \hat{e}^{A + n} \wedge  {e}^{B} + \Alteta_{AB} {e}^{A + n} \wedge  \hat{e}^{B} \;,\notag \\
	dv &= 2\Alteta_{AB} e^{A + n} \wedge e^B \;.		\label{eq:ExtDer}
\end{align}
\el 

Be careful to note the index convention $A = 1,\ldots, n$ and $m,p=1,\ldots,2n$. 
The matrix-valued one-forms $S$ and $\sigma$ were defined on $M$ in the previous subsection and are pulled-back to $P$ in the above expressions. The constant matrices $J$ and $\Alteta$ were also defined in the previous subsection. The appearance of $v$ in the expressions for $d\hat{e}^m, d\hat{u}^1, d\hat{u}^2$ is due to the fact that they are not invariant under the flow of $Z_P$.

\subsubsection{The $\varepsilon$-quaternionic structure}

We now turn our attention to the $\varepsilon$-quaternionic structure. 
Using the connection $v$ we may decompose the tangent space into $TP = {\mathbbm R} Z_P + \ker v$ and the dual tangent space into $T^*P = {\mathbbm R} v + Z_P^0$. The vector space $\ker v$ is dual to $Z_P^0$, which we recall is spanned by $L^*$. There exists a unique basis $L = (e_m,u_{1},u_2,\hat{e}_p, \hat{u}_{1}, \hat{u}_{2})_{m,p=1,\ldots,2n}$ of $\ker v$ dual to $L^*$. In local coordinates this is given by
\begin{align}
	e_m &= {\cal T}_m^{\;\;\;a}\frac{\partial}{\partial q^a}  \;, & 
	\hat{e}_m &=  {\cal T}_m^{\;\;\;a} \left( \frac{\partial}{\partial \hat{q}^{a}}  + 2 \Omega_{ab} \hat{q}^b \frac{\partial}{\partial \tilde{\phi}} \right),  \notag \\
	u_{1} &= q^a \frac{\partial}{\partial q^a} \;, & 
	\hat{u}_{1} &= -q^a \left( \frac{\partial}{\partial \hat{q}^{a}} + 2  \Omega_{ab} \hat{q}^b \frac{\partial}{\partial \tilde{\phi}} \right), \notag \\
	\qquad u_{2} &= - 2H \frac{\partial}{\partial \tilde{\phi}}\;,&
	\hat{u}_{2} &= \tfrac{1}{2} H_a \Omega^{ab}\left( \frac{\partial}{\partial \hat{q}^{b}} + 2  \Omega_{bc} \hat{q}^c \frac{\partial}{\partial \tilde{\phi}} \right) .
	\label{eq:PartialFrame}
\end{align}
Using this basis one may define the three endomorphisms 
		\begin{align}
			J_1 &=  \Alteta^{AB} e_A \wedge  \hat{e}_{B + n} + \Alteta^{AB} \hat{e}_A \wedge {e}_{B + n} + {u}_1 \wedge \hat{u}_{2}  -  \epsilon_2 \hat{u}_1 \wedge u_2 \;, \notag 
			\\ 
			J_2 &= \Alteta^{AB} \hat{e}_A \wedge {e}_B -\epsilon_1 \Alteta^{AB} \hat{e}_{A + n} \wedge {e}_{B + n} + {u}_1 \wedge \hat{u}_1  + \epsilon_1 \epsilon_2  {u}_2 \wedge \hat{u}_2 \;, \label{eq:JJJ}
			\\ 
			J_3 &= - \epsilon_2 \Alteta^{AB} {e}_{A} \wedge e_{B + n}   +  \Alteta^{AB} \hat{e}_{A + n} \wedge \hat{e}_{B}  + u_2 \wedge u_1 + \hat{u}_{1} \wedge \hat{u}_2  \notag \;, 
		\end{align}
where we are using the notation $(X \wedge \,Y)(Z) = g'(Y,Z)X - g'(X,Z)Y$. The endomorphisms $J_1,J_2,J_3$ are skew-symmetric with respect to $g'$, mutually anti-commute and satisfy
\begin{align}
	J_1^2\big|_{\ker v} = -\epsilon_1 \epsilon_2\, ,
	\quad \;\;
	J_2^2\big|_{\ker v} = \epsilon_2 \, ,
	\quad \;\;
	J_3^2\big|_{\ker v} = \epsilon_1\, ,
	\quad \;\;
	J_1 J_2 = J_3 \;,
\end{align}
which is the $\varepsilon$-quaternion algebra up to relabelling. 
Since the expressions for $J_1, J_2, J_3$ are invariant under transformations of the form (\ref{eq:Lchoice}) they are independent of the choice of  frame $L$ of $\ker v$ 
in the class of frames considered above and are therefore globally-defined on $P$.
In Table \ref{table:JJJ} we summarise which of these endomorphisms are almost complex and which are almost para-complex when restricted to $\ker v \subset TP$.
\begin{table}[b]
	\centering
		\begin{tabular}{|c|c|c|}
			\hline
			 & almost complex & almost para-complex.\\
			\hhline{|-|-|-|}
			spatial $c$-map & $J_1, J_2, J_3$ &  \\
			temporal $c$-map & $J_3$ & $J_1, J_2$  \\
			Euclidean $c$-map $(\epsilon_1, \epsilon_2) = (1,1)$ & $J_1$ & $J_2, J_3$ \\
			Euclidean $c$-map $(\epsilon_1, \epsilon_2) = (1,-1)$ & $J_2$ & $J_1, J_3$  \\
			\hline
		\end{tabular}
		\caption{
	Properties of $J_1,J_2,J_3$ when restricted to $\ker v$.} \label{table:JJJ}
\end{table}

It is interesting to define two additional endomorphism fields (which are also independent of the choice of $L$ as above)
\begin{align}
			J_3' &= - \epsilon_2 \Alteta^{AB} {e}_{A} \wedge e_{B + n} + \Alteta^{AB} \hat{e}_{A + n} \wedge  \hat{e}_{B} + u_1 \wedge u_2 + \hat{u}_{1} \wedge \hat{u}_2  \;, \label{eq:Jprime} \\
			\tilde{J} &= -\epsilon_2 \Alteta^{AB}e_{A + n} \wedge  e_B   + \Alteta^{AB} \hat{e}_{A + n} \wedge \hat{e}_{B} + u_2 \wedge u_1 + \hat{u}_2 \wedge \hat{u}_{1} \;, \label{eq:Jtilde}
\end{align}
which satisfy $J'_3{}^2\big|_{\ker v} = \tilde{J}^2\big|_{\ker v} = \epsilon_1$ and are skew-symmetric with respect to $g'$.
The previously defined endomorphism $J_3$ differs from $J'_3$ by sign on the two-dimensional subspace spanned by $(u_1, u_2)$ and from $\tilde{J}$ by sign on the $(2n + 2)$-dimensional subspace spanned by $(e_m, \hat{u}_1, \hat{u}_2)$. Neither $J'_3$ nor $\tilde{J}$ form part of the $\varepsilon$-quaternion algebra.
Using Lemma \ref{lem:Projection},  $\tilde{J}$  can be written in terms of $U_{\epsilon_1} (1)$-invariant vectors as
\begin{align}
	\tilde{J} &=  -\epsilon_2 \Alteta^{AB} e_{A + n} \wedge e_B \notag \\
	&\hspace{1em} + 2e^\phi \left[ \frac{\partial}{\partial \tilde{\phi}} \wedge \frac{\partial}{\partial {\phi}} +   \left( \frac{\partial}{\partial \zeta^I} + \frac12 \tilde{\zeta}_I \frac{\partial}{\partial \tilde{\phi}} \right) 
	\wedge
	\left( \frac{\partial}{\partial \tilde{ \zeta}_I} - \frac12 {\zeta}^I \frac{\partial}{\partial \tilde{\phi}} \right) \right] \;. 
 \label{eq:Jtildeproj}
\end{align}
Here we have used the splitting of the manifold $M$ parametrised by the coordinates $q^a$ into the level sets of the function $\phi$ defined in \re{eq:phi}. In particular,
we can include $\phi$ in a new local coordinate system on $M$ consisting of $\phi$ together with a choice of 
local coordinates on a level set of $\phi$. The coordinates chosen 
on one level set are extended to the other level sets by imposing that the
coordinates are invariant under the flow
of $u_1$. In the resulting new coordinate system, one computes
$u_1 = 2 \frac{\partial}{\partial \phi}$. Next, we note that 
\[
	\hat{\theta}_a : = 2 \left(\begin{array}{c} \frac{\partial}{\partial \zeta^I} + \frac{1}{2} \tilde{\zeta}_I \frac{\partial}{\partial \tilde{\phi}} \\
	\frac{\partial}{\partial \tilde{\zeta}_I} - \frac12 {\zeta}^I \dfrac{\partial}{\partial \tilde{\phi}}  \end{array} \right) \;,
\]
and that
\[
\Alteta^{AB} \hat{e}_{A + n} \wedge \hat{e}_{B} = \frac{1}{2} \Omega^{mn}
\hat{e}_m \wedge \hat{e}_n = \frac{1}{2}
{\cal T}_m^{\;\;a} \Omega^{mn} {\cal T}_n^{\;\;b} \hat{\theta}_a 
\wedge \hat{\theta}_b \;,
\]
where $(\Omega^{mn}) := \left( \begin{array}{cc} 0 & -d \\ d & 0 \end{array} \right)$.
Using Lemma \ref{lem:Projection} one can then show that
\[
\Alteta^{AB} \hat{e}_{A + n} \wedge \hat{e}_{B} = \frac{1}{2} H \Omega^{ab}
\hat{\theta}_a \wedge \hat{\theta}_b - \hat{u}_2 \wedge \hat{u}_1 \;,
\]
from which we finally obtain (\ref{eq:Jtildeproj}). 
Notice that $\tilde{J}$ differs from $\pm J_3$ and $\pm J_3'$ on the $(2n + 4)$-dimensional subspace $\text{span}\left\{ \p/\p \phi, \p/\p \tilde{\phi}, \p/\p\zeta^I,  \p/\p \tilde{\zeta}_I \right\} =  \text{span}\left\{ \hat{e}_m, u_1, u_2,  \hat{u}_1,  \hat{u}_2 \right\}$, unless $n=0$, that is when 
$\bar{M}$ is a point.

The endomorphisms $J_3,J_3',\tilde{J}$ are invariant under the $U_{\epsilon_1} (1)$-action and induce almost $\epsilon_1$-complex structures on the $c$-map target manifold $\bar{N}$. The endomorphisms $J_1,J_2$ transform into one-another under the $U_{\epsilon_1} (1)$-action according to ${\cal L}_{Z_P} J_1 = J_2 \,,\; {\cal L}_{Z_P} J_2 = \epsilon_1 J_1$.

\bt
	\label{thm:IntCxStr}
	\begin{enumerate}[(a)]
		\item On the target manifold of the spatial $c$-map the almost complex structures $J_3$, $J_3'$ are integrable and $\tilde{J}$ is integrable if and only if  the cubic tensor $C$ vanishes. 
		\item On the target manifold of the temporal $c$-map the almost complex structures $J_3$, $J_3'$ are integrable and $\tilde{J}$ is integrable if and only if  the cubic tensor $C$ vanishes.  
		\item On the target manifold of the Euclidean $c$-map the almost para-complex structures $J_3$, $J_3'$ are integrable and $\tilde{J}$  is integrable if and only if the cubic tensor $C$ vanishes.   
	\end{enumerate}
\et
\noindent It was shown in \cite{Cortes:2011ut} that the almost complex structure $J_3$ on the target manifold of the spatial $c$-map is integrable. The other parts of this theorem will be proved on a case-by-case basis in Sections \ref{sec:SpatialCmap} -- \ref{sec:EuclideanCmap}. For easier reference we
stated the result as three cases (a)--(c).

The endomorphisms $J_1, J_2, J_3$ define a (fibre-wise) $\varepsilon$-quaternionic structure $Q_P = \text{span}\{J_1, J_2, J_3\}$ on $\ker v \subset TP$, which is 
skew-symmetric with respect to the metric $g'|_{\ker v }$. Due to the transformation properties under $U_{\epsilon_1}(1)$, this induces an almost $\varepsilon$-quaternionic Hermitian structure $Q_{\bar{N}}$ on $(\bar{N}, g_{\bar{N}})$. In the next section we will show, by explicit calculation, that $Q_{\bar{N}}$ is parallel with respect to the Levi-Civita connection, which proves the following theorem.
\bt
	\label{thm:CmapQK}
	\begin{enumerate}[(a)]
		\item The target manifold of the spatial $c$-map is quaternionic K\"ahler.
		\item The target manifold of the temporal $c$-map is para-quaternionic K\"ahler.
		\item The target manifold of the Euclidean $c$-map is para-quaternionic K\"ahler.
	\end{enumerate}
\et

In all three cases the reduced scalar curvature $\nu = scal/(4(n + 1)(n + 3))$ is equal to $-2$.
This can be seen by comparing the $Sp_\varepsilon(1)$-curvature of the $c$-map target manifold 
\[
	R^H = dp + p\wedge p \;,
\]
with the $Sp_\varepsilon(1)$-curvature of $\varepsilon$-quaternionic projective space ${\mathbbm H}_\varepsilon P^{n + 1}$ 
\[
	R^H_0 = \frac12 \sum_{\alpha = 1,2,3}  J^H_\alpha \omega_\alpha \;.
\]
Here the matrix  $J^H_\alpha$ is given in expression \eqref{eq:JHcmap} and $\omega_\alpha(\cdot, \cdot) := \epsilon_\alpha g'(J_\alpha  \cdot, \cdot)$ is the fundamental two-form associated with the almost $\epsilon_\alpha$-complex structure $J_\alpha$. The matrix $p$ is given for each $c$-map separately in  Sections \ref{sec:SpatialCmap} -- \ref{sec:EuclideanCmap}.
For an $\varepsilon$-quaternionic K\"ahler manifold the above $Sp_\varepsilon(1)$-curvature tensors are related by \cite{Cortes:2004aa}
\[
	R^H = \nu R^H_0 \;.
\]
Computing both sides one finds that $\nu = -2$ in all cases.

\section{Levi-Civita connection and integrable $\varepsilon$-complex structures}
\label{sec:LCint}

In this section we will calculate the Levi-Civita connection on the target manifold $(\bar{N},g_{\bar{N}})$ of the $c$-map for various spacetime signatures. We will also show that the two skew-symmetric almost $\varepsilon$-complex structures $J_3$ and $J_3'$ introduced in the previous section are integrable. 

In order to compute the Levi-Civita connection and to check the integrability of the structures $J_3$ and $J_3'$ one needs to calculate exterior derivatives of an appropriate local co-frame on $\bar{N}$. To do this we will make use of the partial co-frame (\ref{eq:RealViel}) on the $U_{\epsilon_1}(1)$-principal bundle $P \to \bar{N}$. There are two complementary approaches one may take when performing these calculations:
\begin{enumerate}
	\item Use a local section of $P\ra \bar{N}$ to pull-back the partial co-frame from $P$ to a co-frame of $\bar{N}$ and then perform calculations.
	\item Perform calculations directly on $P$ using the partial co-frame and then use a local section to pull-back the results to $\bar{N}$.
\end{enumerate}
We will adopt approach 2 since one only needs to make a choice of local section after all calculations have been performed. There is a slight complication due to the fact that the partial co-frame (\ref{eq:RealViel}) is not invariant under the flow of the fundamental vector field $Z_P$ and therefore not projectable to $\bar{N}$, which we address in Section \ref{sec:CalcP}. The relation between the two approaches is discussed in Section \ref{sec:CalcM}. The explicit calculation of the Levi-Civita connection and 
integrability of the $\varepsilon$-complex structures for various spacetime signatures are presented case-by-case in the remaining sections.

\subsection{Calculating on $P$ using a non-invariant partial co-frame}
\label{sec:CalcP}

In this section we want to discuss the following problem. 
Let $(M,g)$ be a pseudo-Riemannian manifold with almost $\varepsilon$-quaternionic Hermitian structure $Q$ and $\pi: P\rightarrow M$ be a principal bundle with structure group $G$. 
Suppose we are given pointwise linearly independent one-forms $\theta^i$, $i=1,\ldots, n=\dim(M)$ on some open subset $U\subset P$, which are {\em horizontal} in the sense that they vanish on any vertical vector, such that 
\[
\pi^* g = \eta_{ij} \theta^i \theta^j \;,
\]
where $(\eta_{ij}) = \mbox{diag}(\mathbbm{1}_k,-\mathbbm{1}_\ell)$ is the Gram matrix of an orthonormal frame in standard ordering. We will assume that $\pi^* TM$ is trivial on $U$. Systems $(\theta^i)$ as above will be called {\em partial co-frames} of $P$ over $U$. Notice that given a principal connection on $P$ and a basis of
$\mathfrak{g}=Lie\, G$, any partial co-frame of $P$ over $U$ is canonically extended to a co-frame of $P$ over $U$.

The problem is to show that $Q$ is parallel with respect to the Levi-Civita connection, and, hence, the manifold $(M,g)$ is $\varepsilon$-quaternionic K\"ahler. This involves computing the Levi-Civita connection of $g$ in terms of $(\theta^i)$, without assuming that the forms $\theta^i$ are $G$-invariant and, hence, projectable to $M$.

\bp
\label{prop:Connection}
Under the above assumptions,
the system of equations 
\begin{equation}
\label{Connection}
d\theta^i + \sigma^i_j \wedge \theta^j = 0 \;,
\end{equation}
has a unique solution 
$\sigma = (\sigma^i_j) \in \Omega^1(U, \mathfrak{so}(k,\ell))$. 
Given a second system of $n$ linearly independent horizontal one-forms
$(\tilde{\theta}^i)$ on $U\subset P$, 
the solution $\tilde{\sigma}=(\tilde{\sigma}^i_j)$
of the system 
\begin{equation}
\label{Connection2}
d\tilde{\theta}^i + \tilde{\sigma}^i_j \wedge \tilde{\theta}^j = 0 \;,
\end{equation}
is related to $\sigma$ by 
\begin{equation}
\tilde{\sigma} = - (dA) \, A^{-1} + A {\sigma} A^{-1} \;, 
\label{eq:Tildesigma}
\end{equation}
where $A =(A^i_j) \in C^\infty(U,O(k,\ell))$ is the gauge transformation
relating $(\theta^i)$ with $(\tilde{\theta}^i)$, that is
$\tilde{\theta}^i = A^i_j \theta^j$. 
\ep

\pf
\label{PropNonInvariant}
We first prove the uniqueness.
Suppose that $\sigma'$ is a second solution of (\ref{Connection}). Then
the difference 
$\Delta = (\Delta^i_j) = \sigma' - \sigma 
\in \Omega^1(U,\mathfrak{so}(k,\ell))$ satisfies the equations
$\Delta^i_j \wedge \theta^j =0$. For the coefficients
$\Delta^{\;\;i}_{jk}$ in the expansion $\Delta^i_k = \Delta^{\;\;i}_{jk} \theta^j$ 
this implies $\Delta^{\;\;i}_{jk} = \Delta^{\;\;i}_{kj}$. 
Therefore $\Delta_{jik}:= \eta_{il} \Delta^{\;\;l}_{jk}$ is antisymmetric
in $(i,k)$ and symmetric in $(j,k)$, which implies $\Delta=0$. 

One can easily check that given a solution $\sigma$ of (\ref{Connection})
and a  gauge transformation $A\in C^\infty(U, O(k,\ell))$, 
$\tilde{\sigma} = -(dA) A^{-1} + A \tilde{\sigma} A^{-1}$ is 
a solution of (\ref{Connection2}), if we define $\tilde{\theta}^i =
A^i_j \theta^j$.   

Now we prove the existence. Given the above hypothesis on $U$, 
we can assume without restriction of generality that $U=\pi^{-1} (U_0)$
is the preimage of an open subset $U_0\subset M$ on which an 
orthonormal co-frame $(\theta^i_0)$ exists.
It is sufficient to remark that the pullback of
the connection one-form $\sigma_0$ of the Levi-Civita connection of $(M,g)$ 
with respect to the co-frame $(\theta^i_0)$
gives a solution of (\ref{Connection}), where $(\theta^i) = 
( \pi^* \theta^i_0)$. The equation (\ref{Connection}) is in fact 
obtained as the pullback of the equation 
$d\theta_0 + \sigma_0 \wedge \theta_0 = 0$, which expresses the
vanishing of the torsion of the Levi-Civita connection
of $(M,g)$. Here $\theta_0$ is the column vector
with entries $\theta^i_0$. 
\qed

The almost $\varepsilon$-quaternionic Hermitian structure $Q$ on $M$ induces a (fibre-wise) $\varepsilon$-quaternionic Hermitian structure $Q_P$ in the normal bundle ${\cal N} = TP/T^vP$ to the fibres of $P \rightarrow M$, where $T^vP\subset TP$ denotes the vertical distribution. 
The $\varepsilon$-quaternionic structure $Q_P$ is Hermitian in the sense that it consists of endomorphisms which are skew-symmetric with respect to the (fibre-wise) metric $\pi^* g$ in ${\cal N}$. 
By construction $Q_P$ is invariant under the $G$-action on ${\cal N}$ induced by the principal $G$-action on $P$. Conversely, a fibre-wise skew-symmetric $\varepsilon$-quaternionic structure $Q_P$ 
on $({\cal N}, \pi^*g)$, which is invariant under the $G$-action on ${\cal N}$, induces an almost $\varepsilon$-quaternionic Hermitian structure $Q$ on $M$, which may be parallel or not.

\bp
\label{prop:QuaternionicStructure}
Given a $G$-invariant skew-symmetric fibre-wise $\varepsilon$-quaternionic structure $Q_P$ on ${\cal N}$ the induced almost $\varepsilon$-quaternionic Hermitian structure $Q$ on $(M,g)$ is parallel with respect to the Levi-Civita connection if the solution of (\ref{Connection}) takes values in the Lie algebra $\mathfrak{sp}_\varepsilon (1) \oplus \mathfrak{sp}_\varepsilon(k,\ell)$, provided the partial co-frame $(\theta^i)$ is $\varepsilon$-quaternionic. 
\ep

\pf
Consider an open subset  $U \subset P$ on which an $\varepsilon$-quaternionic partial co-frame $(\theta^i)$ is defined. We may assume without restriction of generality that $U=\pi^{-1} (U_0)$ is the preimage of an open subset $U_0\subset M$ on which an $\varepsilon$-quaternionic co-frame $(\theta^i_0)$ exists. This may be pulled back to give another $\varepsilon$-quaternionic partial co-frame $(\tilde{\theta}^i) = (\pi^* \theta_0^i)$.
Since both $(\theta^i)$ and $(\tilde{\theta}^i)$ are $\varepsilon$-quaternionic partial co-frames they are related to one-another by a gauge transformation of the form $A =(A^i_j) \in C^\infty(U,Sp_\varepsilon (1) \cdot Sp_\varepsilon (k,\ell))$. 
Let us denote by $\sigma$ the solution of (\ref{Connection}) in the basis $(\theta^i)$ and by $\tilde{\sigma}$ the solution in the basis $(\tilde{\theta}^i)$. 

Suppose that $\sigma$ takes values in $\mathfrak{sp}_\varepsilon (1) \cdot \mathfrak{sp}_\varepsilon (k,\ell)$. From (\ref{eq:Tildesigma}) it follows that $\tilde{\sigma}$ also takes values in $\mathfrak{sp}_\varepsilon (1) \cdot \mathfrak{sp}_\varepsilon (k,\ell)$. Since $\tilde{\sigma}$ is the pull-back of the Levi-Civita connection one-form in an $\varepsilon$-quaternionic co-frame it follows that the Levi-Civita connection on $M$ takes values in $\mathfrak{sp}_\varepsilon (1) \cdot \mathfrak{sp}_\varepsilon (k,\ell)$ when written in an $\varepsilon$-quaternionic co-frame. 
\qed

Proposition \ref{prop:Connection} shows that in order to compute the Levi-Civita connection of a manifold $(\bar{N},g_{\bar{N}})$ in the image of the $c$-map it is sufficient to solve the equation (\ref{Connection}) locally on $P$ without having to assume that the partial co-frame $(\theta^i)$ is projectable. If the solution to  (\ref{Connection})  takes values in $\mathfrak{sp}_\varepsilon (1) \oplus \mathfrak{sp}_\varepsilon(k,\ell)$ in an $\varepsilon$-quaternionic partial co-frame then the manifold $(\bar{N},g_{\bar{N}})$ is $\varepsilon$-quaternionic K\"ahler by Proposition \ref{prop:QuaternionicStructure}.

\subsection{Alternative approach: calculating on $\bar{N}$ using a co-frame}
\label{sec:CalcM}

Let us now briefly discuss an alternative way of calculating exterior derivatives and the Levi-Civita connection directly on the target manifold $(\bar{N},g_{\bar{N}})$ in the image of the $c$-map.

Let $U \subset P$ be an open set on which the partial co-frame (\ref{eq:RealViel}) is defined. Consider any local section $s: U_0\ra P$ with values in $U$, 
for example the local section defined by the equation $x^0 = 0$. (Recall that $x^0=q^0$ is one of the functions $(q^a,\hat{q}^b,\tilde{\phi})$ on $P$ introduced 
in Section \ref{sec:cmapVar}.) We may use the section $s$ to define a co-frame on $U_0\subset \bar{N}$ given by
\begin{equation}
 \left(e^m_0, u^1_0, u^2_0, \hat{e}^m_0, \hat{u}^1_0, \hat{u}^2_0\right) = s^*\left(e^m, u^1, u^2, \hat{e}^m, \hat{u}^1, \hat{u}^2\right) \;.
	\label{eq:AltBasis}
\end{equation}
It is then possible to calculate the exterior derivatives and the Levi-Civita connection in this local co-frame on $\bar{N}$.

One may relate this approach to that of Section \ref{sec:CalcP} as follows. Since the exterior derivative commutes with the pull-back of a differentiable map we have
\[
	de^m_0 = s^* de^m\;, 	\qquad 	du^1_0 = s^* du^1\;, 	\qquad 	\mbox{etc.}
\]
where the exact expressions on the RHS can be read off from (\ref{eq:ExtDer}). Moreover, from Proposition \ref{prop:Connection} it follows that the Levi-Civita connection $\sigma_0$ on $\bar{N}$ in the basis (\ref{eq:AltBasis}) is given by the pull-back of the unique solution $\sigma$ of equation (\ref{Connection}) in the basis (\ref{eq:RealViel}), which is calculated in the following sections.

\subsection{The spatial $c$-map}
\label{sec:SpatialCmap}

In this section we consider the reduction over space from 3 + 1 to 2 + 1 dimensions. This means that one must set $\epsilon_1 = -1$ and $\epsilon_2 = -1$ in the expressions in Section \ref{sec:cmapVar}. Recall from Section \ref{3dSec} that the Hesse potential $H$ is assumed to be negative. 

In order to expose the quaternionic geometry we define the complex partial co-frame on $P$ 
\begin{align}
	\cu &= i e^{- \phi/2} \left(X^I d \tilde{\zeta}_I - F_I(X) d\zeta^I\right) ,\notag \\
	\cv &= e^{-\phi} \left[ \tfrac{1}{2} d e^{\phi} + i \left(d\tilde{\phi} + \tfrac{1}{2} ( \zeta^I d\tilde{\zeta}_I - \tilde{\zeta}_I d \zeta^I ) \right) \right] ,\notag \\
	\ce^{A} &= P_I^{\;A} dX^I \;,\notag \\
	\cE^{A} &= -ie^{-\phi/2} P_{I}^{\;A} N^{IJ}\left(d \tilde{\zeta}_J - {\cal N}_{JK} d\zeta^K\right)  \;. \label{eq:Scoframe}
\end{align}
Recall that $X^I = e^{-\phi/2} Y^I$, and, 
due to homogeneity, $N_{IJ}(X,\bar{X}) = N_{IJ}(Y,\bar{Y})$
and ${\cal N}_{IJ}(X,\bar{X}) = {\cal N}_{IJ}(Y,\bar{Y})$.
We have locally defined the matrix $(P_I^{\;\;A})$ with entries 
	\begin{equation}
	P_I^{\;\;A} = e^{\phi/2} ({\cal P}_a^{\;\;A} + i {\cal P}_a^{\;\;A + n} ) \Pi_I^{\;\;a} \;, 		\label{eq:Pr}
	\end{equation}
where ${\Pi}_I^{\;\;a}$ represents the holomorphic projection from the special holomorphic coordinates $Y^I$ to special real coordinates $q^a$:
\begin{equation}
	dq^a = \Pi_I^{\;\;a} dY^I + \bar{\Pi}_I^{\;\;a} d\bar{Y}^I \;,
	\qquad
	({\Pi}_I^{\;\;a}) = \frac{\partial q^a}{ \partial {Y}^I}  = \left(\begin{array}{c} \tfrac12 (\delta_I^J) ,\ \tfrac12 ({F}_{IJ}) \end{array}\right).
	\label{eq:HolPr}
\end{equation}
Notice that $P_{I}^{\;\,A} Y^I = 0$, and, hence, $P_{I}^{\;\,A} d X^I = e^{-\phi/2} P_{I}^{\;\,A} d Y^I$. 
Using the local section $s = \{\text{Im}(X^0) = 0 \} \ (= \{ x^0 = 0\}) $  of $P\ra \bar{N}$ discussed in Section \ref{sec:CalcM}
one can pull-back \eqref{eq:Scoframe} to the complex orthonormal co-frame on $(\bar{N},g_{\bar{N}})$ presented in \cite{Ferrara:1989ik}%
\footnote{Recall that we use the D-gauge  $-N_{IJ} X^I \bar{X}^J = 1$ in order to fix the scale transformations of the CASK manifold $M$, whereas in \cite{Ferrara:1989ik} they are fixed by setting $|X^0|= 1$. These two choices are related by the transformation $X^I \mapsto \frac{1}{|X^0|}X^I$.}.

\bp
\label{RealCxBasis}
The complex partial co-frame \eqref{eq:Scoframe} is related to the real partial co-frame \eqref{eq:RealViel} introduced in Section \ref{sec:cmapVar} by 
\emph{
	\begin{align}
			\cu &= \hat{u}^1 + i \hat{u}^2 \;,  & 
			\ce^{A} &= e^A + i e^{A+ n}\;, \notag \\
			\cv &= u^1 + i u^2\;,  & 
			\cE^{A} &= \hat{e}^A + i \hat{e}^{A+ n} \;. \label{eq:SpatialCxCoframe}
	\end{align}
}
The one-form $v$ may be written as
\[
	v = \frac{1}{2i} \left( X^IN_{IJ} d\bar{X}^J - dX^I N_{IJ} \bar{X}^J \right) \;.
\]
\ep
\pf
Using $e^{\phi} = -2H$ and
	\[
		Y^I = x^I + iu^I \;, \qquad  F_I = y_I + i v_I \;, \qquad H_a = (2v_I, -2u^I)^T\;,
	\]
the first two expressions are calculated to be
	\begin{align*}
		\cu &= -i \frac{1}{2H} \left[ x^I d \tilde{\zeta}_I - y_I d\zeta^I + i \left(u^I d \tilde{\zeta}_I - v_I d\zeta^I \right) \right]
			= -i \frac{1}{2H} \left[ 2 q^a \Omega_{ab} d \hat{q}^b - i \left( H_a d\hat{q}^a\right) \right] \;, \\
		\cv &= -\frac{1}{2H} \left[ -dH + i \left(d\tilde{\phi} + 2 \hat{q}^a \Omega_{ab} d\hat{q}^b \right) \right] \;.
	\end{align*}
Comparing with the explicit expressions in \eqref{eq:RealViel} gives the desired result.
Next, we observe that
	\begin{align*}
		e^A + ie^{A + n} &= \frac12 (Id - iJ^*)(e^A + ie^{A + n}) = ({\cal P}_a^{\;\;A} +i{\cal P}_a^{\;\;A +n}) \Pi_I^{\;\;a} dY^I  \\
		&= e^{-\phi/2} P_I^{\;A} dY^I = P_I^{\;A} dX^I = \ce^A \;.
	\end{align*}
Using the fact $P_{I}^{\;\,A} Y^I = 0$ along with \eqref{eq:HolPr} and \eqref{eq:Pr} one may write 
	\begin{align*}
		\cE^A	&= -ie^{-\phi/2} P_{I}^{\;\;A} N^{IJ} \left(d \tilde{\zeta}_J - \bar{F}_{JK} d\zeta^K\right) 
			=-4ie^{-\phi/2} P_{I}^{\;\;A} N^{IJ} \bar{\Pi}_J^{\;\;b}\Omega_{bc}d \hat{q}^c \\
			&=- 4i({\cal P}_a^{\;\;A} +i{\cal P}_a^{\;\;A+n}) \Pi_I^{\;\;a} N^{IJ} \bar{\Pi}_J^{\;\;b}\Omega_{bc} d\hat{q}^c  \;.
	\end{align*}
Making use of the  identity 
$4 \Pi_I^{\;\;a} N^{IJ} \bar{\Pi}_J^{\;\;b} = H^{ab} + \frac{i}{2} \Omega^{ab},$
which can be easily verified using \eqref{eq:Hab} and \eqref{eq:HolPr}, 
along with the expression for $J^*{}_{b}^{\;\;a}$ given by the components of \eqref{eq:ASKCx}, we can write
	\begin{align*}
		\cE^A	&= -\frac{i}{2} ({\cal P}_a^{\;\;A} +i{\cal P}_a^{\;\;A+n}) \left( J^*{}_{b}^{\;\;a} d\hat{q}^b + i d\hat{q}^a \right)  \;.
	\end{align*}
From \eqref{eq:PhiHol} it follows that 
${\cal P}_b^{\;\;m} J^*{}_a^{\;\;b}  = J^*{}_p^{\;\;m} {\cal P}_a^{\;\;p}$, 
and, hence, $\cE^{A} = \hat{e}^A + i \hat{e}^{A + n}$.
Lastly, we calculate
\begin{align*}
	\frac{1}{2i}&\left( X^IN_{IJ} d\bar{X}^J - dX^I N_{IJ} \bar{X}^J \right) \\
	&= e^{-\phi} \text{Im}\left( Y^IN_{IJ} (Y,\bar{Y})d\bar{Y}^J \right)
	= -e^{-\phi} \text{Re}\left( F_I d\bar{Y}^I - Y^I d\bar{F}_I \right)  \\
	&= \frac{1}{2H} \left( y_I dx^I - x^I dy_I - v_I du^I +  u^I dv_I \right) 
	= -\frac{1}{2H} \left( q^a \Omega_{ab} dq^b - \frac14 H_a \Omega^{ab} dH_b \right) \\
	&= -\frac{1}{H} q^a \Omega_{ab} dq^b = v \;,
\end{align*}
where in the second line we used \eqref{g} and in last line \eqref{eq:id3}.
\qed

The exterior derivatives of the one-forms in the complex co-frame may be written as  \cite{Ferrara:1989ik} (see also \cite{Krahe:2001} for the indefinite case)
\begin{align}
	d \cu &= \Big( -u^1 + iv\Big) \wedge \cu - \Alteta_{AB} \bar{\cE}{}^{A} \wedge \ce^{B} \;,\notag \\\ 
	d \cv &= \cu \wedge \bar{\cu} + \cv \wedge \bar{\cv} + \Alteta_{AB} \cE^{A}\wedge \bar{\cE}^{B} \;,\notag\\
	d \ce^A &= -{\sigma}^{{\mathbbm C}}{}^{A}_{\;\;\;B} \wedge \ce^{B} \;,   \label{ext:eq} \\
	d \cE^A &= \Big( - u^1 - iv \Big) \wedge \cE^{A} - \bar{\cu} \wedge \ce^{A}  - \sigma^{\mathbbm C}{}^{A}_{\;\;\;B} \wedge \cE^{B} +S^{{\mathbbm C}A}_{\;\;\;\;\;\;BC}\bar{\ce}^{B} \wedge \bar{\cE}{}^{C} \;,\notag
\end{align}
where $\sigma^{{\mathbbm C}A}_{\;\;\;\;\;B} :=  \sigma^A_{\;\;\,B} + i \sigma^{A + n}_{\phantom{A + n}B}$ and $S^{{\mathbbm C}A}_{\;\;\;\;\;\;BC} := {S}^A_{\;\;\,B C} + i {S}^{A + n}_{\phantom{A + n}BC}$. These expressions may be checked using (\ref{eq:ExtDer}) and the identities (\ref{eq:Scomps}) and (\ref{eq:SigmaComps}).

\vspace{1em}
\noindent {\em Proof of Theorem \ref{thm:IntCxStr} (a):}
The following proof that $J_3$ is integrable was provided in \cite{Cortes:2011ut}: a basis of the $+i$ eigendistribution of $J_3^*$ is given by ${\cal B}^{(1,0)} = \left(\bar{\cu}, \cv, \bar{\ce}^{A}, \cE{}^{A}\right)$. Each term in the exterior derivative of any element in ${\cal B}^{(1,0)}$ contains a one-form in the set ${\cal B}^{(1,0)}$. Therefore the distribution is integrable by the Newlander--Nirenberg theorem, hence the almost-complex structure $J_3$ is integrable.

We now consider the integrability of $J_3'$ and $\tilde{J}$. A basis of the $+i$ eigendistribution of $J_3'{}^*$ is given by ${\cal B}'{}^{(1,0)} = \left(\bar{\cu}, \bar{\cv}, \bar{\ce}^{A}, \cE{}^{A}\right)$, and by the same argument as above $J_3'$ is integrable.
A basis of the $+i$ eigendistribution of $\tilde{J}$ is given by $\left({\cu}, \cv, {\ce}^{A}, \cE{}^{A}\right)$, and therefore $\tilde{J}$ is integrable if and only if $S^A_{\;\;\;BC} =  S^{A + n}_{\phantom{A + n} BC} = 0$, which is the case if and only if the cubic tensor $C$, defined in \re{eq:CS}, vanishes. This is true if and only if the holomorphic prepotential $F$, or, equivalently, the Hesse potential $H$ on the corresponding CASK manifold $M$ is a quadratic polynomial.
\qed

\vspace{1em}
\noindent {\em Proof of Theorem \ref{thm:CmapQK} (a):}
The complex one-forms, along with their conjugates, may be gathered together into the quaternionic vielbein 
\begin{equation}
	{\cal U}^{{\cal A} \mu} = \frac{1}{\sqrt{2}} \left( \begin{array}{cccc} 
		\cu & \ce^{A} & -\bar{\cv} & -\bar{\cE}{}^{A} \\
		\cv & \cE^{A} & \bar{\cu} & \bar{\ce}^{A} \end{array} \right) \;.
	\label{eq:SpatialFrame}
\end{equation}
In this co-frame the Levi-Civita connection one-form decomposes according to (\ref{eq:QK_connection}), where $p,q,t$ are given by \cite{Ferrara:1989ik}
\begin{align*}
	&p = \left( \begin{array}{cc} 
								\frac{i}{2} u^2 - \frac{i}{2} v  & -\hat{u}^1 - i \hat{u}^2 \\ \\
								\hat{u}^1 - i \hat{u}^2 & -\frac{i}{2} u^2 + \frac{i}{2} v
							\end{array} \right) ,\\ \\
	&q = \left( \begin{array}{cc} 
								-\frac{3i}{2} u^2 \;-\; \frac{i}{2} v &  \left(\hat{e}^C - i\hat{e}^{C + n}\right) d_{CB} \\ \\
								-\hat{e}^A - i\hat{e}^{A + n} & \hspace{2em} \sigma^A_{\;\;\,B} + i \sigma^{A + n}_{\phantom{A + n}B} \;-\; \frac{i}{2} (u^2 - v) \delta^A_B 
							\end{array} \right), \\ \\
	&t = \left( \begin{array}{cc}  
								0 & 0 \\ \\
								0 & \hspace{1em} ({S}^A_{\;\;\,B C} + i {S}^{A + n}_{\phantom{A + n}BC})(\hat{e}^{C} - i\hat{e}^{C + n})
							\end{array} \right) \;.
\end{align*}
The quaternionic structure is therefore parallel with respect to the Levi-Civita connection. 
\qed

\vspace{1em}
Let us briefly explain how one may check that the above expression for the Levi-Civita connection is correct. It is obvious
from the formalism that the above expression defines a metric connection so it suffices to check that its torsion is zero. 
In terms of an $\varepsilon$-quaternionic vielbein the latter condition is given by
\[
	d{\cal U}^{{\cal A} \mu} + \Omega^{{\cal A}\mu}_{\phantom{{{\cal A}\mu}}{{\cal B}\nu}} \wedge {\cal U}^{{\cal B} \nu} = 0 \;.
\]
This can be naturally split into two separate sets of equations
\begin{eqnarray}
\label{f:eq}
\begin{aligned} 
	de^{{\cal A} I} &= -p^{\cal A}_{\;\;\;\;{\cal B}} e^{{\cal B} I} - q^I_{\;\;J} e^{{\cal A} J} - t^I_{\;\;J} f^{{\cal A} J}\;,  \\
	df^{{\cal A} I} &= -p^{\cal A}_{\;\;\;\;{\cal B}} f^{{\cal B} I} + s^I_{\;\;J} e^{{\cal A} J} +  (\eta' q^T \eta')^{I}_{\;\;\;J}f^{{\cal A}  J}  \;, 
\end{aligned}
\end{eqnarray}
where we have defined
\[
	e^{{\cal A}I} := \left( \begin{array}{cc} 
		\cu & \ce^{A} \\
		\cv & \cE^{A} \end{array}  \right) \;, \qquad
	f^{{\cal A}I} := \left( \begin{array}{cc} 
		-\bar{\cv} & -\bar{\cE}{}^{A} \\
		\bar{\cu} & \bar{\ce}^{A} \end{array}  \right) \;,
		\qquad
		\eta' := \left( \begin{array}{cc} 1 & 0 \\ 0 & \Alteta \end{array} \right) \;.
\]
In the quaternionic case $s = \bar{t}$ and $f^{{\cal A}I} = \epsilon^{\cal AB}\bar{e}^{{\cal B}J}$, and therefore the second set of equations follows from the first set by complex conjugation. However in the para-quaternionic case, which we will deal with in the following sections, the second set of equations are not implied by the first, and must be checked independently.

Let us end by explicitly checking, for instance, that the formula for $d \cE^{A}$ obtained from \re{f:eq} coincides 
with the exterior derivative of $\cE^{A}$ as given in \re{ext:eq}: 
	\begin{align*}
	&d\cE^{A} = -p^1_{\;\;0} \ce^{A} - p^1_{\;\;1} \cE^{A} - q^A_{\;\;\;0}\cv - q^A_{\;\;B} \cE^{B} - t^A_{\;\;\;0} \bar{\cu} - t^A_{\;\;B} \bar{\ce}^{A} \\
		&= \hat{e}^Au^1 + i \hat{e}^{A + {n}} u^1 + e^A \hat{u}^1   + i e^{A + {n}}  \hat{u}^1 - \hat{u}^2 e^{A + {n}}  + i\hat{u}^2 e^A  + v \hat{e}^{A + {n}}  - i v \hat{e}^A  \\
		&- (\sigma^A_{\;\;\;B} + i \sigma^{A + {n}}_{\phantom{A + {n}} B})(\hat{e}^B + i\hat{e}^{B + {n}}) -({S}^A_{\;\;\,B C} + i {S}^{A + {n}}_{\phantom{A + {n}}BC})(\hat{e}^{C} - i\hat{e}^{C + {n}})({e}^B - i {e}^{B + {n}}) \\
		&= \left( - u^1 - iv \right)  \left(\hat{e}^A  + i\hat{e}^{A + n}\right) - \left(\hat{u}^1 - i\hat{u}^2\right)  \left(e^A  + ie^{A + n}\right)  \\
		&- (\sigma^A_{\;\;\;B} + i \sigma^{A + {n}}_{\phantom{A + {n}} B})\left(\hat{e}^B + i\hat{e}^{B + {n}}\right) +({S}^A_{\;\;\,B C} + i {S}^{A + {n}}_{\phantom{A + {n}}BC})\left({e}^B - i {e}^{B + {n}})(\hat{e}^{C} - i\hat{e}^{C + {n}}\right).
\end{align*}
Here we have omitted writing the symbol for the wedge product.

\subsection{The temporal $c$-map}
\label{sec:TemporalCmap}

We now consider the reduction over time from 3 + 1 to 3 + 0 dimensions. 
In this case we must set $\epsilon_1 = -1$ and $\epsilon_2 = 1$. 
Recall that in our construction of the $c$-map the spatial and temporal $c$-map have the same target manifold but different  metrics. In particular, we may use the same partial co-frame $L^*$ on $P$ defined by \eqref{eq:RealViel} in both cases.

\vspace{1em}
\noindent {\em Proof of Theorem \ref{thm:IntCxStr} (b):}
It follows from \re{eq:GP} and \re{eq:JJJ}  that the almost-complex structures $J_3$ and $J_3'$ in the case $(\e_1,\e_2)= (-1,1)$ 
coincide with $-J_3'$ and $-J_3$, respectively, in the case  $(\e_1,\e_2)= (-1,-1)$. Therefore the integrability of these
structures follows from the proof of Theorem \ref{thm:IntCxStr} (a) given in Section \ref{sec:SpatialCmap}. 

The almost-complex structure $\tilde{J}$ in the case $(\e_1,\e_2)= (-1,1)$ coincides with $-\tilde{J}$ in the case  $(\e_1,\e_2)= (-1,-1)$ except for its action on the two-dimensional subspace spanned by $({u}_1, {u}_2)$, where it acts with opposite sign. Taking this into account, one may use the same argument as in the proof of Theorem \ref{thm:IntCxStr} (a) that $\tilde{J}$ is integrable if and only if $C = 0$.
\qed

\vspace{1em}
Let us define the real partial co-frame 
\begin{align*}
	\cu &= u^2 - \hat{u}^1 \;,					&	\tilde{\cu} &= u^2 + \hat{u}^1\;, \\
	\cv &= u^1 + \hat{u}^2 		\;,			&	\tilde{\cv} &=  u^1 - \hat{u}^2\;, \\
	\ce^A &= -e^{A + n} - \hat{e}^A \;,  &	\tilde{\ce}^{A} &= -e^{A + n} + \hat{e}^A\;, \\
	\cE^A &= -e^A - \hat{e}^{A + n} \;,	&	\tilde{\cE}{}^{A} &= -e^A + \hat{e}^{A + n} \;,
\end{align*}
which we gather together into the para-quaternionic vielbein 
\begin{equation}
	{\cal U}^{{\cal A} \mu} = \frac{1}{\sqrt{2}} \left( \begin{array}{cccc} 
		\cu & \ce^{A} & -\tilde{\cv} & -\tilde{\cE}{}^{A} \\
		\cv & \cE^{A} & \tilde{\cu} & \tilde{\ce}^{A} \end{array} \right) \;. 
		\label{eq:TemporalFrame}
\end{equation}
One may use Proposition \ref{RealCxBasis} to write the vielbein in terms of the real and imaginary parts of the complex co-frame \eqref{eq:Scoframe}. Notice that the above expression for the para-quaternionic vielbein is not related to the expression for the spatial $c$-map quaternionic vielbein by replacing complex coordinates with the para-complex coordinates. However, as we will explain in the next section, such a relationship does exist for the vielbeins of the spatial and Euclidean $c$-maps.

\vspace{1em}
\noindent {\em Proof of Theorem \ref{thm:CmapQK} (b):}
In the frame (\ref{eq:TemporalFrame}) the Levi-Civita connection one-form decomposes according to (\ref{eq:QK_connection}), where $p,q,t,s$ are given by
\begin{align*}
	&p = \left( \begin{array}{cc} 
								\hat{u}^2  & -\frac{1}{2}(u^2 + v) + \hat{u}^1 \\ \\
								\frac{1}{2}(u^2 + v) + \hat{u}^1  & -\hat{u}^2
							\end{array} \right),  \\ \\
	&q = \left( \begin{array}{cc} 
								0 & \hat{e}^{C + n} \Alteta_{CB} \\ \\
								\hat{e}^{A + n}  & \hspace{2em} \sigma^A_{\;\;\;B} \;\; - \;\; {S}^{A + n}_{\phantom{A + n}Bm}\hat{e}^m  
							\end{array} \right),  \\ \\
	&t = \left( \begin{array}{cc}  
								\frac{3}{2}u^2 - \frac{1}{2}v & -\hat{e}^C \Alteta_{CB} \\ \\
								-\hat{e}^A & -\sigma^{A + n}_{\phantom{{A + n}}B} \;\;+\;\;{S}^{A }_{\phantom{A }Bm}\hat{e}^m  \;\;-\;\; \frac{1}{2}(u^2 + v)\delta^A_B  
							\end{array} \right),  \\ \\
	&s = \left( \begin{array}{cc}  
								\frac{3}{2}u^2 - \frac{1}{2}v & \hat{e}^C \Alteta_{CB} \\ \\
								\hat{e}^A & -\sigma^{A + n}_{\phantom{{A + n}}B} \;\;-\;\;{S}^{A }_{\phantom{A }Bm}\hat{e}^m \;\;-\;\; \frac{1}{2}(u^2 + v)\delta^A_B  
							\end{array} \right) .
\end{align*}
The para-quaternionic structure is therefore parallel with respect to the Levi-Civita connection. 
\qed

\subsection{The Euclidean $c$-map}
\label{sec:EuclideanCmap}

We now consider the reduction from 4 + 0 to 3 + 0 dimensions. 
In this case we make the choice $\epsilon_1 = 1$ but $\epsilon_2$ may be left arbitrary.

Let us define the real partial co-frame on $P$
\begin{align}
	\cu &= \hat{u}^1 - \epsilon_2 \hat{u}^2 \;,&	\tilde{\cu} &= -\epsilon_2 \hat{u}^1 - \hat{u}^2 \;,	\notag \\
	\cv &=  u^1 + u^2				\;,								&	\tilde{\cv} &= u^1 - u^2  \;, \notag \\
	\ce^{A} &= e^A - \epsilon_2 e^{A + n} \;,	&	\tilde{\ce}^{A} &= e^A + \epsilon_2 e^{A + n}\;, \notag \\
	\cE^{A} &= -\epsilon_2\hat{e}^A + \hat{e}^{A + n} \;,	&	\tilde{\cE}{}^{A} &= \hat{e}^A + \epsilon_2 \hat{e}^{A + n} \;. \label{eq:Recoframe}
\end{align}
\bp
	The following para-complex partial co-frame (and its para-complex conjugate) is related to the above real partial co-frame by replacing the para-complex unit $i_{\epsilon_1}$ with $1$:
	\begin{align}
			\hat{u}^1 + i_{\epsilon_1} \hat{u}^2 &= i_{\epsilon_1} e^{-\phi/2}\left(X^I d\tilde{\zeta}_I - F_I d\zeta^I\right), \notag \\
			u^1 + i_{\epsilon_1} u^2 &= e^{-\phi} \left[ \frac12 de^\phi + i_{\epsilon_1}(d\tilde{\phi} + \frac12 (\zeta^I d\tilde{\zeta}_I - \tilde{\zeta}_I d \zeta^I)) \right], \notag \\
		e^A + i_{\epsilon_1} e^{A + n} &= {P}_I^{\;\;A} dX^I \;, \notag  \\
		\hat{e}^A + i_{\epsilon_1} \hat{e}^{A + n} &= -i_{\epsilon_1} e^{-\phi/2} P_I^{\;\;A} N^{IJ} \left(d\tilde{\zeta}_J - \bar{F}_{JK} d\zeta^K\right) ,
	\end{align}
where $dq^a =: \Pi_I^{\;\;a} dY^I + \bar{\Pi}_I^{\;\;a} d\bar{Y}^I \;$ and $P_I^{\;\;A} := e^{\phi/2} ({\cal P}_a^{\;\;A} + i_{\epsilon_1} {\cal P}_a^{\;\;A + n} ) \Pi_I^{\;\;a}$.
The one-form $v$ may be written as
\[
	v = \frac{1}{2i_{\epsilon_1}} \left( X^IN_{IJ} d\bar{X}^J - dX^I N_{IJ} \bar{X}^J \right) \;.
\]
\label{prop:Euc}
\ep
\vspace{-1.5em}
\begin{proof}
	The proof is analogous to the proof of Proposition \ref{RealCxBasis}. In the para-complex case one must use the  identities ${\textbf e}^A = \frac12(Id + \epsilon_1 i_{\epsilon_1} J^*){\textbf e}^A$ and $4 \Pi_I^{\;\;a} N^{IJ} \bar{\Pi}_J^{\;\;b} = H^{ab} -\epsilon_1 \frac{i_{\epsilon_1}}{2} \Omega^{ab}$.
\end{proof}
The exterior derivatives of the one-forms in the real partial co-frame can be computed from Lemma \ref{lem:extder}  
\begin{align*}
	d \cu &= \Big( -u^1 -\epsilon_2 v\Big) \wedge \cu - \delta_{AB} \tilde{\cE}^{A} \wedge \ce^{B} \;, \\
	d \cv &= \cu \wedge \tilde{\cu} + \cv \wedge \tilde{\cv} + \delta_{AB} \cE^{A}\wedge \tilde{\cE}{}^{B}\;,\\
	d \ce^{A} &= -\left(\sigma^A_{\;\;\;B} -\epsilon_2  \sigma^{A + n}_{\phantom{{A + n}}B}\right) \wedge {\ce}^{B} \;, \\
	d \cE{}^{A} &= \Big( - u^1 + \epsilon_2 v \Big) \wedge {\cE}^{A} - \tilde{\cu} \wedge {\ce}^{A}  - \left(\sigma^A_{\;\;\;B} -\epsilon_2 \sigma^{A + n}_{\phantom{{A + n}}B} \right) \wedge {\cE}^{B}  \\ 
		&\hspace{2em} + \left(-\epsilon_2{S}^A_{\;\;\,BC} +  {S}^{A + n}_{\phantom{{A + n}}BC}\right) \tilde{\ce}^{B} \wedge \tilde{\cE}{}^{C}\;,
 \end{align*}
\vspace{-1em}
\begin{align*}
	d \tilde{\cu} &= \Big( -u^1 + \epsilon_2 v\Big) \wedge \tilde{\cu} - \delta_{AB} {\cE}^{A} \wedge \tilde{\ce}^{B}\;,  \\
	d \tilde{\cv} &= \tilde{\cu} \wedge \cu + \tilde{\cv} \wedge \cv + \delta_{AB} \tilde{\cE}^{A}\wedge \cE^{B} \;,\\
	d \tilde{\ce}^{A} &= -\left( \sigma^A_{\;\;\;B} + \epsilon_2 \sigma^{A + n}_{\phantom{{A + n}}B}\right)\wedge \tilde{\ce}^{B} \;,\\
	d \tilde{\cE}{}^{A} &= \Big( - u^1 - \epsilon_2 v \Big) \wedge \tilde{\cE}{}^{A} - {\cu} \wedge \tilde{\ce}^{A}  - \left( \sigma^A_{\;\;\;B} + \epsilon_2 \sigma^{A + n}_{\phantom{{A + n}}B}\right)\wedge \tilde{\cE}^{B}  \\ 
		&\hspace{2em} + \left( -\epsilon_2 {S}^A_{\;\;\,BC} - {S}^{A + n}_{\phantom{{A + n}}BC}\right) {\ce}^{B}\wedge{\cE}^{C} \;.
\end{align*}

\noindent {\em Proof of theorem \ref{thm:IntCxStr} (c):}
We first consider $J_3$. A basis of the $+1$ eigendistribution of $J_3^*$ is given by ${\cal B}^{+} = (\tilde{\cu}, {\cv}, \tilde{\ce}^{A}, {\cE}{}^{A})$. Each term in the exterior derivative of any element in ${\cal B}^{+}$ contains a one-form in the set ${\cal B}^{+}$. Therefore the distribution is integrable by Frobenius' theorem. A basis of the $-1$ eigendistribution of $J_3^*$ is given by ${\cal B}^{-} = ({\cu}, \tilde{\cv}, {\ce}^{A}, \tilde{\cE}^{A})$, and by the same argument it is also an integrable distribution. Therefore the almost-para-complex structure $J_3$ is integrable.

Let us now consider $J_3'$ and $\tilde{J}$. A basis of the +1 eigendistribution of $J_3'{}^*$ is given by ${\cal B}'{}^{+} = (\tilde{\cu}, \tilde{\cv}, \tilde{\ce}^{A}, {\cE}{}^{A})$ and a basis of the $-1$ eigendistribution by ${\cal B}'{}^{-} = ({\cu}, {\cv}, {\ce}^{A}, \tilde{\cE}{}^{A})$. By the same argument as above $J_3'$ is integrable.
A basis of the $+1$ eigendistribution of $\tilde{J}$ is given by $({\cu}, \cv, {\ce}^{A}, \cE{}^{A})$ and the $-1$ eigendistribution by $(\tilde{\cu}, \tilde\cv, \tilde{\ce}^{A}, \tilde\cE{}^{A})$. Therefore $\tilde{J}$ is integrable if and only if the cubic tensor $C$ vanishes. 
\qed

\vspace{1em}
One may gather together the elements of the real partial co-frame \eqref{eq:Recoframe} into the para-quaternionic vielbein
\begin{equation}
	{\cal U}^{{\cal A} \mu} = \frac{1}{\sqrt{2}} \left( \begin{array}{cccc} 
		\cu & \ce^{A} & -\tilde{\cv} & -\tilde{\cE}{}^{A} \\
		\cv & \cE^{A} & \tilde{\cu} & \tilde{\ce}^{A} \end{array} \right) \;. 
		\label{eq:EuclideanFrame}
\end{equation}
Proposition \ref{prop:Euc} shows that one may replace the complex unit $i$ and holomorphic coordinates in the formal expression for the spatial $c$-map quaternionic vielbein \eqref{eq:SpatialFrame} with the para-complex unit $i_{\epsilon_1}$ and para-holomorphic coordinates in order to obtain the above expression for the para-quaternionic vielbein in the Euclidean $c$-map with $(\epsilon_1,\epsilon_2) = (1,-1)$.

The three endomorphisms $J_1,J_2,J_3$ defined in (\ref{eq:JJJ}) 
correspond to the following three 2-by-2 matrices $J_1^H,J_2^H,J_3^H$:
\begin{equation}
			J^*_\alpha  {\cal U}^{{\cal A} \mu} = \left(J^H_\alpha\right)^{\cal A}_{\;\;\,{\cal B}}{\cal U}^{{\cal B} \mu}  \;, \notag
			\end{equation}
			\begin{equation} 
			(J^H_\alpha ) = 
			\begin{cases}
					(-i\sigma_\alpha) &\mbox{if } \; (\e_1,\e_2)=(-1,-1)\\
					(\tau_\alpha) &\mbox{if } \;  (\e_1,\e_2)=(-1,+1)\\
					(\tau_3,-\tau_2,-\tau_1) &\mbox{if } \;  (\e_1,\e_2)=(+1,+1)\\
					(-\tau_2,-\tau_3,-\tau_1) &\mbox{if } \;  (\e_1,\e_2)=(+1,-1)\;.
			\end{cases}
			\label{eq:JHcmap}
		\end{equation}
Notice that in the last three cases we could have used the same basis $(\tau_\a)$. The reason not
do so was to allow for the unified expression (\ref{eq:JJJ}) for $(J_\alpha)$ in terms of the orthonormal basis.

\vspace{1em}
\noindent {\em Proof of theorem \ref{thm:CmapQK} (c):}
	In the basis (\ref{eq:EuclideanFrame}) the Levi-Civita connection one-form decomposes according to (\ref{eq:QK_connection}), where $p,q,t,s$ are given by
	\begin{align*}
		&p = \left( \begin{array}{cc} 
									\frac12 u^2 + \epsilon_2 \frac12 v  & -\hat{u}^1 + \epsilon_2\hat{u}^2 \\ \\
									-\epsilon_2\hat{u}^1 - \hat{u}^2  & -\frac12 u^2 - \epsilon_2 \frac12 v
								\end{array} \right) , \\ \\
		&q = \left( \begin{array}{cc} 
									-\frac32 u^2 + \epsilon_2 \frac12 v &  \left(\hat{e}^C + \epsilon_2 \hat{e}^{C + n}\right) \delta_{CB} \\ \\
									\epsilon_2 \hat{e}^A - \hat{e}^{A + n}   & \hspace{2em} \sigma^A_{\;\;\;B} - \epsilon_2 \sigma^{A + n}_{\phantom{{A + n}}B} \;\; - \;\; \frac12 (u^2 + \epsilon_2 v) \delta^A_B  
								\end{array} \right) , 
	\end{align*}
	\begin{align*}
		&t = \left( \begin{array}{cc}  
									0 & 0 \\ \\
									0 & (-\epsilon_2 {S}^A_{\;\;\,BC} + {S}^{A + n}_{\phantom{{A + n}}BC})(\hat{e}^C + \epsilon_2  \hat{e}^{C + n}) 
								\end{array} \right) , \\ \\
		&s = \left( \begin{array}{cc}  
									0 & 0 \\ \\
									0 & ({S}^A_{\;\;\,BC} + \epsilon_2 {S}^{A + n}_{\phantom{{A + n}}BC})(\hat{e}^C - \epsilon_2 \hat{e}^{C + n})
								\end{array} \right) .
	\end{align*}
	The Levi-Civita connection is therefore compatible with the para-quaternionic structure.
\qed

\section{$c$-map spaces as fibre bundles with bundle metrics
\label{Sec:fibre_geometry}}

In Section \ref{sec:cmapVar} we have described $c$-map spaces in terms
of the $U_{\epsilon_1}(1)$-principal bundle  
$P = TM \times \mathbbm{R} \rightarrow \bar{N}$ 
equipped with the degenerate symmetric tensor field $g'$, see \re{eq:GP}, which pushes
down to the $\varepsilon$-quaternionic K\"ahler metric $g_{\bar{N}}$. 
We now turn to a complementary
point of view, where $c$-map spaces are locally described as
product manifolds
\[
\bar{N} = \bar{M} \times G \;,
\]
where $\bar{M}$ is the original PS$\epsilon_1$K manifold, 
which is locally a PS$\epsilon_1$K domain, and where
$G$ is the Iwasawa subgroup of $SU(1,n+2)$. The $\varepsilon$-quaternionic
K\"ahler metric can then be written in the form of a `bundle metric'
\begin{equation}
\label{bundle_metric}
g_{\bar{N}} = \bar{g} + g_G(p) \;,
\end{equation}
where $\bar{g}$ is the PS$\epsilon_1$K metric, and where $g_G(p)$ is
a family of left invariant metrics on $G$ which is parametrised by
$p \in \bar{M}$. We will show that for fixed $p\in \bar{M}$ 
the metrics $g_G(p)$ are among the symmetric $\epsilon_1$-K\"ahler
metrics of constant $\epsilon_1$-holomorphic sectional curvature that were
discussed in Sections \ref{sec:symmetric_spaces} and 
\ref{sec:solvable_Lie_group}, and give explicit expressions 
for the metric, $\epsilon_1$-complex structure and 
$\epsilon_1$-K\"ahler potential.

\subsection{The bundle metric}

We start from (\ref{3dLagrangian}), where we re-write
the expression $\left. g_{IJ} \partial_m X^I \partial^m 
\bar{X}^J \right|_D$ in terms
of the physical four-dimensional scalars $z^A$. 
Explicitly, the metric $g_{\bar{N}}$ now takes the form
(\ref{bundle_metric})
where $\bar{g} = \bar{g}_{AB} dz^A d\bar{z}^B$, see \re{eq:gbarg}, 
and where
\begin{eqnarray}
&& g_G(p) = \frac{1}{4} d\phi^2 - \epsilon_1 e^{-2\phi} \left(d\tilde{\phi}
+ \frac{1}{2} (\zeta^I d \tilde{\zeta}_I - \tilde{\zeta}_I d\zeta^I)
\right)^2  \nonumber \\
& & + \frac{\epsilon_2}{2} e^{-\phi} \left( {\cal I}_{IJ}(p) d \zeta^I d\zeta^J
- \epsilon_1 {\cal I}^{IJ}(p) ( d\tilde{\zeta}_I - {\cal R}_{IK}(p) d\zeta^K)
( d\tilde{\zeta}_J - {\cal R}_{JL}(p) d\zeta^L) \right) \;, \nonumber \\ 
\label{bundle_metric_decomp}
\end{eqnarray}
which as indicated depends on $p\in\bar{M}$. Taking $\bar{M}$ to be
a PS$\epsilon_1$K domain, we find that $\bar{N}$ is a product
$\bar{N} = \bar{M} \times L \rightarrow \bar{M}$ with fibre
$L=\mathbbm{R}^{2n+4}$. 
The fields $z^A$ provide holomorphic coordinates on  $\bar{M}$ and 
$(\zeta^I, \tilde{\zeta}_I, \tilde{\phi},\phi)$ are real coordinates
on $L$.

For $\epsilon_1=\epsilon_2=-1$, the metric (\ref{bundle_metric_decomp})
agrees with the expression in
\cite{Cortes:2011aj} upon making the following field redefinitions
\begin{eqnarray}
&& 
\zeta'^I = \frac{1}{\sqrt{2}} \zeta^I \;,\;\;\;
\tilde{\zeta}'_I = \frac{1}{\sqrt{2}} \tilde{\zeta}_I \;,\;\;\; 
\tilde{\phi}' = \tilde{\phi} \;,\;\;\;
\phi' = \frac{1}{2} e^\phi \;,\;\;\;
\nonumber \\
&&{\cal I}'_{IJ} = - {\cal I}_{IJ} \;,\;\;\;
{\cal R}'_{IJ} =- {\cal R}_{IJ} \;, \nonumber
\end{eqnarray}
where the `primed' coordinates are those used in \cite{Cortes:2011aj}.

Following \cite{Cortes:2011aj}
we define the following one-forms:
\begin{eqnarray}
\eta^I = \sqrt{2} e^{-\phi/2} d \zeta^I \;,\;\;\;
\xi_I = \sqrt{2} e^{-\phi/2} \left( d \tilde{\zeta}_I - 
{\cal R}_{IK} d \zeta^K \right) \;, &&  \nonumber \\
\eta^{n+1} = 2 e^{-\phi} \left( d \tilde{\phi} + \frac{1}{2}
(\zeta^I d\tilde{\zeta}_I - \tilde{\zeta}_I d \zeta^I ) \right) \;,\;\;\;
\xi_{n+1} = d\phi \;,  & & \label{eq:XYcoordbasis}
\end{eqnarray}
where $I=0,\ldots, n$. 
n this co-frame the fibre metric is
\[
4 g_G = (\xi_{n+1})^2 - \epsilon_1 (\eta^{n+1})^2 + \epsilon_2
{\cal I}_{IJ} \eta^I \eta^J + \epsilon \,{\cal I}^{IJ} \xi_I \xi_J \;.
\]
where $\epsilon := - \epsilon_1 \epsilon_2$. 
Since ${\cal I}_{IJ}$ is symmetric and invertible, by a linear change of coordinates, we
assume 
\[
{\cal I}_{IJ} = - \eta_{IJ} \;,
\]
where $(\eta_{IJ}) = \mathrm{diag}(-\e_1,1,\ldots ,1)$. Here 
we used the information about the signature of the matrix $({\cal I}_{IJ})$ 
provided in Section \ref{sec:3.1}. 

Thus pointwise with respect to $p\in \bar{M}$ we can bring
the fibre metric to the standard form
\be \label{eq:stnd}
4 g_G = \xi_{n+1}^2 - \epsilon_1 (\eta^{n+1})^2 - \epsilon_2\,
\eta_{IJ} \eta^I \eta^J - \epsilon\, \eta^{IJ} \xi_I \xi_J \;.
\ee
The one-forms are invariant under the following group of 
affine transformations depending on $2n+4$ real parameters
$(v^I, \tilde{v}_I, \alpha, \lambda)$:
\begin{eqnarray}
\zeta^I & \rightarrow & e^{\lambda/2} \zeta^I + v^I\;, \nonumber \\
\tilde{\zeta}_I & \rightarrow & e^{\lambda/2} \tilde{\zeta}_I + \tilde{v}_I\;,
\nonumber \\
\tilde{\phi} & \rightarrow & e^{\lambda} \tilde{\phi} + 
\frac{1}{2} e^{\lambda/2} (\tilde{v}^T \zeta - v^T \tilde{\zeta} ) + \alpha \;,
\nonumber \\
\phi & \rightarrow & \phi + \lambda \;. \label{AffineTransf}
\end{eqnarray}
The Lie group structure underlying the above affine transformations is
\begin{equation}
\label{group_law}
(v, \tilde{v}, \alpha, \lambda ) \cdot
(v', \tilde{v}', \alpha', \lambda' ) =
\end{equation}
\[
(v + e^{\lambda/2} v' ,
\tilde{v} + e^{\lambda/2} \tilde{v}',
\alpha + e^{\lambda} \alpha' + \frac{1}{2} e^{\lambda/2} 
(\tilde{v}^T  v' -v^T \tilde{v}' ), \lambda + \lambda') \;.
\]
Thus $\mathbbm{R}^{2n+4}$, considered as a Lie group $G$ with the
above  multiplication, 
acts on $L=\mathbbm{R}^{2n+4}$ by the affine
transformations \eqref{AffineTransf}. Using this group action we can 
identify the Lie group $G$ with 
the $G$-orbit of the point $(0,0,0,0)$, which is all of $L$. The
affine transformation (\ref{AffineTransf}) is then given by the 
left action of $G$ on itself.

The differentials of the one-forms $(\theta^a)=(\eta^I, \xi_I, 
\eta^{n+1}, \xi_{n+1})$ are linear combinations of wedge products
of the $\theta^a$  with constant coefficients:
\[
d \eta^I = -\frac{1}{2} \xi_{n+1} \wedge \eta^I \;,\;\;\;
d \xi_I = - \frac{1}{2} \xi_{n+1} \wedge \xi_I \;,\;\;\;
d \eta^{n+1} = - \sum_{A=0}^{n+1} \xi_A \wedge \eta^A \;,\;\;\;
d\xi_{n+1}=0 \;.
\]
These coefficients are, in fact,  the structure constants of the Lie algebra $\mathfrak{g}$ of the group $G$. 
This is clear since the forms $(\theta^a)$ can be considered as a basis of the space of left-invariant forms on the group $G$. 

The left-invariant vector fields $(V_a)= (X_I, Y^I, Z_0,D)$
dual
to the one-forms $(\theta^a) = (\eta^I, \xi_I, \eta^{n+1}, \xi_{n+1})$ are 
given explicitly by 
\be \label{eq:frame} 
X_I = \frac{1}{\sqrt{2}} e^{\phi/2} \left( \frac{\partial}{\partial
\zeta^I} + \frac{1}{2} \tilde{\zeta}_I \frac{\partial}{\partial \tilde{\phi}}
\right) \;,\;\;\;
Y^I = \frac{1}{\sqrt{2}} e^{\phi/2} \left( \frac{\partial}{\partial
\tilde{\zeta}_I} - \frac{1}{2} {\zeta}^I \frac{\partial}{\partial \tilde{\phi}}
\right) \;,\;\;\;
\ee 
\[
Z_0 = \frac{1}{2} e^\phi \frac{\partial}{\partial \tilde{\phi}} \;,\;\;\;
D = \frac{\partial}{\partial \phi} \;.
\]
The non-trivial commutators between these vector fields are
\begin{equation}
[Y^I, X_J ] = \delta^I_J Z_0 \;,\;\;\
[D,Y^I] = \frac{1}{2} Y^I \;,\;\;\;
[D,X_I] = \frac{1}{2} X_I \;,\;\;\;
[D, Z_0] = Z_0 \;.
	\label{eq:FibreCommutators}
\end{equation}
This is a solvable Lie algebra, and looking back at Section 
\ref{sec:solvable_Lie_group} we recognise it as 
the Iwasawa Lie algebra $\mathfrak{g}$ of $SU(1,n+2)$.
Therefore  \eqref{group_law} is the group multiplication of the
Iwasawa group $G$. 
Thus $g_G(p)$ is a family of  left-invariant metrics on the fibres $L\simeq G$ 
of the product $\bar{N}=\bar{M} \times G \rightarrow \bar{M}$.

We saw in Section \ref{sec:solvable_Lie_group} that the natural $\varepsilon$-complex structure $J_G$ (setting $\varepsilon=\epsilon_1$) on $\mathfrak{g}$  is given by its action on the basis of vector fields $Z_0,D,X_I,Y^I$ via 
\[
J_GD=-Z_0, \quad J_GZ_0=-\epsilon_1 D, \quad 
J_GY^I=-\tilde{\eta}^{IJ}X_J, \quad J_GX_I=-\epsilon_1\tilde{\eta}_{IJ}Y^J, 
\]
where $\tilde{\eta}^{IJ}= \langle Y^I,Y^J\rangle$.  Identifying $4g_G$ as given in \re{eq:stnd} with the scalar product considered in Section \ref{sec:solvable_Lie_group}
we get $\tilde{\eta}_{IJ}= \e_1\e_2\eta_{IJ}$. Comparing this with \re{eq:Jtildeproj} we see that the almost $\epsilon_1$-complex structure $\tilde{J}$ on $\bar{N}$ 
obtained by projecting the tensor field $\tilde{J}$ from $P$ to $\bar{N}$ can be written as\footnote{Note that $J_G = D \wedge Z_0 + Y^I \wedge X_I$, when
evaluating the endomorphism $J_G$ using the scalar product $\langle \cdot, \cdot \rangle$. In
\re{eq:Jtildeproj} the metric $g'$ is used instead, which restricts to $\frac{1}{4}
\langle \cdot, \cdot \rangle$ on the fibre. This leads to an additional
factor 4 in \re{eq:Jtildeproj}.}
\[
	\tilde{J} = -\epsilon_2 J_{\bar{M}} - J_G \;.
\]
In particular, this shows that $J_G$ is different from  the restriction of the structures $\pm J_3$ and $\pm J_3'$ to the fibres of the projection $\bar{N} = \bar{M} \times G \ra \bar{M}$, with the exception of the case when $\bar{M}$ is a point
and therefore $G$ is $4$-dimensional. In the latter case $J_3'$ coincides with $-J_G$, see the end of the next section for a discussion of 
this special case.

\subsection{K\"ahler potentials for the fibre metrics}

We will now identify the $\varepsilon$-K\"ahler potentials for the
metrics on the fibres $G\simeq L$ of $c$-map spaces, and thus show 
that they 
are among the $\varepsilon$-K\"ahler metrics described in Section
\ref{sec:symmetric_spaces}, where now $\varepsilon=\e_1$. 
We treat all three cases of the $c$-map simultaneously. 
Along the fibre the matrix $\mathcal{N}_{IJ}=\mathcal{R}_{IJ}+i_{\epsilon_1}\mathcal{I}_{IJ}$ is constant.
Let us introduce the $\epsilon_1$-complex coordinates $(C_I,S)$ via
\[
C_I:= \tilde{\zeta}_I -\mathcal{N}_{IJ}\zeta^J, \quad
S:= e^\phi -\epsilon\left(2i_{\epsilon_1}\tilde{\phi}-\frac{1}{2}C_I\mathcal{I}^{IJ}\bar{C}_J\right).
\]
One can show that
\[
\sqrt{2}e^{-\phi/2} dC_I= \xi_I -i_{\epsilon_1}\mathcal{I}_{IJ}\eta^J,
\]
\[
e^{-\phi}\left(dS -\e \, dC_I\mathcal{I}^{IJ}\bar{C}_J\right) =\xi_{n+1}-\e i_{\epsilon_1}\eta^{n+1},
\]
where differentials are restricted to the fibre, whilst the Kaluza-Klein scalar $\phi$ can be expressed in terms of the fields $(C_I,S)$ as
\[
2e^\phi =S+\bar{S}-\epsilon\, C_I\mathcal{I}^{IJ}\bar{C}_J.
\]
Hence, the metric on $G$ is given by
\[
g_G =\frac{\left|dS -\e \, dC_I\mathcal{I}^{IJ}\bar{C}_J\right|^2}{(S+\bar{S}-\epsilon\, C_I\mathcal{I}^{IJ}\bar{C}_J)^2} +\epsilon\, \frac{dC_I\mathcal{I}^{IJ}d\bar{C}_J}{S+\bar{S}-\epsilon\, C_I\mathcal{I}^{IJ}\bar{C}_J}.
\]

In order to compare to the 
parametrisation used in Section \ref{sec:symmetric_spaces},
we introduce the $\epsilon_1$-complex variables $(u,u^I)$ via
\[
2S=\frac{1-u}{1+u}, \quad C_I=i_{\epsilon_1}\frac{\mathcal{I}_{IJ}u^J}{1+u},
\]
in terms of which the metric on $G$ becomes
\[
g_G =\frac{\left|\bar{u} du+\epsilon_2\bar{u}^I\mathcal{I}_{IJ}du^J\right|^2}{\left(1-|u|^2-\epsilon_2 u^I\mathcal{I}_{IJ}\bar{u}^J\right)^2}
+\frac{|du|^2 +\epsilon_2 du^I\mathcal{I}_{IJ}d\bar{u}^J}{1-|u|^2-\epsilon_2 u^I\mathcal{I}_{IJ}\bar{u}^J},
\]
as can be checked by a straightforward but long calculation. A simple calculation shows that the metric $g_G$ is $\epsilon_1$-K\"{a}hler with potential
\[
K=-\log\left(1-|u|^2-\epsilon_2 u^I\mathcal{I}_{IJ}\bar{u}^J\right).
\]

Since ${\cal I}_{IJ}$ is constant, non-degenerate and symmetric, 
we can make a linear coordinate 
transformation to make it
diagonal with entries $\pm 1$. Setting
\[ z^1 = u\;,\;\;\; 
z^a = u^{a-1} \;,\;\;\; a=2, \ldots, N = n+2 \;,\;\;\;
\psym{\eta}_{ab} = \mbox{diag}(1, \epsilon_2 {\cal I}_{IJ}) \;,
\]
we obtain
\[
K_L = - \log \left( 1 - \psym{\eta}_{ab} z^a \bar{z}^b \right) = - \psym{K}\;,\;\;\;
\]
and 
\[
g_L = - \psym{g} = 
(1 - \psym{\eta}_{ab} z^a \bar{z}^b )^{-2}
\left( (1-\psym{\eta}_{ab} z^a \bar{z}^b) \psym{\eta}_{ab} +  \psym{\eta}_{ac}\bar{z}^c
\psym{\eta}_{bd} z^d \right) dz^a d\bar{z}^b \;.
\]
These metrics $g_L$ with $\e_1$-K\"ahler potential $K_L$ are, up to an overall sign, 
among the $\varepsilon$-K\"ahler metrics $\bar{g}$ with  $\varepsilon$-K\"ahler potential $\bar{K}$  
introduced in Section \ref{sec:symmetric_spaces}, where now $\varepsilon= \e_1$. 
Since the choice of the initial special $\varepsilon$-K\"ahler
metric on $\bar{M}$ determines the signature of the metric on $G$, only a
subset of the metrics considered in Section 
\ref{sec:symmetric_spaces} can be realised by the $c$-map. 
In particular the Fubini-Study metric on 
\[
\mathbbm{C}P^{n+2} \simeq U(n+3)/(U(1)\times U(n+2)) \;,
\]
cannot be realised. To obtain the negative of the 
Fubini-Study metric, we would need to take
$\epsilon_1=-1$, and  
$\psym{\eta}_{ab} = - \delta_{ab}$ which gives
\[
K_L  = - \log ( 1 + \delta_{ab} z^a \bar{z}^b) \;,\;\;\;
g_L = - \frac{ (1 + \delta_{cd} z^c \bar{z}^d) \delta_{ab} - \bar{z}^a z^b }{
(1 + \delta_{cd} z^c \bar{z}^d)^2} dz^a d\bar{z}^b \;.
\]
However, since $\mbox{diag}(1, \epsilon_2 {\cal I}_{IJ})
\not=-\delta_{ab}$ it is not possible to obtain this geometry using the
$c$-map, even if we were to allow for four-dimensional 
vector fields with negative 
kinetic energy.

We now discuss the geometries realised by the three $c$-maps.
In order to interpret the resulting signatures in terms of
dimensional reduction, we recall that the coordinates $z^a$
encode the followings fields:
the Kaluza-Klein scalar $\phi$ , 
the dualised Kaluza-Klein vector $\tilde{\phi}$, 
the components of the four-dimensional vector fields along
the direction we reduce over, $\zeta^I$, and
the scalars dual to the three-dimensional vector fields, $\tilde{\zeta}_I$.
The signs in front of the kinetic terms of these fields can be
read off from the three-dimensional Lagrangian (\ref{3dLagrangian}).
Equivalently, they are determined in terms of the signs in the
four-dimensional Lagrangian (\ref{VMLagrangian}) through the following
general properties of dimensional reduction and Hodge dualisation:
(i) spacelike reduction preserves all signs, while timelike reduction
reverses the sign for the Kaluza-Klein vector and for the scalars 
obtained by reducing vector fields; (ii) dualisation of three-dimensional
vector fields preserves the sign in Lorenzian signature and reverses
it in Euclidean signature. Now we list the cases which can be realised by the 
different versions of the $c$-map.
\begin{enumerate}
\item
If we take $\epsilon_1=-1$ and
$\psym{\eta}_{ab} = \delta_{ab}$, we obtain
\[
K_L = - \log ( 1 - \delta_{ab} z^a \bar{z}^b) \;,\;\;\;
g_L = \frac{ (1 - \delta_{cd} z^c \bar{z}^d) \delta_{ab} + \bar{z}_a z_b }{
(1 - \delta_{cd} z^c \bar{z}^d)^2} dz^a d\bar{z}^b\;,
\]
where $z_a =  z^a$. 
This is a positive definite K\"ahler metric on the complex 
hyperbolic space 
\[
\mathbbm{C}H^{n+2} = U(n+2,1) / (U(n+2) \times U(1)) \;,
\] 
which has constant 
holomorphic sectional curvature $-1$. It is realised, up to a factor $4$,  as the 
fibre geometry $g_G=\frac{1}{4}g_L$ of the spatial $c$-map, $\epsilon_1=\epsilon_2=-1$, which has
holomorphic sectional curvature $-4$.
Here we assume that we start with a four-dimensional theory
of vector multiplets with positive definite kinetic terms. This
implies that ${\cal I}_{IJ}$ is negative definite. Dimensional reduction
over a spacelike direction then results in a three-dimensional theory
with positive definite kinetic terms.
\item
Next we take $\epsilon_1=-1$ and $(\psym{\eta}_{ab}) = \mbox{diag}(\mathbbm{1}_{\ell}
\;, -\mathbbm{1}_{k-1})$.
Then we obtain
\begin{equation}
\label{g_L}
K_L = - \log ( 1 - \psym{\eta}_{ab} z^a \bar{z}^b) \;,\;\;\;
g_L = \frac{ (1 - \psym{\eta}_{cd} z^c \bar{z}^d) \psym{\eta}_{ab} + \psym{\eta}_{ac}\bar{z}^c 
\eta_{bd} z^d }{
(1 - \psym{\eta}_{cd} z^c \bar{z}^d)^2} dz^a d\bar{z}^b\;,
\end{equation}
which are
pseudo-K\"ahler metrics of complex signature $(\ell ,k-1)$ 
with constant holomorphic
sectional curvature $-1$ on the indefinite complex hyperbolic spaces
$\mathbbm{C}H^{(\ell ,k-1)}$. This case is realised by the temporal $c$-map,
$\epsilon_1=-1, \epsilon_2 = 1$ with the specific value $\ell=1$.
Here  we use again
that ${\cal I}_{IJ}$ is negative definite, if we start with a theory
in Minkowski space-time signature with positive kinetic energy. 
Upon timelike reduction all scalars resulting from the four-dimensional
vector fields have a negative sign, while $\phi$ and $\tilde{\phi}$
have a positive sign, resulting in kinetic terms with signature
$(2,2n+2)$, or complex signature $(1,n+1)$, which corresponds to $\ell =1$.
We thus obtain the following indefinite version of the complex hyperbolic space
\[
\mathbbm{C}H^{(1,n+1)} \simeq U(1,n+2)/(U(1,n+1) \times U(1)) \;,
\]
with complex signature $(1,n+1)$. 
\item
Finally we take $\epsilon_1=1$ and obtain the same expressions as in 
(\ref{g_L}) with complex fields replaced by para-complex fields. 
This is a para-K\"ahler metric of constant para-holomorphic
sectional curvature $-1$ on the para-complex hyperbolic space
\[
CH^{n+2} \simeq SL(n+3)/S(GL(1) \times GL(n+2)) \;.
\]
The (real) signature is $(n+2,n+2)$ 
irrespective of the signature of $\psym{\eta}_{ab}$. 
This geometry is realised as a fibre geometry 
for the Euclidean $c$-map, $\epsilon_1=1$, $\epsilon_2=\pm1$.
The result is independent of the signature of ${\cal I}_{IJ}$,
and hence of $\psym{\eta}_{ab}$, since the metric is para-Hermitian
and has split-signature. In terms of dimensional reduction,
$\phi$ and $\tilde{\phi}$, and $\zeta^I$ and $\tilde{\zeta}_I$
have opposite signs irrespective of the signs in the four-dimensional
Lagrangian. From \cite{Cortes:2009cs} we know that if we obtain
the Euclidean theory by reduction of five-dimensional supergravity
with vector multiplets over time, then ${\cal I}_{IJ}$ has signature
$(1,n)$, which reflects the fact that the Kaluza-Klein vector of the 5d/4d reduction
has a negative kinetic term.

\end{enumerate}

We remark that by matching the $\varepsilon$-K\"ahler potentials 
obtained by the $c$-map to those found in Section \ref{sec:symmetric_spaces}
we have now proved that the solvable Lie groups presented in 
Section \ref{sec:solvable_Lie_group} do indeed provide local realisations
of the symmetric spaces discussed in Section \ref{sec:symmetric_spaces}.
We further remark that for the non-compact symmetric spaces of indefinite
signature, that is for $\mathbbm{C}H^{(l,k-1)}$ and $CH^{n+2}$,  
the Iwasawa subgroup does not act transitively, though one can find an Iwasawa subgroup which acts with open orbit. In these cases the fibre cannot be
identified globally with the corresponding symmetric space, since 
the fibre has trivial topology, while the symmetric space has non-trivial
topology.  This is different for $\mathbbm{C}H^{n+2}$, where the Iwasawa 
group acts transitively, so that the fibre is globally isometric
to $U(1,n+2)/U(1)\times U(n+2)$.

The simplest examples of $c$-map spaces are obtained by taking
the initial special $\varepsilon$-K\"ahler manifold to be trivial,
$\bar{M}=\{ \mbox{pt} \}$. This corresponds to starting with
pure supergravity, and gives rise to a single hypermultiplet, 
often referred to as the universal hypermultiplet. The corresponding
real four-dimensional
$\varepsilon$-quaternionic K\"ahler manifolds\footnote{The definition we gave
for $\varepsilon$-quaternionic K\"ahler spaces is only valid if the real
dimension is larger than four. As is well known, for the quaternionic case,
in four dimensions this definition is not satisfactory, as it only
implies orientability. One then 
takes as a definition that in addition the manifold is Einstein and that 
the curvature tensor is invariant under
the quaternionic structure,  
a property that in higher dimensions follows from the definition
in terms of holonomy.}
are rather special as
they only consist of the fibre, and are therefore locally symmetric spaces
which are simultaneously $\varepsilon$-K\"ahler and $\varepsilon$-quaternionic
K\"ahler. Here the $\varepsilon$-complex structure $J_G$ compatible with
the $\varepsilon$-K\"ahler metric coincides with the additional integrable $\varepsilon$-complex structure $-J_3'$, which is not part of the
$\varepsilon$-quaternionic structure. The three different $c$-maps
give rise to three different universal hypermultiplets, for which we discuss below the 
corresponding globally symmetric space.
\begin{enumerate}
\item
$\epsilon_1 = -1, \epsilon_2 = -1$. The symmetric space
\[
\mathbbm{C}H^2 = U(2,1) / U(2) \times U(1)\;,
\]
is simultaneously K\"ahler and quaternionic K\"ahler. 
This space is the simplest hypermultiplet geometry occurring in 
supergravity and appears naturally in various constructions. In 
particular the classical moduli spaces of M-theory and type-II
superstrings on Calabi-Yau threefolds contain this space as a
subspace, with the scalar $\phi$ being related to the Calabi-Yau
volume and the type-II dilaton, respectively. 
\item
$\epsilon_1=-1, \epsilon_2=1$. The symmetric space
\[
\mathbbm{C}H^{(1,1)} \simeq U(2,1) / (U(1,1) \times U(1))\;,
\]
is simultaneously pseudo-K\"ahler and para-quaternionic K\"ahler.
It occurs in the timelike reduction of pure four-dimensional 
${\cal N}=2$ supergravity \cite{Gunaydin:2005mx}.
\item
$\epsilon_1=1, \epsilon_2=-1$. The symmetric space
\[
CH^2 \simeq SL(3)/S(GL(1)\times GL(2))\;,
\]
is simultaneously para-K\"ahler and para-quaternionic K\"ahler.
In \cite{Theis:2002er} it was observed that this geometry is realised
by reduction of pure Euclidean supergravity \cite{Theis:2001ef}, 
and by dualising the double-tensor multiplet in Euclidean signature. 
\end{enumerate}

\subsection*{Acknowledgements}

The work of T.M. is supported in part by the STFC consolidated grant
ST/L000431/1. He thanks the Department of Mathematics of the University 
of Hamburg and the SFB 676 for hospitality and support during several
stages of this work. The work of V.C.\ and O.V.\ is supported by the German Science
Foundation (DFG) under the Collaborative Research Center (SFB) 676 
``Particles, Strings and the Early Universe.''
The work of P.D.\ is supported by  National Research Foundation of Korea grants 2005-0093843, 2010-220-C00003 and 2012K2A1A9055280. We would like to thank Malte Dyckmanns for useful discussions.

\providecommand{\href}[2]{#2}\begingroup\raggedright\endgroup

\end{document}